\documentclass[12pt,preprint]{aastex}

\usepackage{amsmath}
\usepackage{appendix}
\usepackage{epsfig}
\usepackage{subcaption}
\usepackage{caption}
\usepackage{physics}
\usepackage{comment}
\usepackage{color}
\usepackage{xcolor}
\usepackage{mathrsfs}
\usepackage{textcomp}
\usepackage{gensymb}

\usepackage{lineno}

\usepackage{epsfig}

\usepackage{lineno}
\usepackage{verbatim}

\def\greenphoton{\dot N_\epsilon^\text{G}}

\newcommand{\vel}{\upsilon}

\def\green{f_{_\text{G}}}

\newcommand{\gapprox}{\lower.4ex\hbox{$\;\buildrel >\over{\scriptstyle\sim}\;$}}
\newcommand{\lapprox}{\lower.4ex\hbox{$\;\buildrel <\over{\scriptstyle\sim}\;$}}
\newcommand{\begeq}{\begin{equation}}
\newcommand{\fineq}{\end{equation}}
\newcommand{\msun}{M_\odot} 
\newcommand{\epsmin}{\epsilon_\text{min}}
\newcommand{\epsmax}{\epsilon_\text{max}}
\newcommand{\epsilonabs}{\epsilon_\text{abs}}
\newcommand{\chiabs}{\chi_\text{abs}}

\newcommand{\sigmaT}{\sigma_{_\text{T}}}
\def\sig{\sigma_{_\text{T}}}
\def\sigpar{\sigma_{_{||}}}
\def\sigperp{\sigma_\perp}
\def\sigbar{\overline\sigma}

\def\green{f_{_\text{G}}}

\newcommand{\xlum}{L_\text{X}}
\newcommand{\LE}{L_\text{E}}
\newcommand{\thetamax}{\Theta_1}
\newcommand{\thetamin}{\Theta_2}

\def\sigmapar{\sigma_{||}}

\shorttitle{SPECTRAL FORMATION IN X-RAY PULSARS}
\shortauthors{Gibson \& Becker}

\begin{document}

\title{A New Relativistic Model for Spectral Formation in Accretion-Powered X-ray Pulsars: Pulse Profiles and Phase-Averaged Spectra}

\author{Ethan J. Gibson\altaffilmark{1} and Peter A. Becker\altaffilmark{2}}

\affil{Department of Physics and Astronomy,
George Mason University,
Fairfax, VA 22030-4444, USA}

\vfil

\altaffiltext{1}{egibso3@gmu.edu}
\altaffiltext{2}{pbecker@gmu.edu}

\begin{abstract}

We develop a new analytical model describing the radiative and dynamical structure of an accretion-powered X-ray pulsar, including relativistic effects and a detailed representation of the rotational and magnetic geometry of the neutron star and the two accretion columns. The model provides for the first time a simultaneous calculation of both the phase-averaged spectrum and the pulse profile for an accretion-powered X-ray pulsar. The X-ray continuum spectrum is calculated using the analytical model of Becker \& Wolff (2022), which assumes a conical accretion column geometry. The trajectory of the radiation escaping from the two columns is tracked through the curved spacetime using the Schwarzschild metric. The angular distribution of the radiation escaping from the surfaces of the columns (the beaming pattern) is represented using a set of ``laser-like'' emission directions, with associated amplitudes, called weight coefficients, that each contribute ``sub-profiles'' to the observed pulse profile. The sub-profiles provide basis functions that are used to fit the observed pulse profile. This yields a set of weight coefficients that determine the beaming pattern of the emission from the accretion column. We use the new model to analyze {\sl NuSTAR} data for Her X-1, allowing the determination of the temperature, accretion rate, and magnetic field strength, as well as the rotational inclination angle and the latitudes of the two magnetic poles. The method also yields the beaming pattern of the emission, hence providing for the first time a self-consistent phenomenological description of the physical and radiative structures of the two accretion columns.
\end{abstract}

\section*{\sl Accepted for publication in the Astrophysical Journal}



\section{INTRODUCTION}
\label{sec:intro}

Accretion-powered X-ray pulsars (XRPs) are fascinating natural laboratories for probing the physics of highly magnetized plasmas, subject to extreme gravitational and radiation fields. These systems consist of a neutron star orbiting in a binary configuration with a main sequence companion. The neutron star siphons matter from the atmosphere of the companion via Roche lobe overflow in the case of low-mass X-ray binaries, or via stellar winds in the case of high-mass X-ray binaries \citep[see][]{verbunt1982}. The fully-ionized plasma is subsequently accelerated to quasi-relativistic velocities as it is guided by the strong ($B \sim 10^{12}\,$G) magnetic field onto the surface of the neutron star, forming luminous accretion columns over both magnetic poles. Due to the conversion of gravitational potential energy into thermal energy, the plasma within the accretion columns can reach temperatures $T \sim 10^{7-8}\,$K \citep[for a review, see][]{MushtukovEtal2023}.

The physics governing the structure and properties of the plasma inside a pulsar accretion column is quite complex and involves a number of processes operating in an environment characterized by intense radiation and magnetic fields and strong gravity. Collisions between electrons and ions inside the column result in the generation of radiation due to bremsstrahlung and cyclotron emission. These photons are reprocessed via thermal and bulk Comptonization occurring in the hot infalling plasma before eventually escaping through the walls and top of the accretion column to form the emergent X-ray spectrum \citep{BeckerandWolff2007}. Additionally, at the bottom of the column, where the density is high enough to allow for the establishment of full thermodynamic equilibrium, blackbody radiation with temperature $\sim 10^{7}\,$K is produced, which is subsequently reprocessed via electron scattering before escaping to contribute to the observed X-ray spectrum.

\subsection{Critical Luminosity}

The X-rays observed from accretion-powered XRPs are generated by the conversion of gravitational potential energy into radiation. The observed X-ray luminosity, $L_\text{X}$, is therefore given by the fundamental expression
\begin{equation}
L_\text{X} = \frac{G M_\text{star} \dot M_\text{obs}}{R_\text{star}}  \ ,
\label{eq:mdotFirst}
\end{equation}
where $M_\text{star}$ and $R_\text{star}$ denote the mass and radius of the neutron star, respectively,  and $\dot M_\text{obs}$ represents the accretion rate in the observer's frame. The accretion flows in XRPs are channeled toward the magnetic poles of the neutron star by the strong magnetic field, but the nature of the flow also depends on the X-ray luminosity. 
For low-luminosity sources, such as X Per, with $\xlum \sim 10^{34}\,{\rm erg\,s}^{-1}$, radiation pressure is insufficient to halt the flow before the plasma collides with the stellar surface. On the other hand, in high-luminosity sources, such as Her X-1 and LMC X-4, with $\xlum \sim 10^{37-38}\,{\rm erg\,s}^{-1}$, the dynamics are more complicated, due to the presence of a radiation-dominated, radiative shock within the accretion column, which is a crucial consideration when building a suitable formalism to describe the formation of the emergent radiation spectrum in luminous sources \citep{BaskoandSunyaev1975,BaskoandSunyaev1976,Davidson1973}.

In situations involving spherically-symmetric accretion, the relevant luminosity characterizing the dynamical role of radiation pressure is the spherical Eddington limit, which for fully ionized hydrogen is given by
\begin{equation}
    \LE \equiv \frac{4 \pi G M_\text{star} m_p c}{\sigmaT} = 1.26 \times 10^{38} \left(\frac{M_\text{star}}{\msun}\right) \, {\rm erg \ s}^{-1} \ ,
\end{equation}
where $m_p$ is the proton mass, $c$ is the speed of light, $G$ is the gravitational constant, $\sigmaT$ denotes the Thomson cross section, and $M_\text{star}$ is the stellar mass. In the case of the X-ray pulsar accretion column, the flow is of course not spherically symmetric, and in addition, the scattering cross section can be significantly below the Thomson value for the photons propagating in the outward direction, which is aligned with the magnetic field. We therefore find that the critical luminosity for X-ray pulsars is given by \citep{Becker_etal2012}
\begin{equation}
    L_\text{crit} = L_\text{E}
\, \frac{\pi r_\text{col}^2}{4 \pi R_\text{star}^2} \, \frac{\sigmaT}{\sigmapar} \ ,
\label{critlum}
\end{equation}
where $r_\text{col}$ is the radius of the column at the stellar surface, $R_\text{star}$ is the radius of the neutron star, and $\sigma_{||}$ is the electron scattering cross section for photons whose propagation direction is aligned with the magnetic field. Here, we focus on the treatment of high-luminosity sources (with $\xlum>L_\text{crit}$) such as Her X-1, LMC X-4, and Cen X-3.

\subsection{Previous Models}

The category (1) models comprise those that focus on the application of fundamental quantum electrodynamics (QED) calculations for the computation of the exact energy and angular dependence of the electron scattering cross section in a strong magnetic field, with applications usually confined to plasmas with a static slab or static cylindrical geometry. These models do not include the dynamical effects associated with the accretion flow, and they also do not treat the rotation of the star or the geometry of the magnetic poles. Category (1) involves some of the earliest attempts to fit spectral data for pulsating neutron stars, which emerged 12 years after their discovery. The earliest theoretical studies of XRP emission involved application of QED within a highly magnetized static plasma \citep[see][]{Ventura1979,Nagel1981b} to study scattering cross sections. Incorporating simple ``hot spot'' geometries, these early works calculated an intensity distribution emitted from a static ``spot'' on the star. This led to early models for the radiation distributions emitted from accretion-powered X-ray pulsars, such as those produced by \citet{MeszarosandNagel1985a,MeszarosandNagel1985b}. These models are able to roughly reproduce the general morphology of observed X-ray pulsar emission spectra. However, they are unable to fit the observational data for any specific source. The inability to fit the observational data is due to the failure to include dynamical effects stemming from the quasi-relativistic plasma motion, and also the lack of treatment of thermal Comptonization, as discussed by \citet{BeckerandWolff2007}.

The models in category (2) incorporate detailed treatments of the effects of general relativity (GR) on the propagation of radiation emitted from ``hot spots'' or ``hot columns'' on the stellar surface. These models do account for the rotational and magnetic geometry of the neutron star, but they do not include detailed treatments of the hydrodynamics of the accretion flows, and typically assume that the emitting plasma is in a static configuration. Furthermore, these models generally do not incorporate detailed calculations of the continuum spectrum generated in the frame of the star, and instead make the simplistic assumption that the escaping spectrum has a blackbody shape. Class (2) emission models for XRPs incorporate GR effects into analysis of XRP spectra. The work of \citet[][hereafter {\tt RM88}]{RiffertandMeszaros1988} represented a landmark development in the field because it facilitated the tracking of rays emitted at a fixed angle from a conical accretion surface as they propagated through the Schwarzschild spacetime around the neutron star to reach a distant observer. The general relativistic approach of {\tt RM88} stimulated subsequent work by \citet{Kraus_etal_1995,Kraus_etal_1996,Kraus_etal_2003}, \citet{Leahy_1991}, \citet{Blum_Kraus_2000}, \citet{Sasaki_etal_2010}, and \citet{choudhury_etal_2024}. These papers calculated pulse profiles from accretion-powered X-ray pulsars, by focusing on modeling the emission generated from hot columns and hot spots or hot rings, placed asymmetrically on the star. While these models were able to produce pulse profiles in rough agreement with the data, they did not include simultaneous calculations of the phase-averaged spectrum, nor do they treat the dynamics of the accreting gas. Hence these models do not provide a comprehensive description of all the essential physical processes occurring in the accretion flows powering X-ray pulsars such as Her X-1.

The category (3) models focus on providing a detailed treatment of the radiation hydrodynamics of the infalling plasma, coupled with a self-consistent calculation of the continuum X-ray spectrum emitted from the walls and top of the accretion column. In order to render the mathematical derivations feasible, this category of models generally employs an approximate treatment of the electron scattering cross section. Furthermore, these models provide a calculation of the X-ray continuum spectrum in the frame of the neutron star, and do not account for the effects of the rotational and magnetic geometrical configurations. Works such as \citet{BeckerandWolff2007} and \citet[][hereafter {\tt BW22}]{BeckerandWolff2022} fall into class (3), as they provide an extremely detailed analysis of the radiative transport equation in geometries which are reasonable approximations of the accretion columns formed by the magnetic poles. These models, most notably {\tt BW22}, analyze the Comptonization of photons within the accretion flow of the infalling matter.

By solving the transport equation describing the Comptonization of bremsstrahlung, cyclotron, and blackbody seed photons in a conical accretion column, the model of {\tt BW22} is able to reproduce the observed phase-averaged X-ray spectrum for Her X-1 with great success (see Figure~\ref{fig:BW22HerX1_spect}a). However, the {\tt BW22} model and other class (3) models do not include the gravitational lensing and redshifting effects of GR, or the rotational and magnetic geometry of the neutron star. Hence these models cannot be used to compute pulse profiles by themselves, but they can be used to compute the continuum emission components in the frame of the neutron star, which can then be cast in a proper geometric framework and linked with a relativistic formalism to simulate the propagation of the radiation through the curved spacetime. The resulting unified model can then be used to calculate pulse profiles and true phase-averaged spectra measured in the frame of a distant observer. The development of such a generalized model is the main goal of the present paper.

The category (4) models focus on investigating the multidimensional structure of X-ray pulsar accretion columns. Examples include \citet{Gornostaev2021}, \citet{Zhang_2022}, and \citet{Zhang_2025}, who simulate the column structure in multiple spatial dimensions based on detailed magnetohydrodynamic (MHD) simulations. These models include detailed analysis of the radiation-dominated shock, which is also present in our model. However, they do not compute either pulse profiles or phase-averaged X-ray spectra that can be directly compared with observational data. Category (4) models also include sophisticated time-dependent simulations that investigate instabilities in the accretion flow onto the neutron star. The goal of these models is to investigate the formation of the quasi-periodic oscillations (QPO) observed from some sources such as Cen X-3 \citep{JerniganEtal2000} and V0332+53 \citep{QuEtal2005}. For example, \citet{Abolmasov2023_shock_dynamics_column} and \citet{Mushtukov2024_magnetospheric_flows_I} perform MHD simulations in multiple spatial dimensions to simulate the variability of the accretion rate distribution within the column. Furthermore, \citet{Zhang2023_split_monopole_columns} investigate the variability of the location of the radiation-dominated shock in ultraluminous sources. However, these models do not calculate pulse profiles or phase-averaged spectra and hence it is difficult to compare them with observational data.

The various models described above each focuses on a particular part of the overall physical problem of the relationship between the structure of an X-ray pulsar accretion flow and the production of the emergent X-ray spectrum. However, only the category (3) models provide a detailed calculation of the X-ray continuum spectrum based on a prescribed dynamical structure. Hence these are the only models that make direct contact with the observational data. In the present paper, we will therefore focus on exploiting the most advanced category (3) model, developed by {\tt BW22}, to simultaneously calculate both phase-averaged X-ray spectra and pulse profiles that can be compared with the observational data for a wide variety of sources spanning a broad range of luminosity.

\subsection{Development of Generalized Relativistic Model}

Our goal in this paper is to develop a new model that combines the detailed dynamical and continuum calculations of models in category (3) with the comprehensive treatment of the effects of GR and the rotational and magnetic geometries, which comprise the main strengths of the category (2) models. To date, no model has combined the strengths of both of these classes; that is, no model has provided a detailed treatment of the flow dynamics along with the thermal and bulk Comptonization of the seed photons, while also tracking the emergent emission spectrum along null geodesics toward the distant detector. The central goal of this new formalism is to model the X-ray emission spectra and pulse profiles of XRPs by considering these fundamental mechanisms, and using GR to account for the gravitational lensing and redshifting of the emission spectrum as it propagates through space. In our work, we will model the emission from neutron stars with two conical accretion columns. The assumption of a conical geometry is a reasonable approximation of the dipole magnetic field in the vicinity of the stellar magnetic poles, where the accretion columns form. Following the work of \citet{Blum_Kraus_2000} and \cite{Leahy_1991}, we will also allow for the possibility that the two columns do not form an exactly dipolar pair. Indeed, in the case of certain millisecond pulsars, such as PSR J0030+0451 and PSR J0740+6620, analysis of {\sl NICER} data suggests that the neutron star may have two hot spots occupying the same hemisphere \citet{Petri_etal_2023,Miller_etal_2019,Miller_etal_2021}.

The new model developed here provides a number of crucial enhancements to previous work. Namely, our new model: (1) employs a true continuum model that is dependent on the radius measured from the center of the star and also on the photon energy; (2) uses GR to identify the null geodesics that reach the distant detector; (3) uses GR to account for potential redshifting effects on certain parameters from previous continuum models; (4) preserves the dependence of the emission model on the geometry parameters of the star, namely the rotational inclination of the star, the rotational latitudes of the accretion columns with respect to the spin axis, and the separation angle of the accretion columns; and (5) allows for computation of pulse profiles by tracking the stellar emission over the course of the rotational phase.

The remainder of the paper is organized as follows. In Section~\ref{sec:continuummodel}, we will briefly review the dynamical and continuum model of {\tt BW22}, and how we compute the spectrum in the frame of the star using reprocessed blackbody, cyclotron, and bremsstrahlung emission. In Section~\ref{sec:RelForm}, we discuss the coordinate system and calculation of the flux in the frame of the detector using relativistic coordinate transformations. In Section~\ref{sec:UniMod}, we discuss the unitary emission model, which essentially expands the angular dependence of the intensity field in the local stellar frame using quasi-orthogonal basis functions that are connected with specific ``laser-like'' beams of intensity radiating in specific directions. We also show how the relativistic formalism introduced in Section~\ref{sec:RelForm} can be used to transform the flux produced by the unitary emission model from the frame of the star into the frame of the distant observer. In Section~\ref{sec:pulseProfile}, we develop the mathematical framework for computing the pulse profile and the phase-averaged spectrum based on the flux formalism, and in Section~\ref{sec:results} we use the new model to analyze the pulse profile and the phase-averaged spectrum for Her X-1 using {\sl NuSTAR} data. The fitting procedure allows the determination of the temperature, accretion rate, magnetic field strength, and magnetic and rotational geometries of the source, in addition to the beaming pattern of the emitted radiation field. In Section~\ref{sec:conc} we present our main conclusions and compare our findings with previous work.

\section{THE CONTINUUM MODEL}
\label{sec:continuummodel}

\begin{figure}[htbp]
    \centering
\includegraphics[width=0.8\linewidth]{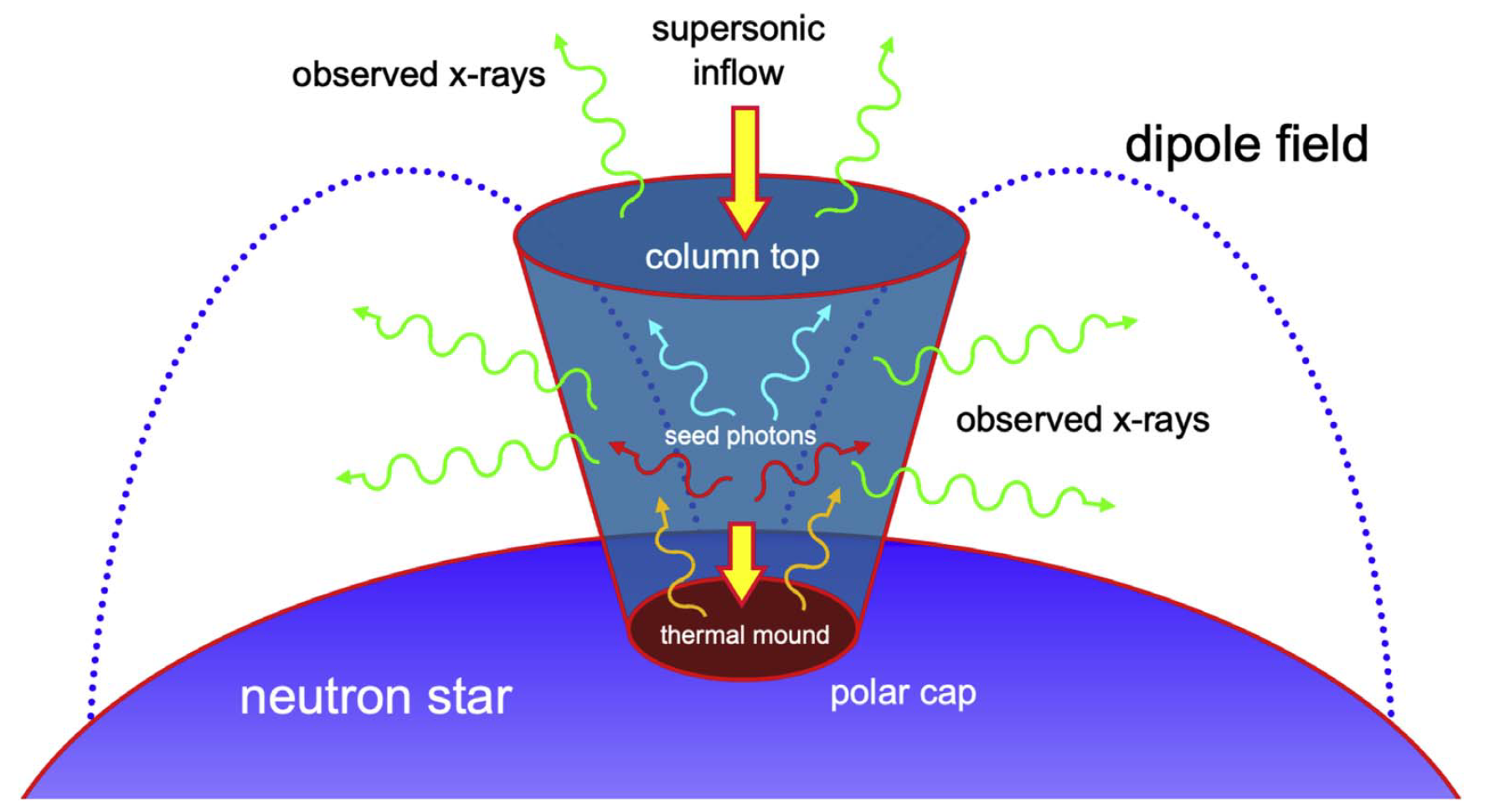}
    \caption{Schematic representation of the continuum model described in \citet{BeckerandWolff2022}. Seed photons produced via blackbody, cyclotron, and bremsstrahlung processes are injected into the conical accretion column, are reprocessed via electron scattering, and escape through the column walls and top to form the observed X-ray spectrum.}
    \label{fig:BW22Model}
\end{figure}

The {\tt BW22} model provides a detailed description of the velocity distribution in an X-ray pulsar accretion column, as well as the emergent X-ray spectrum resulting from the reprocessing of bremsstrahlung (free-free), cyclotron, and blackbody seed photons. Fundamentally, the reprocessing occurs via the thermal and bulk Comptonization of the seed photons, which results from repeated collisions between the photons and the hot electrons in the accreting plasma. The geometry of the model is depicted in Figure~\ref{fig:BW22Model}. In the case of the highly luminous X-ray pulsars treated here, such as Her X-1, LMC X-4, the effect of radiation pressure is so strong that the flow velocity is expected to vanish at the surface of the neutron star, where the accreting gas merges with the star's crust.

The conical geometry of the {\tt BW22} continuum model is depicted in Figure~\ref{fig:BW22Model}. Following {\tt BW22}, we assume here that the quasi-dipolar stellar magnetic field lines can be reasonably approximated as radial lines near the magnetic poles on the stellar surface. Since the fully-ionized plasma is constrained to follow the magnetic field lines, we will therefore model an accretion column which obeys a conical geometry. As seen in Figure~\ref{fig:BW22Model}, photons escape both from the wall of the column and the top of the column. In the development of our relativistic model, we will distinguish between the wall and the top as separate sources of radiation, as both obey different equations describing their boundary conditions and emission spectra in the conical geometry.

\begin{figure}[htbp]
    \centering
    \includegraphics[width=1.0\linewidth]{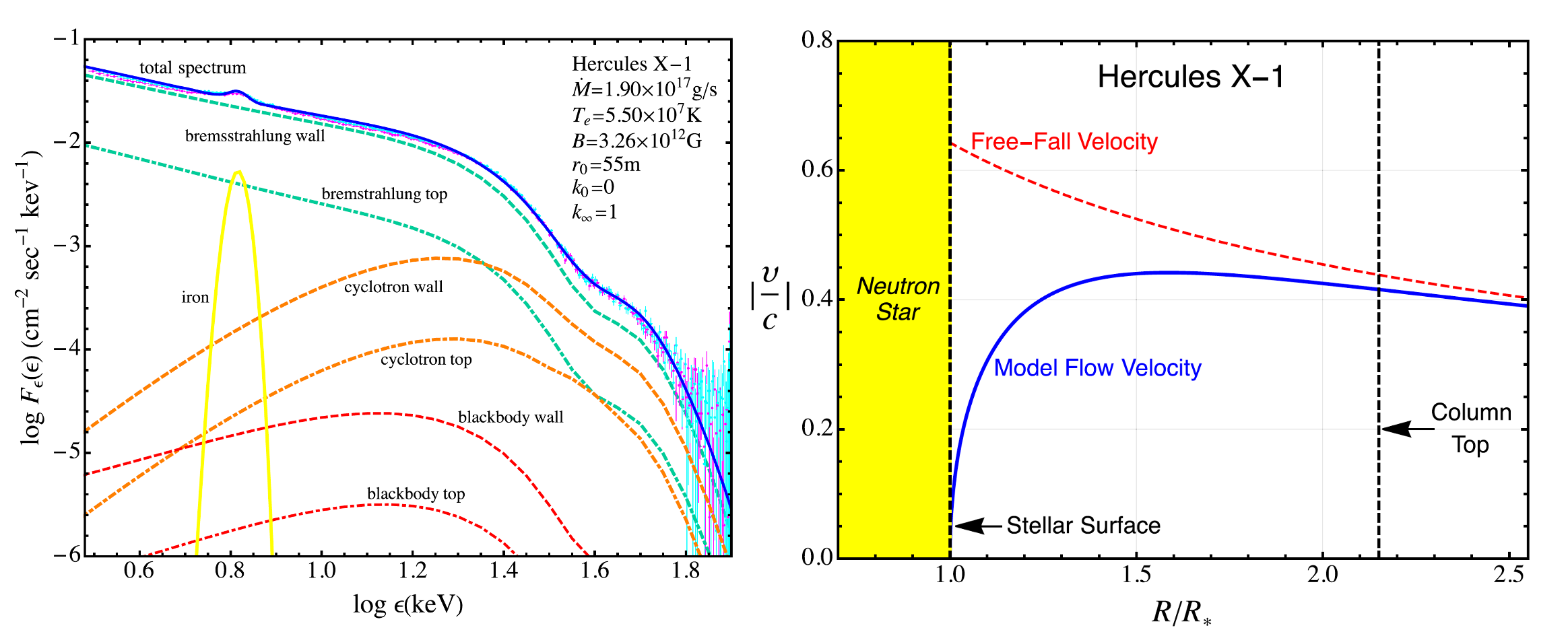}
    \caption{The phase-averaged theoretical X-ray spectrum of Her X-1 computed using the {\tt BW22} model is plotted in panel (a). Note that the spectrum is dominated by reprocessed bremsstrahlung emission escaping through the column walls. In panel (b), the dimensionless flow velocity, $|\vel|/c$, for the Her X-1 accretion column, evaluated using the {\tt BW22} model, is plotted as a function of the dimensionless radius, $y=R/R_\text{star}$ (blue line). For comparison, the red dashed line indicates the local Newtonian free-fall velocity profile.}
    \label{fig:BW22HerX1_spect}
\end{figure}

{\tt BW22} approximated the phase-averaged X-ray spectrum of an accretion-powered pulsar by simply adding together the emission components generated by the walls and top of a single conical accretion column, without regard to the rotational or magnetic geometry of the neutron star. The resulting theoretical spectral fit in the case of Her X-1, depicted in Figure~\ref{fig:BW22HerX1_spect}a, provides good agreement with the observational data for this source. The corresponding dynamical structure of the Her X-1 accretion flow is plotted in Figure~\ref{fig:BW22HerX1_spect}b. In the new model developed here, we will diverge from the basic {\tt BW22} model by incorporating the emission from two separate accretion columns, which are expected to be located in a roughly dipolar configuration on the stellar surface. Our model will also employ an explicit treatment of the rotational and magnetic geometry of the neutron star, as discussed in Section~\ref{sec:parameterspacesearch}.

\begin{figure}[htbp]
    \centering
    \includegraphics[width=0.7\linewidth]{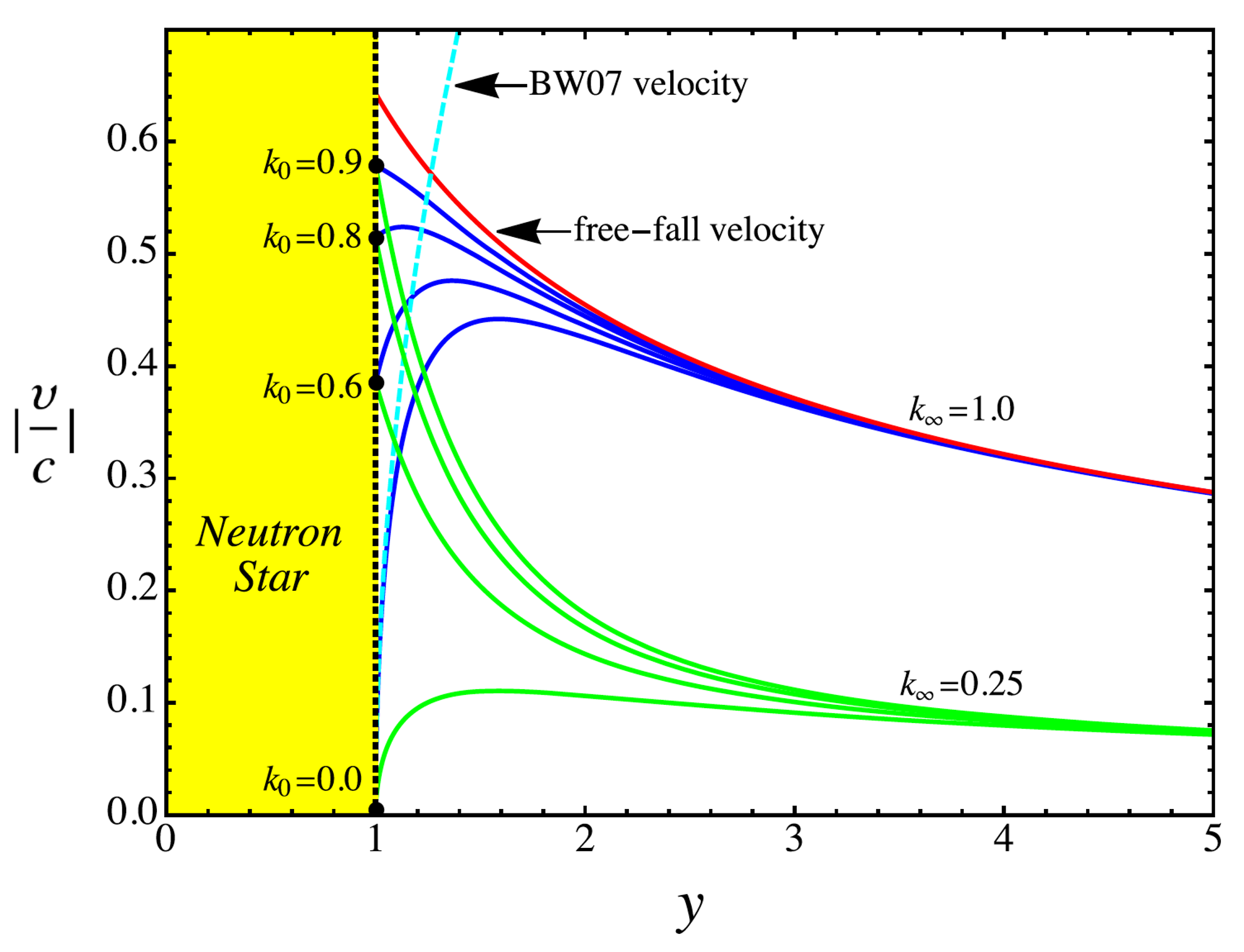}
    \caption{Dimensionless flow velocity, $u=|\vel|/c$, of the accreting plasma near the surface of a neutron star, plotted as a function of the dimensionless radius, $y=R/R_\text{star}$. The values of the parameters $k_0$ and $k_\infty$ determine the relation between the flow velocity and the local Newtonian free-fall velocity near the stellar surface ($y=1$) and as $y \to \infty$, respectively.}
    \label{fig:velocity}
\end{figure}

The detailed topology of X-ray pulsar magnetospheres has been studied by many previous authors, most notably \citet{Leahy_1991} and \citet{Blum_Kraus_2000}. Typically, these models find that the two magnetic poles of an X-ray pulsar are not exactly dipolar, but rather that the second pole will exhibit a nonzero deflection angle from the anti-dipodal position of the first pole. Usually the deflection angle is relatively small, so that the overall topology of the neutron star magnetic field can be considered to be ``quasi-dipolar.'' However, in extreme situations, the models suggest that the two magnetic poles may even occupy the same hemisphere of the star \citep[see][]{Leahy_1991,Petri_etal_2023,Miller_etal_2019,Miller_etal_2021}. In the case of Her X-1, there is precedent from prior works which claim that the offset angle from the anti-dipodal position between the two columns is in the range $\sim 5^\circ-15^\circ$ \citep[see][]{Leahy_1991,Blum_Kraus_2000}.

In this section, we will review the formalism required to compute the continuum spectrum emitted from a single accretion column. This will form the basis for later work where we will develop the procedure for computing the total spectrum emitted from a rotating neutron star with two accretion columns that have arbitrary locations relative to the star's spin axis. In Section~\ref{sec:spectrum}, we will compute the total flux from the neutron star as the sum of the fluxes from each individual column. In the next section, we briefly review the methodology used by {\tt BW22} to derive the solution for the continuum spectral components emitted from the accretion column.

\subsection{The Transport Formalism}
\label{sec:transporteq}

Transport equations are often used to describe the evolution of particle and radiation distributions in high-energy astrophysical environments. In the X-ray pulsar application, the transport equation must include specialized terms that are used to model the advection, diffusion, energization, and escape of radiation propagating through the fully-ionized plasma inside the accretion column. The intensity of the radiation field in the accreting plasma will generally depend on the photon energy, the location within the column, and the time, and we therefore need to specify a transport equation that governs the variation of the radiation distribution function, $f(\vec r,\epsilon,t)$, with respect to each of these variables. The transport equation utilized in the {\tt BW22} model can be written in the vector form
\begin{eqnarray}
\frac{\partial f}{\partial t} + \vec \vel \cdot \vec\nabla f
&=& (\vec\nabla \cdot \vec \vel) \, \frac{\epsilon}{3} \,
\frac{\partial f}{\partial\epsilon}
+ \vec\nabla \cdot \left(\frac{c}{3 n_e \sigpar} \, \vec\nabla f \right)
\nonumber
\\
&+& \frac{n_e \sigbar c}{m_e c^2} \frac{1}{\epsilon^2}
\frac{\partial}{\partial\epsilon}\left[\epsilon^4\left(f
+ k T_e \, \frac{\partial f}{\partial\epsilon}\right)\right]
+ Q
\ ,
\label{transporteq}
\end{eqnarray}
where $f(\vec r,\epsilon,t)$ is normalized such that $\epsilon^2f(\vec r,\epsilon,t)d\epsilon$ gives the number density of photons with energy between $\epsilon$ and $\epsilon+d\epsilon$ at position $\vec r$ and at time $t$. In Equation~(\ref{transporteq}), the quantities $\vec\vel$, $\sigbar$, $k$, $c$, and $n_e$, $m_e$ and $T_e$ represent the flow velocity, the angle-averaged electron scattering cross section, Boltzmann's constant, the speed of light, and the electron number density, mass, and temperature, respectively.

A thorough discussion of the definitions and physical motivations for each term in Equation~(\ref{transporteq}) is presented in {\tt BW22}, but in summary, the left-hand side represents the co-moving time derivative of $f$, and the terms on the right-hand side represent first-order Fermi energization (“bulk Comptonization”), spatial diffusion, thermal Comptonization, and the radiation source term, $Q$, respectively. The electron temperature, $T_e$, appearing in Equation~(\ref{transporteq}) is assumed to maintain a constant value throughout the accretion column, reflecting the Compton thermostat effect that is expected to operate in radiation-dominated flows \citep{BeckerandWolff2007,BeckerandWolff2022,LyubarskiiandSyunyaev1982,SunyaevandTitarchuk1980}. We note that the radiation source term $Q$ in Equation~(\ref{transporteq}) is related to the photon emissivity, $\dot n_\epsilon$, via
\begin{equation}
\epsilon^2 Q(\vec r,\epsilon) = \dot n_\epsilon(\vec r,\epsilon)
\ ,
\label{eq:Qsource}
\end{equation}
where $\dot n_\epsilon(\vec r,\epsilon) \, d\epsilon$ gives the number of photons injected into the accretion column per unit time per unit volume at location $\vec r$ with energy between $\epsilon$ and $\epsilon+d\epsilon$. It should also be emphasized that the ``lab'' frame for Equation~(\ref{transporteq}) is the frame of the star, and in this frame, the radiation spectrum has a steady-state shape, although in the frame of a distant observer, the spectrum will exhibit time variation due to the spin of the neutron star. Hence we will set the time derivative $\partial f/\partial t=0$ in Equation~(\ref{transporteq}).

{\tt BW22} presents the adaptation of Equation~(\ref{transporteq}) into a conical geometry with a source term, $Q$, corresponding to the continual injection of monoenergetic seed photons at a single radius. They demonstrated that the general solution corresponding to an arbitrary source term can be obtained from the Green's function solution via convolution. Since the {\tt BW22} model is developed in the frame of the neutron star, we must adapt our notation to allow for the discussion of measurements made with respect to the reference frame of the observer on Earth. We will use a standardized notation to distinguish among parameters in the frame of the star, parameters in the frame of the observer on Earth, and parameters associated with the injection of  seed photons within the column. For these purposes, in our work, parameters measured in the local frame that is stationary with respect to the neutron star will be indicated using the ``0'' subscript. In our work, quantities measured in the frame of an observer on Earth will not be given a subscript. For example, the energy of the photon for an observer on Earth is $\epsilon$ and the distance from the center of the neutron star to the observer is $R$. These quantities will obey the transformations described by the relativistic characteristics outlined in Section~\ref{sec:relchar}.

Finally, in the context of the Green's function solution to the transport equation, a subscript ``*'' will be used to indicate quantities describing the injection of seed photons. For example, one would say that a seed photon was injected with energy $\epsilon_*$ at radius $R_*$. Hence the seed photon injection quantities $(\epsilon_*,R_*)$ are distinct from the general coordinate values $(\epsilon_0,R_0)$, although both sets are measured in the same frame of reference. The quantities $(\epsilon_*,R_*)$ in our work are directly analogous to the quantities $(\epsilon_0,R_0)$ in {\tt BW22}. This distinction is subtle, yet crucial for establishing the basis for the new theoretical model developed throughout the remainder of the paper.

With this clarification of the notation, we are now ready to present the transport equation governing the Green's function. By adapting Equation~(\ref{transporteq}) into a conical geometry with a monochromatic source term corresponding to the injection of $\dot N_*$ photons per unit time with energy $\epsilon_*$ at radius $R_*$, we obtain in a steady state
\begin{eqnarray}
\vel \, \frac{\partial \green}{\partial R_0}
&=& \frac{1}{R_0^2}\frac{\partial}{\partial R_0}\left(R_0^2 \, \vel\right)\,\frac{\epsilon_0}{3} \,
\frac{\partial \green}{\partial\epsilon_0}
+ \frac{1}{R_0^2} \frac{\partial}{\partial R_0}
\left(\frac{R_0^2 \, c}{3 n_e \sigpar}\,\frac{\partial \green}{\partial R_0}\right)
- \frac{\green}{t_\text{esc}}
\nonumber
\\
&+& \frac{n_e \sigbar c}{m_e c^2} \frac{1}{\epsilon_0^2}
\frac{\partial}{\partial\epsilon_0}\left[\epsilon_0^4\left(\green
+ k T_e \, \frac{\partial \green}{\partial\epsilon_0}\right)\right]
+ \frac{\dot N_* \, \delta(\epsilon_0-\epsilon_*) \, \delta(R_0-R_*)}{\Omega R_\text{star}^2 \, \epsilon_*^2}
\ ,
\label{transporteqgreen}
\end{eqnarray}
where the solid angle of the conical accretion column, $\Omega$, is related to the outer opening angle $\thetamax$ and the inner opening angle $\thetamin$ via
\begin{equation}
\Omega = 2 \pi \, (\cos\thetamin - \cos\thetamax) \ ,
\label{eq:Omega}
\end{equation}
and $t_\text{esc}$ represents the ``escape time,'' which is the mean time for photons to reside in the plasma before diffusing through the walls of the conical accretion column, modeled using an escape-probability formalism.

\subsection{Emission from Column Walls}

The value of the mean escape time, $t_\text{esc}$, appearing in Equation~(\ref{transporteqgreen}) is computed using the expressions
\begin{equation}
t_\text{esc}(R_0) = \frac{r_{\perp}}{\vel_\text{diff}} \ , \qquad
\vel_\text{diff}(R_0) = \frac{c}{\tau_\perp} \ ,
\label{eqTesc1}
\end{equation}
where $\vel_\text{diff}$ is the radiation diffusion velocity at radius $R_0$ associated with the random walk of the photons as they diffuse through the plasma toward the surface of the column. The perpendicular scattering optical thickness, $\tau_\perp$, and the perpendicular radius of the accretion column, $r_\perp$, appearing in Equation~(\ref{eqTesc1}) are given by
\begin{equation}
\tau_\perp(R_0) = n_e \, \sigma_\perp \, r_\perp \ , \qquad
r_\perp(R_0) = b \, R_0 \ ,
\label{eqTauPerp}
\end{equation}
respectively, where $\sigma_\perp$ represents the scattering cross section for photons propagating perpendicular to the magnetic field. In Equation~(\ref{eqTauPerp}), the constant $b$ reflects the conical geometry of the accretion column, with outer opening angle $\thetamax$ and inner opening angle $\thetamin$, and is computed using
\begin{equation}
b = \tan\thetamax - \tan\thetamin \ .
\label{eq:bCons}
\end{equation}
In a conical accretion column filled with fully-ionized hydrogen gas, the accretion rate, $\dot M$, is given by
\begin{equation}
\dot M = \Omega \, R_0^2 \, n_e \, m_p \, |\vel|
\ .
\label{eq:MdotDyn}
\end{equation}
By combining Equations~(\ref{eqTesc1}), (\ref{eqTauPerp}), and (\ref{eq:MdotDyn}), we find that the mean escape timescale can also be expressed as
\begin{equation}
t_\text{esc}(R) = \frac{b^2 \sigma_\perp \dot M}{\Omega \, m_p \, c \, |\vel|}
\ .
\label{eq:tEsc}
\end{equation}

\subsection{Velocity Profile and Dynamical Boundary Conditions}
\label{sec:VelBC}

The velocity profile adopted by {\tt BW22}, resulting from the application of the separability principle to the fundamental transport equation given by Equation~(\ref{transporteqgreen}) in a conical geometry, is given by
\begin{equation}
    \vel(R_0) = c\,u(R_0) = - c \left(\frac{2R_g}{R_\text{star}}\right)^{1/2}\left[\frac{k_\infty^2(y_0^3-1)+k_0^2}{y_0^4}\right]^{1/2} \ ,
    \label{eq:velocity}
\end{equation}
where $u = \vel/c$ is the dimensionless flow velocity, $R_\text{star}$ is the neutron star radius, $R_g = GM_\text{star}/c^2$ is the gravitational radius for a neutron star with mass $M_\text{star}$, and $y_0$ represents the dimensionless radius, which is related to $R_0$ via
\begin{equation}
    y_0 = \frac{R_0}{R_\text{star}} \ .
        \label{dimrad}
\end{equation}
We note that the velocity defined in Equation~(\ref{eq:velocity}) is a negative quantity, which is appropriate because we are dealing with inflowing gas. The parameters $k_0$ and $k_\infty$ in Equation~(\ref{eq:velocity}) are constants of integration that are used to set the boundary conditions for the velocity variation.

Next we briefly review the physical interpretation of the velocity parameters $k_0$ and $k_\infty$. If the infalling matter falls unimpeded towards the stellar surface, then the conversion of gravitational potential energy to kinetic energy dictates that the velocity at radius $R$ from the center of the star will be given by the standard Newtonian relation
\begin{equation}
    \vel_\text{Newt}(R) \equiv \left(\frac{2GM_\text{star}}{R}\right)^{1/2} \ .
    \label{eq:NewtVel}
\end{equation}
The parameter $k_0$ in the {\tt BW22} model represents the ratio of the impact velocity at the stellar surface divided by the local Newtonian free-fall velocity, so that
\begin{equation}
    k_0 \equiv \frac{\vel(R_\text{star})}{\vel_\text{Newt}(R_\text{star})} \ .
    \label{eq:k0par}
\end{equation}
Conversely, the constant $k_\infty$ denotes the ratio of the inflow velocity divided by the Newtonian free-fall velocity in the asymptotic domain far from the star, and therefore
\begin{equation}
    k_\infty \equiv \lim_{R\to\infty}\frac{\vel(R)}{\vel_\text{Newt}(R)}  \ .
    \label{eq:kinfpar}
\end{equation}
The validity of the velocity profile given by Equation~(\ref{eq:velocity}) was established by {\tt BW22} via comparisons with exact velocity profiles obtained by solving the coupled set of hydrodynamical equations for pulsar accretion flows. The family of {\tt BW22} velocity profiles for sources with different dynamical boundary conditions is plotted in Figure~\ref{fig:velocity}.

It is important to understand how the variation of the dynamical parameters $k_0$ and $k_\infty$ affects the shape of the X-ray spectrum emitted from the column. This issue was explored in detail by {\tt BW22}, and we summarize the main results here. First we address the role of the surface impact velocity parameter, $k_0$, which describes the ratio of the impact velocity divided by the Newtonian free fall velocity at the stellar surface (see Equation~(\ref{eq:k0par})). {\tt BW22} showed that when $k_0 = 0$, so that the flow precises stagnates at the stellar surface, the $PdV$ work done on the radiation by the compressing gas is maximized. This process naturally produces a flatter (harder) X-ray spectrum with a smaller power-law index, as observed in luminous sources such as Her X-1. Furthermore, in this situation, the kinetic energy of the accreting gas in the thermal mound is reduced, which minimizes the contribution of blackbody seed photons relative to bremsstrahlung seed photons. On the other hand, when $k_0 > 0$, less $PdV$ work is done on the radiation by the gas, the power-law shape is steeper (softer spectrum), and the injection of blackbody seed photons is enhanced relative to bremsstrahlung injection due to the larger kinetic energy flux in the thermal mound. This situation tends to produce spectra with shapes typified by X Per. Next we focus on the physical effect of variation of the asymptotic velocity parameter, $k_\infty$, which describes the ratio of the inflow velocity divided by the Newtonian free fall velocity in the asymptotic region far from the star (see Equation~(\ref{eq:kinfpar})). In luminous sources such as Her X-1, it is appropriate to use the value $k_\infty = 1$ because at large distances, the gas flows towards the neutron with the Newtonian free-fall velocity. However, in low-luminosity sources such as X Per, we expect the flow to pass through a gas-mediated discontinuous standing shock at the top of the column, which causes the velocity to drop by a factor of four \citep{LangerandRappaport1982}. In this situation, it is appropriate to set $k _\infty \sim 0.25$ \citep[][]{BeckerandWolff2022}.

In luminous sources such as Her X-1, Cen X-3, and LMC X-4, the dynamical structure of the accretion column just above the surface of the neutron star leads to the deceleration of the accreting material to rest at the stellar surface. However, the deceleration to rest is the result of a limiting process in which the radiation pressure gradient gradually diminishes as the flow decelerates. At the stellar surface, both the radiation pressure gradient and the flow velocity vanish simultaneously. The fine-tuning between these two processes is not arbitrary, but is a result of the fact that the kinetic energy of the accreting gas has to escape from the flow in the form of radiation before the gas can settle onto the neutron star surface and merge with the stellar crust. In the establishment of a steady state for a given input accretion rate, the timescales for radiation diffusion and velocity deceleration adjust until a balance is achieved and the gas comes to rest at the stellar surface \citep{Becker1998}. Hence, at the stellar surface, the pressure gradient vanishes, and the downward gravitational force is balanced by the pressure of the degenerate neutrons in the stellar interior, establishing the hydrostatic structure of the star \citep{LattimerAndPrakash2000}.

It is worth noting that the constants $k_0$ and $k_\infty$ are set a priori, and are not treated as free parameters of the fit. Our respective assignation of 0 and 1 to these parameters is rooted in the physical assumptions we make based on the observed X-ray luminosity of Her X-1. As discussed in Section \ref{sec:intro}, it is highly likely that Her X-1 surpasses the critical luminosity for the presence of a radiation-dominated shock. These values are therefore derived from the assumed picture of infalling matter being decelerated from the Newtonian free-fall velocity far from the stellar surface and ultimately coming to rest at the stellar surface. For that reason, the physics rooted in physical observation allows us to assign these values a priori. For sources that are not supercritically luminous, further parameter studies will be necessary to constrain the velocity profiles (see {\tt BW22} for discursive application of these studies to the source X Per.

Though the flow velocity approaches zero at the stellar surface for sources with radiation-dominated shock (which are the exclusive focus of this work), lower-luminosity sources like X Per would have a finite velocity at the stellar surface. In these cases, the infalling flow strikes the surface, yielding a finite radiation pressure gradient, meaning that a fraction of the accretion power is absorbed by the surface. This energy is then re-emitted as thermalized radiation. {\tt BW22} reprocesses this thermalized radiation and adds it to the blackbody component of radiation, which is emitted from the thermal mound.

The velocity of the accretion flow described by Equation~(\ref{eq:velocity}) and plotted in Figure~\ref{fig:velocity} is quasi-relativistic, approaching a maximum speed of $\sim 0.6\,c$ (see Figure~\ref{fig:BW22HerX1_spect}b). For $k_\infty = 1$, the velocity profile approaches the Newtonian free-fall velocity for the infalling plasma as $y\to\infty$. For $k_0 = 0$, the flow velocity approaches zero at the stellar surface. This is indicative of deceleration of the infalling plasma due to high radiation pressure. This limit is characteristic of sources with luminosities exceeding the critical luminosity given by Equation~(\ref{critlum}), leading to domination of radiation pressure slowing the flow velocity. In the application of our model to Her X-1, we will utilize $k_\infty = 1$ and $k_0 = 0$, which is consistent with the expected behavior of the flow velocity profile in a high-luminosity source. 

\subsection{Transformation of Spatial Variable}

For convenience, {\tt BW22} also transformed the spatial variable in the transport equation from the radius $R_0$ to the electron scattering optical depth $\tau_0$ measured along the magnetic field direction, and they also ignored the time derivative by considering only steady-state solutions in the star frame. The scattering optical depth is related to the dimensionless radius $y_0$ via
\begin{equation}
    d\tau_0=n_e(R_0)\,\sigma_{||}\,dR_0
    = n_e(y_0)\,\sigma_{||}\,R_\text{star}\,dy_0 \ .
    \label{eq:dtau0}
\end{equation}
In a conical accretion column, the accretion rate, $\dot M$, is given by Equation~(\ref{eq:MdotDyn}), which can be combined with Equation~(\ref{eq:dtau0}) to show that
\begin{equation}
d\tau_0 = \frac{\dot M \sigpar}{\Omega R_\text{star} y_0^2 m_p |\vel(y_0)|} \, dy_0
\ .
    \label{eq:dtau0b}
\end{equation}
It was demonstrated by {\tt BW22} that Equation~(\ref{eq:dtau0b}) can be rewritten in the equivalent form
\begin{equation}
d\tau_0 = \frac{3 k_\infty^2 R_g}{\alpha R_\text{star} y_0^2 |u(y_0)|} \, dy_0
\ ,
\label{eq:dtau0c}
\end{equation}
where $u = \vel/c$ is the dimensionless flow velocity, and the dimensionless constant $\alpha$ is related to the accretion rate $\dot M$ via
\begin{equation}
\alpha = \frac{3 k_\infty^2 \Omega \, R_g \, m_p c}{\sigpar \dot M} \ .
\label{eq:alpha}
\end{equation}
The optical depth relationship expressed in Equation~(\ref{eq:dtau0c}) has an exact analytical solution given by
\begin{equation}
    \tau_0(y_0)=\frac{3\sqrt{2}\,k_\infty}{\alpha}\left(\frac{R_g}{R_\text{star}}\right)^{1/2}\left[G(k_0,k_\infty,1)-G(k_0,k_\infty,y_0)\right] \ ,
    \label{tautoy}
\end{equation}
where
\begin{equation}
G(k_0,k_\infty,y_0) \equiv \, _2F_1\left(\frac{1}{6},\frac{1}{2};\frac{7}{6};
\frac{1-k_0^2/k_\infty^2}{y_0^3}\right) \frac{1}{\sqrt{y_0}}
\ ,
\label{hypergeometric}
\end{equation}
and $_2F_1$ denotes the hypergeometric function \citep{AbramowitzandStegun1970}. Equation~(\ref{tautoy}) will allow the calculation of $\tau_0$ for a given value of $y_0=R_0/R_\text{star}$. It is also convenient to introduce a transformation to the dimensionless energy $\chi_0$ and the dimensionless injection energy $\chi_*$, defined by
\begin{equation}
\chi_0 \equiv \frac{\epsilon_0}{k T_e} \ , \qquad \chi_* \equiv \frac{\epsilon_*}{k T_e} \ .
\label{eq:DimEn}
\end{equation}
Combining relations, we find that Equation~(\ref{transporteqgreen}) can now be written in the form
\begin{eqnarray}
u(\tau_0) \, \frac{\partial \green}{\partial \tau_0}
&=& \frac{1}{y_0^2}\,\frac{d}{d\tau_0}\left[y_0^2\,u(\tau_0)\right]\,\frac{\chi_0}{3} \,
\frac{\partial \green}{\partial\chi_0}
+ \frac{1}{3 y_0^2}\,\frac{\partial}{\partial\tau_0} \left(y_0^2 \, \frac{\partial\green}{\partial\tau_0}\right)
- \xi^2 u^2(\tau_0) y_0^2 \green
\nonumber
\\
&+& \frac{\sigbar}{\sigpar} \, \frac{k T_e}{m_e c^2}
\frac{1}{\chi_0^2}\frac{\partial}{\partial\chi_0}\left[\chi_0^4\left(\green
+ \frac{\partial\green}{\partial\chi_0}\right)\right]
+ \frac{\dot N_* \, \delta(\chi_0-\chi_*) \, \delta(\tau_0-\tau_*)}{\Omega \, c \, R_\text{star}^2 \chi_*^2 (k T_e)^3}
\ ,
\label{eq:transEq}
\end{eqnarray}
where the dimensionless radius $y_0$ is an implicit function of the scattering optical depth $\tau_0$ via Equation~(\ref{tautoy}), and the dimensionless flow velocity $u$ is evaluated using Equation~(\ref{eq:velocity}), and the dimensionless parameter $\xi$ is defined by
\begin{equation}
\xi \equiv \frac{\Omega \, R_\text{star} m_p c}{b \, \dot M (\sigpar \sigperp)^{1/2}}
\ .
\label{eq:xi}
\end{equation}
Equation~(\ref{eq:transEq}) can be solved exactly for the Green's function $\green$ by employing separation of variables. The resulting solution is further discussed in Section~\ref{sec:transporteqsolutions}.

\subsection{Green's Function Solution}
\label{sec:transporteqsolutions}

Next we will review the formalism developed by {\tt BW22} for solving Equation~(\ref{eq:transEq}) for the Green's function, $\green$, representing the radiation distribution inside the accretion column resulting from the continual injection of monoenergetic seed photons from a source located at one specific injection radius. This fundamental solution will then be used to calculate the continuum spectral components emitted through the walls and top of a conical accretion column, via integral convolution of the Green's function with a physical source term. Separate results will be obtained for the cases treating the injection of bremsstrahlung, cyclotron, and blackbody seed photons. In order to obtain the solution to Equation~(\ref{eq:transEq}), we must also impose a set of rigorous physical boundary conditions. Following {\tt BW22}, we require that the radiation flux vanishes at the stellar surface, and we also impose a free-streaming boundary condition at the column top, located at radius $R_\text{top}$, or equivalently at optical depth $\tau_\text{top}$. As discussed in Section~\ref{sec:VelBC}, the radiation pressure gradient vanishes at the surface of the neutron star as a consequence of the deceleration of the accreting gas to rest at the stellar surface. This implies that both the advective and diffusive components of the radiation energy flux also vanish at the stellar surface \citep{Becker1998}. We can interpret this behavior as a consequence of energy conservation based on Equation~(\ref{eq:mdotFirst}), which gives the relationship between the X-ray luminosity, $L_\text{X}$, and the observer frame accretion rate, $\dot M_\text{obs}$. Equation~(\ref{eq:mdotFirst}) states that the change in the gravitational potential energy of the infalling matter must be balanced by the escape of radiation energy before the gas merges with the stellar crust. This provides an additional physical argument for the vanishing of the radiation energy flux at the stellar surface.

Equation~(\ref{eq:transEq}) can be solved to determine the Green's function, $\green(\tau_*,\tau_0,\epsilon_*,\epsilon_0)$, using an expansion over the spatial eigenfunctions, $g_n(\tau)$. This results in a rapidly converging series given by (see {\tt BW22} for complete details)
\begin{eqnarray}
\green(\tau_*,\tau_0,\epsilon_*,\epsilon_0)
&=& \frac{3 \, \psi \dot N_* \, \omega(\tau_*) \, k T_e
\, \epsilon_0^{\kappa-4} \, e^{(\epsilon_*-\epsilon_0)/(2kT_e)}}
{\alpha \, \Omega \, c \, R_\text{star}^2 \, y_*^2 \, \epsilon_*^\kappa}
\sum_{n=0}^\infty \
\frac{\Gamma(\mu-\kappa+1/2)}{{\cal I}_n \, \Gamma(1+2\mu)}
\nonumber
\\
&\times& g_n(\tau_*) \, g_n(\tau_0)
\, M_{\kappa,\mu}\left(\frac{\epsmin}{k T_e}\right)
W_{\kappa,\mu}\left(\frac{\epsmax}{k T_e}\right)
\ ,
\label{greensfuncsolution}
\end{eqnarray}
where
\begin{equation}
    \epsilon_\text{max} \equiv \text{Max}(\epsilon_0,\epsilon_*) \ ,
    \qquad\epsilon_\text{min} \equiv \text{Min}(\epsilon_0,\epsilon_*) \ ,
    \label{eq:epsmax}
\end{equation}
and the symbols $M_{\kappa,\mu}$ and $W_{\kappa,\mu}$ denote the Whittaker functions \citep{AbramowitzandStegun1970}. The parameter $\mathcal{I}_n$ appearing in Equation~(\ref{greensfuncsolution}) represents the $n$th quadratic normalization integral of the spatial eigenfunctions, defined by
\begin{equation}
    \mathcal{I}_n \equiv \int_0^{\tau_\text{top}}g^2_n(\tau_*) \, \omega(\tau_*) \, d\tau_*
    \label{scriptIeq} \ ,
\end{equation}
where $\tau_\text{top}$ is the optical depth at the column top. The indices $\kappa$ and $\mu$ of the Whittaker functions are computed using
\begin{equation}
\kappa \equiv \frac{1}{2} \, (\psi+4) \ ,
\qquad\qquad
\mu \equiv \frac{1}{2} \left[(3-\psi)^2 + 4 \, \psi \lambda \right]
^{1/2}
\ ,
\label{eq:KappaMu}
\end{equation}
where $\lambda_n$ denotes the $n$th eigenvalue, associated with spatial eigenfunction $g_n(\tau)$, and the constant $\psi$ is defined by
\begin{equation}
\psi \equiv \frac{\alpha}{3} \,
\frac{\sigpar}{\sigbar} \, \frac{m_e c^2}{kT_e}
\ .
\label{eq:psi}
\end{equation}
Finally, the weight function $\omega(\tau_*)$ is given by
\begin{equation}
\omega(\tau_*) = y^2(\tau_*) \exp\left\{\frac{9 R_g k_\infty^2}{\alpha R_\text{star}} \left[1-\frac{1}{y(\tau_*)}\right]\right\} \ .
\label{eq4.10j}
\end{equation}
The Green's function solution given by Equation~(\ref{greensfuncsolution}) represents the radiation distribution inside the accretion column resulting from the continual injection of $\dot N_*$ monochromatic seed photons per unit time with energy $\epsilon_*$ at optical depth $\tau_*$, which is the fundamental solution to the radiation transport problem. The next step is to understand how Equation~(\ref{greensfuncsolution}) is used to compute the components of the spectrum resulting from the escape of radiation through the wall and top of the accretion column.

\subsection{Green's Function for Column Wall Emission}

The Green's function describing the photon number distribution for the radiation escaping through the walls of the conical accretion column is given by
\begin{equation}
\greenphoton(\tau_*,\tau_0,\epsilon_*,\epsilon_0) \equiv \frac{\Omega \, R_0^2
\epsilon_0^2}{t_\text{esc}(\tau_0)} \, \green(\tau_*,\tau_0,\epsilon_*,\epsilon_0)
\ ,
\label{ndotwallfromgreens}
\end{equation}
where $\green$ is computed using Equation~(\ref{greensfuncsolution}), the photon escape timescale $t_\text{esc}$ is given by Equation~(\ref{eq:tEsc}), and $\greenphoton$ is normalized so that $\dot N^\text{G}_\epsilon \, d\epsilon_0 \, dR_0$ gives the number of photons escaping through the column wall with energy between $\epsilon_0$ and $\epsilon_0 + d\epsilon_0$ from the disk-shaped volume between radii $R_0$ and $R_0+dR_0$ due to the continual injection of seed photons with energy $\epsilon_*$ at optical depth $\tau_*$. We note that the variables $R_0$ and $\tau_0$ are related via Equation~(\ref{tautoy}).
We can now substitute Equation~(\ref{greensfuncsolution}) into Equation~(\ref{ndotwallfromgreens}) to derive the closed-form solution for the Green's function representing the photon number spectrum escaping through the column wall. The result obtained is
\begin{align}
    \dot{N}_\epsilon^\text{G}(\tau_*,\tau_0,\epsilon_*,\epsilon_0)&=\frac{9k_\infty^2y_0^2\xi^2R_g\abs{v_0}\psi\dot{N}_* \, \omega(\tau_*)kT_e\epsilon_0^{\kappa-2}e^{(\epsilon_*-\epsilon_0)/(2kT_e)}}{\alpha^2cR_\text{star}^2y_*^2\epsilon^{\kappa}_*}
    \nonumber\\
    &\times\sum_{n=0}^\infty\frac{\Gamma(\mu-\kappa+1/2)}{\mathcal{I}_n\Gamma(1+2\mu)}g_n(\tau_*)g_n(\tau_0)M_{\kappa,\mu}\left(\frac{\epsilon_\text{min}}{kT_e}\right)W_{\kappa,\mu}\left(\frac{\epsilon_\text{max}}{kT_e}\right) \ ,
    \label{eq:NdotGeq}
\end{align}
where ${\cal I}_n$ is computed using Equation~(\ref{scriptIeq}), and $\epsilon_\text{min}$ and $\epsilon_\text{max}$ are defined in Equation~(\ref{eq:epsmax}).

\subsection{Green's Function for Column Top Emission}

Next we compute the Green's function describing the number distribution for the photons escaping through the top of the conical column, at radius $R_\text{top}$. We can accomplish this by utilizing the free-streaming upper boundary condition to write
\begin{equation}
\dot {\cal N}^\text{\,G}_\epsilon(\tau_*,\epsilon_*,\epsilon_0) = c \,  \Omega \, R^2_\text{top}
\, \epsilon_0^2 \green(\tau_*,\tau_\text{top},\epsilon_*,\epsilon_0)
\ ,
\label{ndottopfromgreens}
\end{equation}
where $\dot {\cal N}^\text{\,G}_\epsilon d\epsilon_0$ gives the number of photons escaping from the column top with energy between $\epsilon_0$ and $\epsilon_0+d\epsilon_0$ due the continual injection of monoenergetic seed photons with energy $\epsilon_*$ at optical depth $\tau_*$. The relationship between the quantities $R_\text{top}$ and $\tau_\text{top}$ is given by Equation~(\ref{tautoy}). We can also combine Equations~(\ref{greensfuncsolution}) and (\ref{ndottopfromgreens}) to derive the closed-form solution for the column-top photon number distribution, obtaining
\begin{eqnarray}
\dot {\cal N}^\text{\,G}_\epsilon(\tau_*,\epsilon_*,\epsilon_0) &=&
\frac{3 \, \psi \dot N_* \, \omega(\tau_*) \, k T_e
\, R^2_\text{top} \,
\epsilon_0^{\kappa-2} \, e^{(\epsilon_*-\epsilon_0)/(2kT_e)}}
{\alpha R_\text{star}^2 \, y_*^2 \, \epsilon_*^\kappa}
\sum_{n=0}^\infty \
\frac{\Gamma(\mu-\kappa+1/2)}{{\cal I}_n \, \Gamma(1+2\mu)}
\nonumber
\\
&\times& g_n(\tau_*) \, g_n(\tau_\text{top})
\, M_{\kappa,\mu}\left(\frac{\epsmin}{k T_e}\right)
W_{\kappa,\mu}\left(\frac{\epsmax}{k T_e}\right)
\ ,
\label{eq:GreenPhot}
\end{eqnarray}
where $\epsilon_\text{min}$ and $\epsilon_\text{max}$ are defined in Equation~(\ref{eq:epsmax}), and ${\cal I}_n$ is defined in Equation~(\ref{scriptIeq}).

\subsection{Photon Number Spectrum for General Source}
\label{numspecfromsource}

To obtain the photon number spectrum describing the emission due to a given photon source mechanism, one must perform a convolution of the Green's function with the respective source function. The detailed expressions depend on whether the radiation is escaping through the walls or the top of the accretion column. For the column wall component, the number spectrum of the escaping radiation is computed using the integral convolution
\begin{equation}
\dot N_\epsilon(R_0,\epsilon_0) = \int_{R_\text{star}}^{R_\text{top}}\int_0^\infty
\frac{\greenphoton(R_*,R_0,\epsilon_*,\epsilon_0)}{\dot N_*} \ \epsilon_*^2
\, \Omega R_*^2 \, Q(R_*,\epsilon_*) \, d\epsilon_* \, dR_*
\ ,
\label{eq:WallConvo}
\end{equation}
where the Green's function $\greenphoton$ is computed using Equation~(\ref{eq:NdotGeq}) and the quantity $\dot N_\epsilon \, d\epsilon_0 \, dR_0$ represents the number of photons escaping per unit time with energy between $\epsilon_0$ and $\epsilon_0 + d\epsilon_0$ from the disk-shaped region between radii $R_0$ and $R_0 + dR_0$. Making a change of variable from $dR_*$ to $d\tau_*$ using Equation~(\ref{eq:dtau0c}) yields the equivalent form
\begin{equation}
    \dot{N}_\epsilon(\tau_0,\epsilon_0)=\int_0^{\tau_{\text{top}}}\int_0^\infty\frac{\dot{N}_\epsilon^\text{G}(\tau_*,\tau_0,\epsilon_*,\epsilon_0)}{\Dot{N}_*}\,\ \epsilon_*^2 \, \Omega \, Q(\tau_*,\epsilon_*)\,\frac{\alpha R_\text{star}^4y_*^4\abs{u(\tau_*)}}{3k_\infty^2R_g}\,d\epsilon_*\,d\tau_* \ ,
    \label{convolutionwall}
\end{equation}
 where $Q(\tau_*,\epsilon_*)$ represents the source function for a given radiative process.

Next we calculate the emission component escaping through the column top resulting from an arbitrary seed photon injection process. In this case, the required integral convolution is given by
\begin{equation}
\dot {\cal N}_\epsilon(\epsilon_0) = \int_{R_\text{star}}^{R_\text{top}}\int_0^\infty
\frac{\dot {\cal N}^\text{\,G}_\epsilon(R_*,\epsilon_*,\epsilon_0)}{\dot N_*} \ \epsilon_*^2
\, \Omega R_*^2 \, Q(R_*,\epsilon_*) \, d\epsilon_* \, dR_*
\ ,
\label{eq:TopConvo}
\end{equation}
where $\dot{\cal N}^\text{\,G}_\epsilon$ is evaluated using Equation~(\ref{eq:GreenPhot}). It is convenient to transform the spatial variable of integration in Equation~(\ref{eq:TopConvo}) from $R_*$ to $\tau_*$ using Equation~(\ref{eq:dtau0c}), which yields
\begin{equation}
\dot {\cal N}_\epsilon(\epsilon_0) = \int_{0}^{\tau_\text{top}}\int_0^\infty
\frac{\dot {\cal N}^\text{\,G}_\epsilon(\tau_*,\epsilon_*,\epsilon_0)}{\dot N_*} \ \epsilon_*^2
\, Q(\tau_*,\epsilon_*) \,
\frac{\alpha \Omega R_\text{star}^4 \, y_*^4 |u(\tau_*)|}{3 k_\infty^2 R_g} \, d\epsilon_* \, d\tau_*
\ ,
\label{convolutiontop}
\end{equation}
where $\tau_\text{top}$ is related to $R_\text{top} = R_\text{star} \, y_\text{top}$ via Equation~(\ref{tautoy}). In this paper, we wish to compute the total radiation spectrum emitted through the walls and top of the accretion column resulting from the reprocessing of cyclotron, blackbody, and bremsstrahlung seed photons. We must therefore evaluate Equations~(\ref{convolutionwall}) and (\ref{convolutiontop}) for all three of these injection sources in order to obtain the total emission components radiated through the column walls and top.

\subsubsection{Column Wall Spectrum due to Cyclotron Seed Photons}
\label{cyclotronwall}

We begin by considering the source function corresponding to cyclotron seed photon emission, which is given by
\begin{equation}
    Q^\text{cyc}(\tau_*,\epsilon_*) = 2.10\times10^{36}\rho(\tau_*)^2B_{12}^{-7/2}H\left(\frac{\epsilon_c}{kT_e}\right)\epsilon_*^{-2}e^{-\epsilon_c/(kT_e)}\delta(\epsilon_*-\epsilon_c) \ ,
    \label{cyclotronsourcefun}
\end{equation}
where the cyclotron energy, $\epsilon_c$, and the dimensionless magnetic field strength, $B_{12}$ are defined by
\begin{equation}
    \epsilon_c \equiv 11.57 \, B_{12}\,{\rm keV} \ ,\qquad B_{12} \equiv \frac{B}{10^{12} \ {\rm G}} \ ,
\end{equation}
and the function $H$ is defined by
\begin{equation}
    H\left(\frac{\epsilon_c}{kT_e}\right)\equiv
    \begin{cases}
        0.41 \ ,&\epsilon_c/kT_e > 7.5 \\
        0.15 \sqrt{\epsilon_c/kT_e} \ ,&\epsilon_c/kT_e < 7.5 
    \end{cases} \ .
\end{equation}
The delta function appearing in Equation~(\ref{cyclotronsourcefun}) reflects the resonant nature of the cyclotron emission mechanism around $\epsilon_c$. We can compute the continuum spectrum resulting from the reprocessing of the cyclotron seed photons by substituting Equation~(\ref{cyclotronsourcefun}) into Equation~(\ref{convolutionwall}). The energy integration is trivial due to the delta function in Equation~(\ref{cyclotronsourcefun}), and the remaining spatial integration can be carried out analytically. The result obtained for the photon number spectrum emitted through the column walls due to the reprocessed cyclotron seed photons is given by
\begin{align}
    \Dot{N}_\epsilon^\text{cyc}(\tau_0,\epsilon_0)&=2.10\times10^{36}B_{12}^{-3/2}H\left(\frac{\epsilon_c}{kT_e}\right)e^{-\epsilon_c/(kT_e)}\sum_{n=0}^\infty\frac{\Gamma(\mu-\kappa+1/2)}{\mathcal{I}_n\Gamma(1+2\mu)}g_n(\tau_0)
    \nonumber\\
    &\times\frac{3\Dot{M}^2y_0^2\xi^2\abs{u(\tau_0)}\psi kT_e\epsilon_0^{\kappa-2}e^{-\epsilon_0/(2kT_e)} }{\alpha\Omega R_\text{star}^2c^2}\left(\frac{R_\text{star}}{2R_g}\right)^{1/2}J_n\epsilon_c^{-\kappa}e^{\epsilon_c/(2kT_e)}
        \nonumber\\
        &\times M_{\kappa,\mu}\left[\frac{\text{min}(\epsilon_0,\epsilon_c)}{kT_e}\right]W_{\kappa,\mu}\left[\frac{\text{max}(\epsilon_0,\epsilon_c)}{kT_e}\right] \ ,
    \label{eq:NdotCyc}
\end{align}
where ${\cal I}_n$ is defined in Equation~(\ref{scriptIeq}) and we have introduced the new quantity $J_n$, defined by
\begin{equation}
    J_n\equiv\int_{\tau_\text{th}}^{\tau_\text{top}}\frac{\omega(\tau_*) \, g_n(\tau_*)}{\sqrt{k_\infty^2(y(\tau_*)^3-1)+k_0^2}} \, d\tau_*
    \label{Jneq} \ ,
\end{equation}
which appears as a result of the integration of the emission region over the volume of the accretion column.

\subsubsection{Column Wall Spectrum due to Blackbody Seed Photons}
\label{sec:blackbodywall}

As discussed by \citet{BeckerandWolff2007,BeckerandWolff2022}, the emission spectrum of Her X-1 is dominated by reprocessed bremsstrahlung radiation, with a significant contribution made by reprocessed cyclotron emission, which appears in the vicinity of the cyclotron energy, $\epsilon_c$ (see Figure~\ref{fig:BW22HerX1_spect}). The contribution made by reprocessed blackbody seed radiation is relatively unimportant in this source, but it can become dominant in low-luminosity X-ray pulsars such as X Per. In the case of blackbody emission, the seed photons are primarily injected from the thermal mound, located at the base of the flow, where the accreting gas achieves full thermodynamic equilibrium. The radius at the top of the thermal mound is denoted by $R_\text{th}$, with the associated altitude from the stellar surface, $z_\text{th}$, defined by
\begin{equation}
    z_\text{th} \equiv R_\text{th} - R_\text{star} \ .
        \label{eq:zth}
\end{equation}
In the {\tt BW22} model, the thermal mound radius, $R_\text{th}$, is determined by setting the free-free absorption optical depth across the column equal to unity. The corresponding temperature at the top of the thermal mound is denoted by $T_\text{th}$, and the associated electron scattering optical depth, $\tau_\text{th} = \tau(R_\text{th})$, is computed using Equation~(\ref{tautoy}).

The source function corresponding to blackbody emission is given by \citep{RybickiandLightman1979,BeckerandWolff2007,BeckerandWolff2022}
\begin{equation}
    Q^{\text{bb}}(R_*,\epsilon_*) = \frac{2\pi}{c^2h^3}\frac{\delta(R_*-R_\text{th})}{e^{\epsilon_*/(k T_\text{th})}-1}\label{blackbodysource} \ .
\end{equation}
In order to model the contribution to the spectrum made by reprocessed blackbody photons escaping through the column walls, we substitute Equation~(\ref{blackbodysource}) into Equation~(\ref{convolutionwall}), which yields
\begin{align}
    \Dot{N}_\epsilon^\text{bb}(\tau_0,\epsilon_0)&=18y_0^2\xi^2\abs{u(\tau_0)}\ \psi\omega(\tau_\text{th})(kT_e)^{4-\kappa}\epsilon_0^{\kappa-2}e^{-\epsilon_0/(2kT_e)}\nonumber\\&\times\Omega \frac{\pi k_\infty^2R_g}{\alpha^2 c^2h^3}\sum_{n=0}^\infty\frac{\Gamma(\mu-\kappa+1/2)}{\mathcal{I}_n\Gamma(1+2\mu)}\, g_n(\tau_\text{th}) \, g_n(\tau_0) \, G_n(\chi_0) \ ,
    \label{eq:NdotBB}
\end{align}
where
\begin{equation}
    G_n(\chi_0)\equiv \int_0^\infty\frac{\chi_*^{2-\kappa}e^{\chi_*/2}}{e^{\chi_*T_e/T_\text{th}}-1}\,M_{\kappa,\mu}[\min(\chi_0,\chi_*)] \, W_{\kappa,\mu}[\max(\chi_0,\chi_*)] \, d\chi_* \ ,
    \label{Gndef}
\end{equation}
with $\chi_0=\epsilon_0/(kT_e)$.

\subsubsection{Column Wall Spectrum due to Bremsstrahlung Seed Photons}
\label{sec:bremwallspec}

Next we consider the source function corresponding to bremsstrahlung seed photon emission, given by
\begin{equation}
    Q_\text{ff}(\tau_*,\epsilon_*)=3.68\times10^{36}\rho^2(\tau_*) T_e^{-1/2}\epsilon_*^{-3}e^{-\epsilon_*/(kT_e)} \ ,
    \label{BremSource}
\end{equation}
where $\rho = n_e \, m_p$ is the mass density for a plasma composed of fully-ionized hydrogen.
By substituting Equation~(\ref{BremSource}) into Equation~(\ref{convolutionwall}) and carrying out the two integrations, we can compute the photon number spectrum for the radiation escaping through the column walls due to the reprocessed bremsstrahlung seed photons generated within the column. The result obtained is
\begin{align}
    \Dot{N}_{\epsilon}^\text{ff}(\tau_0,\chi_0)=3.68\times10^{36}\left(\frac{R_\text{star}}{2R_g}\right)^{1/2}&\frac{3y_0^2\abs{u_0}\Dot{M}^2\xi^2\psi kT_e^{1/2}e^{-\chi_0/2}\chi_0^{\kappa-2}}{\alpha \Omega R_\text{star}^2 c^2 (kT_e)^2} \nonumber \\
    &\times \sum_{n=0}^\infty\frac{\Gamma(\mu-\kappa+1/2)}{\mathcal{I}_n\Gamma(1+2\mu)}\, g_n(\tau_0) \, J_n \, B_n \ ,
    \label{numSpecBrem}
\end{align}
where ${\cal I}_n$ and $J_n$ are computed using Equations~(\ref{scriptIeq}) and (\ref{Jneq}), respectively. The quantity $B_n$ appearing in Equation~(\ref{numSpecBrem}), defined by
\begin{equation}
B_n \equiv \int_{\chiabs}^\infty \chi_*^{-1-\kappa}
\, e^{-\chi_*/2} \, M_{\kappa,\mu}[\min(\chi_0,\chi_*)] \,
W_{\kappa,\mu}[\max(\chi_0,\chi_*)]
\, d\chi_*
\ ,
\label{BnEq}
\end{equation}
expresses the integration over the injection energy for the bremsstrahlung source term, where $\chi_0 = \epsilon_0/(kT_e)$ and $\chiabs = \epsilonabs/kT_e$ represent the dimensionless injection energy and the dimensionless self-absorption cutoff, respectively. For values of the photon energy below $\epsilonabs$, the plasma is optically thick and the bremsstrahlung emission makes a transition to blackbody emission. We set $\chi_{\text{abs}} = 0.05$, which is a reasonable approximation for luminous accretion-powered X-ray pulsars \citep{BeckerandWolff2007,BeckerandWolff2022}. In practice, we find that the expansion in Equation~(\ref{numSpecBrem}) and is rapidly convergent, so that taking the first $\sim 10$ provides about 3 decimal digits of accuracy, which is sufficient for the astrophysical applications of interest here.

\subsubsection{Column Top Spectrum due to Cyclotron Seed Photons}

By substituting Equation~(\ref{cyclotronsourcefun}) into Equation~(\ref{convolutiontop}), we can obtain the number spectrum for cyclotron photons escaping from the column top. This yields
\begin{align}
\dot {\cal N}_\epsilon^\text{cyc}(\epsilon_0)
&= \frac{5.76 \times 10^{-13} \dot M^2 \, T_e \, y_\text{top}^2 \, \psi \sqrt{R_g R_\text{star}} \, H[\epsilon_c/(kT_e)] \, \epsilon_0^{\kappa-2}}
{\Omega R_g^2 \, k_\infty^2
\, e^{(\epsilon_0+\epsilon_c)/(2kT_e)} \, \epsilon_c^{\kappa+3/2}} \nonumber \\
&\times \sum_{n=0}^\infty \frac{\Gamma(\mu - \kappa +1/2)}{\Gamma(1 + 2 \mu)}
\frac{J_n \, g_n(\tau_\text{top})}{{\cal I}_n} 
M_{\kappa,\mu}\left[\frac{\min(\epsilon_0,\epsilon_c)}{kT_e}\right]
W_{\kappa,\mu}\left[\frac{\max(\epsilon_0,\epsilon_c)}{kT_e}\right]
\ ,
\phantom{\left(\frac{\epsmin}{k T_e}\right)}
\label{topspeccyc}
\end{align}
where ${\cal I}_n$ and $J_n$ are defined in Equations~(\ref{scriptIeq}) and (\ref{Jneq}), respectively.

\subsubsection{Column Top Spectrum due to Blackbody Seed Photons}

Next, we will compute the number spectrum for blackbody photons escaping from the column top. By substituting Equation~(\ref{blackbodysource}) into Equation~(\ref{convolutiontop}), we find that
\begin{equation}
\dot{\cal N}_\epsilon^\text{bb}(\epsilon_0) = \frac{2 \pi \, \Omega R_\text{th}^2}{c^2 h^3} \int_0^\infty
\frac{\dot {\cal N}^\text{\,G}_\epsilon(R_\text{th},\epsilon_*,\epsilon_0)}{\dot N_*} \ 
\, \frac{\epsilon_*^2}{e^{\epsilon_*/k T_\text{th}} - 1} \ d\epsilon_*
\ ,
\label{eq7.11b}
\end{equation}
which, upon simplification, reduces to
\begin{align}
    \Dot{\mathcal{N}}_\epsilon^\text{bb}(\epsilon_0) &= 2\pi\Omega\frac{3\psi \omega(\tau_\text{th})(kT_e)^{4-\kappa}R_\text{top}^2\epsilon_0^{\kappa-2}e^{-\epsilon_0/(2kT_e)}}{\alpha c^2h^3} \nonumber\\
    &\times\sum_{n=0}^\infty\frac{\Gamma(\mu-\kappa+1/2)}{\mathcal{I}_n\Gamma(1+2\mu)}g_n(\tau_\text{top})g_n(\tau_\text{th})G_n(\chi_0) \ ,
    \label{eq:BB2}
\end{align}
where $G_n(\chi_0)$ is defined in Equation~(\ref{Gndef}).

\subsubsection{Column Top Spectrum due to Bremsstrahlung Seed Photons}

Finally, we can substitute Equation~(\ref{BremSource}) into Equation~(\ref{convolutiontop}) to obtain the number spectrum for bremsstrahlung photons escaping from the column top. This yields
\begin{align}
\dot {\cal N}_\epsilon^\text{ff}(\epsilon_0)
&= \frac{3.40 \times 10^{7} \, \dot M^2 \, y_\text{top}^2 \, \psi \, \epsilon_0^{\kappa-2} \,
e^{-\epsilon_0/(2 k T_e)}\sqrt{R_g R_\text{star}}}{k_\infty^2 R_g^2 \, \Omega 
\, (k T_e)^{\kappa-1/2}}
\nonumber
\\
&\times \sum_{n=0}^\infty \ \frac{\Gamma(\mu-\kappa+1/2) }
{\Gamma(1+2\mu)} \ \frac{g_n(\tau_\text{top}) \, J_n}{{\cal I}_n} \, B_n
\ ,
\label{topSpecbrem}
\end{align}
where ${\cal I}_n$, $J_n$, and $B_n$ are computed using Equations~(\ref{scriptIeq}), (\ref{Jneq}), and (\ref{BnEq}), respectively.

\subsubsection{Total Number Spectra from Column Wall and Column Top}

The photon number spectrum components emitted from the column wall and the column top will form the basis for the relativistic formalism developed in subsequent sections, which will allow us to link the continuum spectrum emitted in the frame of the neutron star with the flux measured by a distant observer via propagation through the Schwarzschild metric. As a final note in this section, we will present expressions for the total number spectra emitted from the column wall and the column top, obtained by summing over the individual components due to bremsstrahlung, cyclotron, and blackbody seed photons.

The total number spectrum due to all sources emitting from the column wall, in ${\rm photons}$ \, ${\rm erg}^{-1} \, {\rm s}^{-1} \, {\rm cm}^{-1}$, is given by
\begin{equation}
    \Dot{N}_\epsilon^\text{tot}(\tau_0,\epsilon_0) = \Dot{N}_\epsilon^\text{ff}(\tau_0,\epsilon_0) + \Dot{N}_\epsilon^\text{cyc}(\tau_0,\epsilon_0) + \Dot{N}_\epsilon^\text{bb}(\tau_0,\epsilon_0) \ ,
    \label{totalwallspec}
\end{equation}
where $\dot N_\epsilon^\text{ff}$, $\dot N_\epsilon^\text{cyc}$, and $\dot N_\epsilon^\text{bb}$ are computed using Equations~(\ref{numSpecBrem}), (\ref{eq:NdotCyc}), and (\ref{eq:NdotBB}), respectively.
Likewise, the total number spectrum due to all sources emitting from the column top, in ${\rm photons}$ \, ${\rm erg}^{-1} \, {\rm s}^{-1}$, is given by
\begin{equation}
    \Dot{\mathcal{N}}_\epsilon^\text{tot}(\epsilon_0) = \Dot{\mathcal{N}}_\epsilon^\text{ff}(\epsilon_0) + \Dot{\mathcal{N}}_\epsilon^\text{cyc}(\epsilon_0) + \Dot{\mathcal{N}}_\epsilon^\text{bb}(\epsilon_0) \ ,
    \label{totaltopspec}
\end{equation}
where $\dot{\mathcal{N}}_\epsilon^\text{ff}$, $\dot{\mathcal{N}}_\epsilon^\text{cyc}$, and $\dot{\mathcal{N}}_\epsilon^\text{bb}$ are computed using Equations~(\ref{topSpecbrem}), (\ref{topspeccyc}), and (\ref{eq:BB2}), respectively. We will utilize Equations~(\ref{totalwallspec}) and (\ref{totaltopspec}) in our calculations of simulated pulse profiles and phase-averaged spectra. In the remainder of this section, we will derive the connection between the photon spectra computed using Equations~(\ref{totalwallspec}) and (\ref{totaltopspec}) and the local intensity measured in the local frame that is stationary with respect to the neutron star. The local intensity will then be linked with the {\tt RM88} formalism to compute the intensity measured by a distant observer after the photons propagate through the Schwarzschild metric.

\subsection{Total Continuum Flux in the Local Frame}
\label{intensityderiv}

Our central goal in this paper is to compute pulse profiles and phase-averaged spectra resulting from the application of the {\tt BW22} continuum model in the context of a relativistic neutron star, with the propagation of the radiation modeled using the formalism developed by {\tt RM88}. In order to accomplish this, we must calculate the intensity distribution in the local frame that is stationary with respect to the star, which is hereafter referred to as the local {\tt RM88} frame. The intensity specified in this frame provides the input required in order to utilize the relativistic formalism developed by {\tt RM88}. This requires us to connect the {\tt BW22} continuum model with the local intensity measured in the frame of the star. In carrying this out, it is convenient to use the locally measured energy flux as an intermediate concept.

\subsubsection{Local Flux for Radiation Escaping from Column Wall}
\label{sec:fluxfromcolwall}

\begin{figure}[htbp]
    \centering
    \includegraphics[width=0.5\linewidth]{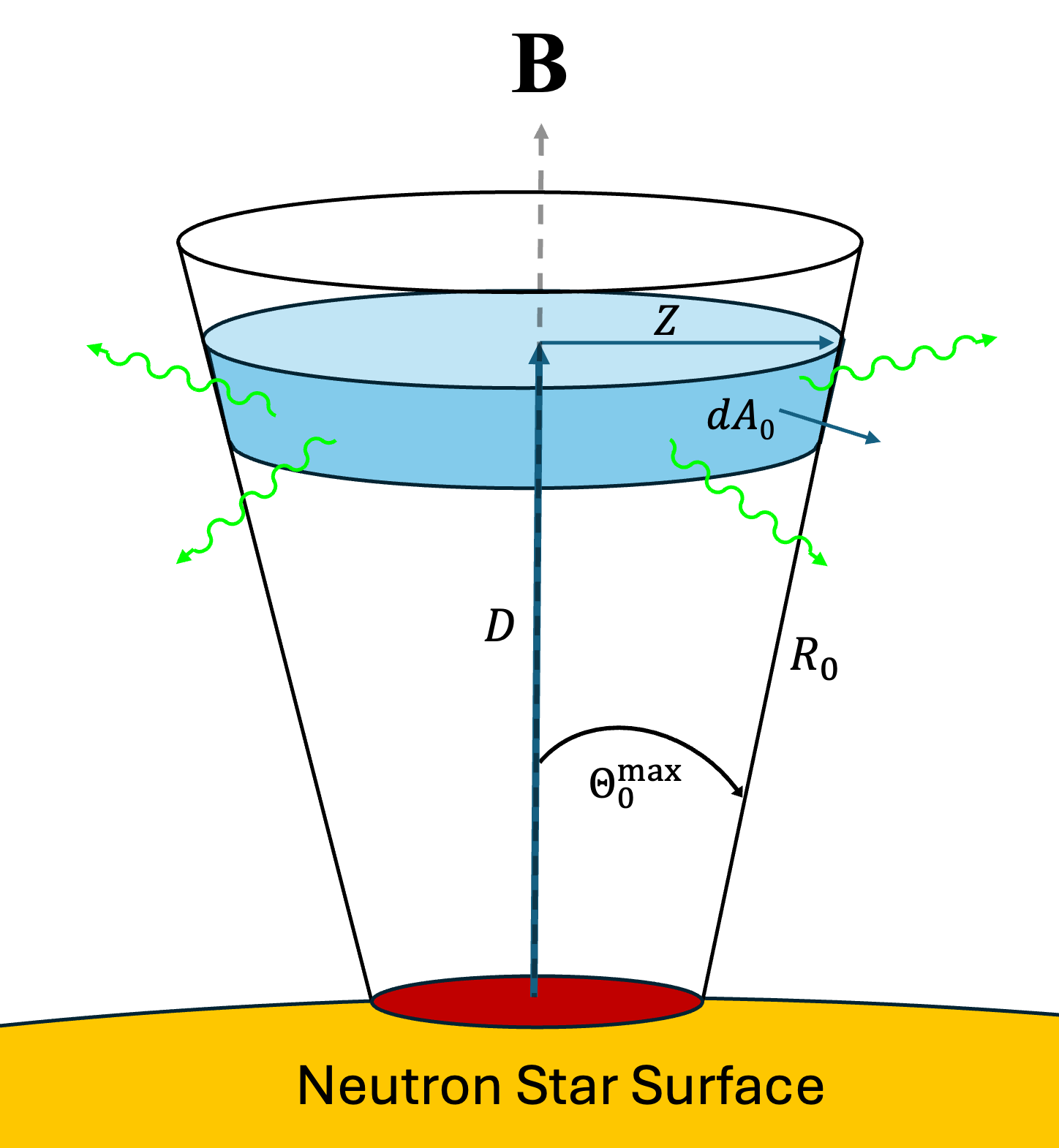}    
        \caption{Diagram illustrating the geometry for the escape of photons through band area $dA_0$ on the column wall, with $D = R_0 \cos \thetamax$ and $Z = R_0 \sin \thetamax$. See the discussion in the text.}
    \label{fig:intensitydiagram}
\end{figure}

The specific energy flux measured in the local frame due to photons escaping through the column wall, computed using the {\tt BW22} continuum model, is denoted by $F_{\epsilon,{\rm cont}}^\text{wall}$, with units of $\rm erg\,s^{-1}\,cm^{-2}
\,erg^{-1}$. It is important to relate this quantity to the photon number spectrum for emission through the column wall, $\dot N_\epsilon$, with units of $\rm s^{-1}\,cm^{-1} \,erg^{-1}$, which is computed using Equation~(\ref{convolutionwall}) for a general source term. Referring to the geometry indicated in Figure~\ref{fig:intensitydiagram}, we find that
\begin{equation}
    F_{\epsilon,{\rm cont}}^\text{wall} \, dA_0 = \epsilon_0\Dot{N}^\text{tot}_{\epsilon} \, dR_0 \ ,
    \label{fluxnumspecrelation}
\end{equation}
where
\begin{equation}
    dA_0 = 2\pi R_0\sin\thetamax \, dR_0 \ ,
    \label{dAdef}
\end{equation}
represents the differential area element on the surface of the column wall corresponding to a thin ring at radius $R_0$ with width $dR_0$. Solving Equation~(\ref{fluxnumspecrelation}) for the wall flux, $F_{\epsilon,{\rm cont}}^\text{wall}$, and substituting for $dA_0$ using Equation~(\ref{dAdef}), we obtain
\begin{equation}
    F_{\epsilon,{\rm cont}}^\text{wall}(R_0,\epsilon_0) = \frac{\epsilon_0\Dot{N}^\text{tot}_\epsilon(R_0,\epsilon_0)}{2\pi R_0\sin\thetamax} \ ,
    \label{fluxdefwall}
\end{equation}
where $\dot N^\text{tot}_\epsilon(R_0,\epsilon_0)$ is computed using Equation~(\ref{totalwallspec}). We will use Equation~(\ref{fluxdefwall}) to make the connection between the column wall continuum spectrum computed using the {\tt BW22} model and the flux and intensity measured in the local {\tt RM88} frame, which is the primary input for the relativistic {\tt RM88} formalism.

\subsubsection{Local Flux for Radiation Escaping from Column Top}
\label{sec:fluxfromcoltop}

The specific energy flux due to emission through the column top, computed using the {\tt BW22} continuum model, is denoted by $F_{\epsilon,{\rm cont}}^\text{top}$, with units $\rm erg\,s^{-1}\,cm^{-2}
\,erg^{-1}$. We must relate this quantity to the photon number spectrum for emission through the column top, $\dot{\mathcal{N}}^\text{tot}_\epsilon$, computed for a general source term using Equation~(\ref{convolutiontop}). We have
\begin{equation}
    F_{\epsilon,{\rm cont}}^\text{top} A_\text{top} = \epsilon_0 \, \Dot{\mathcal{N}}^\text{tot}_\epsilon\ ,
   \label{fluxtondottop}
\end{equation}
where $A_\text{top} = \Omega R_\text{top}^2$ is the area of the upper surface of a conical accretion column with solid angle $\Omega$. Solving Equation~(\ref{fluxtondottop}) for the specific energy flux at the column top yields
\begin{equation}
    F_{\epsilon,{\rm cont}}^\text{top}(\epsilon_0) = \frac{\epsilon_0 \, \Dot{\mathcal{N}}^\text{tot}_\epsilon(\epsilon_0)}{\Omega R_\text{top}^2} \ ,
    \label{fluxdeftop}
\end{equation}
where $\dot{\mathcal{N}}^\text{tot}_\epsilon(\epsilon_0)$ is computed using Equation~(\ref{totaltopspec}). In the following section, we will introduce the relativistic formalism used to trace the emission from the wall and top of the accretion column through the Schwarzschild metric to a detector located in the frame of a distant observer.

\section{RELATIVISTIC FORMALISM}
\label{sec:RelForm}

The formalism developed by {\tt BW22} provides a rigorous framework for the computation of the X-ray continuum spectrum in the frame of the neutron star that yields good agreement with the observational data for a wide variety of accretion-powered X-ray pulsars. However, in order to compute pulse profiles and true phase-averaged X-ray spectra, this continuum model must be cast in a relativistic framework so that the photon propagation through the Schwarzschild metric is properly followed, including the effects of gravitational lensing and the gravitational redshift. In addition, we must also implement a rigorous model for the rotational and magnetic geometry of the neutron star. In this section we will develop the required coordinate system and solve the relativistic characteristic equations for a photon produced at an emission point located on either the wall or top of the accretion column. We will then be able to calculate the corresponding spectral flux in the frame of the distant observer by making use of the appropriate coordinate transformations.

\subsection{Coordinate System}
\label{sec:coordsys}

\begin{figure}[htbp]
    \centering
    \includegraphics[width=0.9\linewidth]{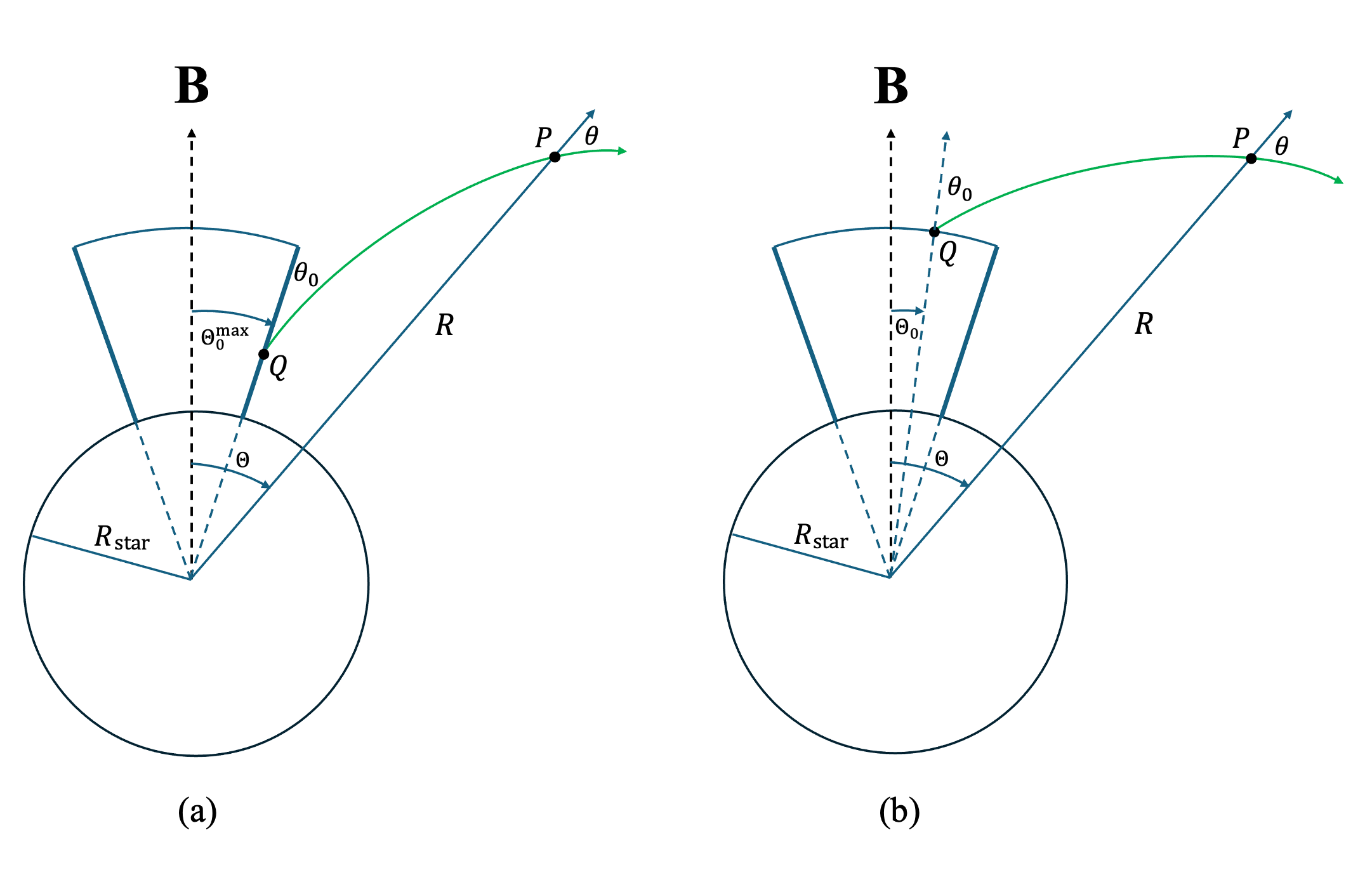}    \caption{Diagram of the RM88 coordinate system, depicting the curved path of a photon emitted from point $Q$ (green line), located on either (a) the column wall, or (b) the column top. The observer at point $P$ is at radius $R$ and has angle $\Theta$ relative to the magnetic axis.}
    \label{fig:rm88diagram}
\end{figure}

\begin{figure}[htbp]
    \centering
    \includegraphics[width=0.5\linewidth]{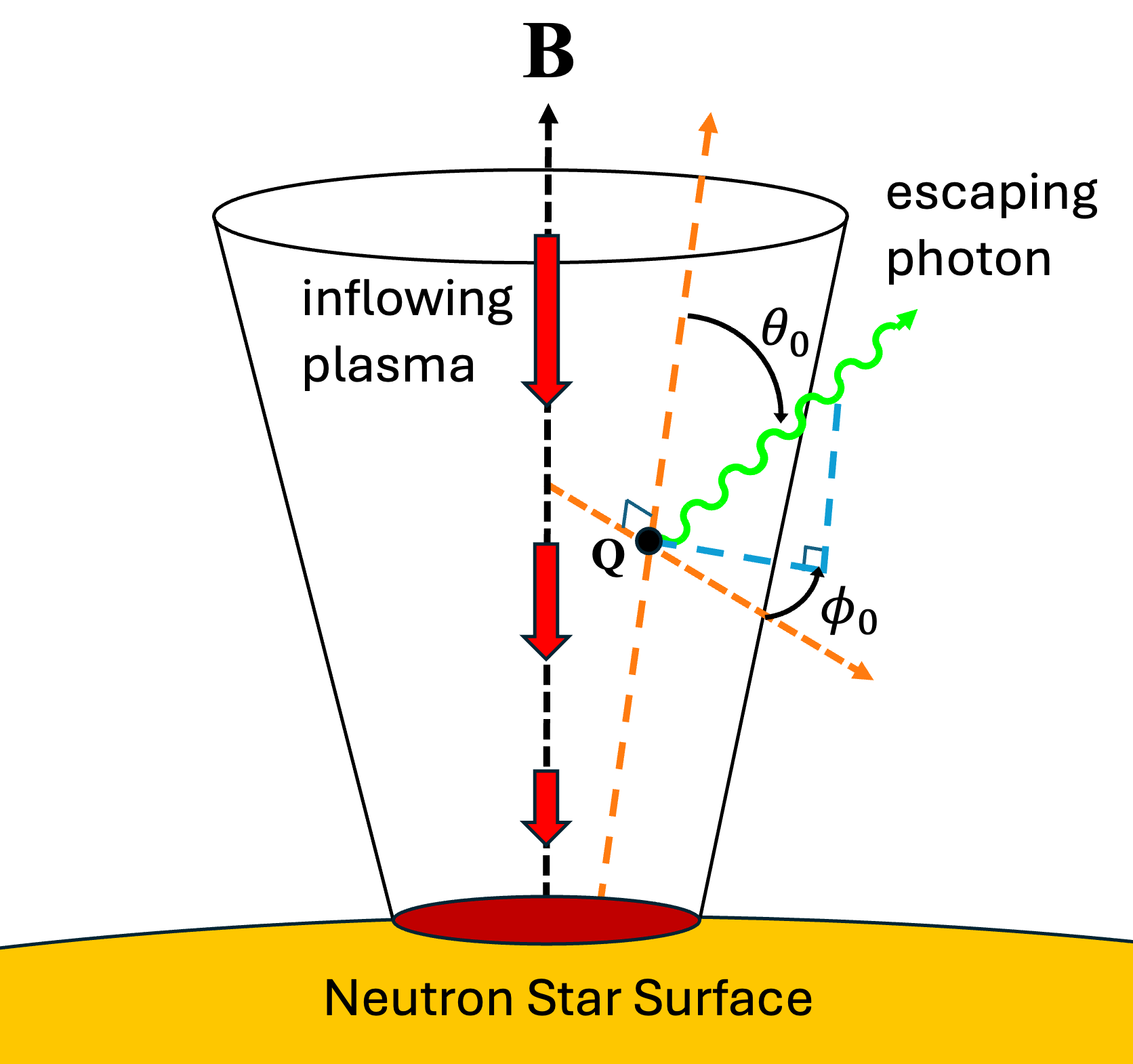}
    \caption{Diagram of the {\tt RM88} coordinate system used to describe the photon emission direction in the local frame that is stationary with respect to the star. The two dashed orange axes represent the polar axis (aligned with the column wall) and the axis normal to the wall. The angles $(\theta_0,\phi_0)$ describe the emission direction with respect to this polar coordinate system.}
    \label{fig:coordinatesys}
\end{figure}

We will follow the propagation of the photons produced in the XRP frame as they travel along null geodesics to the frame of the distant observer using the formalism developed by {\tt RM88}. The emission geometry is represented using a spherical-polar coordinate system centered at the emission point, $Q$, which is located at radius $R_0$ from the neutron star's center, on either the wall or the upper surface of the conical accretion column, as depicted in Figure~\ref{fig:rm88diagram}. The photon propagates along a curved path to the distant observer at point $P$, which is located at radius $R$ from the center of the neutron star and has polar angle $\Theta$ relative to the axis of the accretion column, which is also the magnetic axis. As illustrated in Figure~\ref{fig:coordinatesys}, the angle between the photon propagation direction and the radial vector measured in the frame of the distant observer at point $P$ is denoted by $\theta$. The corresponding propagation angle measured in the {\tt RM88} frame at the emission point $Q$ is $\theta_0$, and the associated azimuthal angle is $\phi_0$. Hence the angles $(\theta_0,\phi_0)$ indicate the propagation direction of a photon as viewed in the local {\tt RM88} frame at the emission point $Q$. The distance from the center of the neutron star to the emission point in the general coordinate system is represented by $r_0$, which is measured in gravitational units, so that
\begin{equation}
    r_0 \equiv \frac{R_0}{R_\text{S}} \ ,
\end{equation}
where $R_0$ is the dimensional emission radius, and $R_\text{S} = 2 G M_\text{star}/c^2$ denotes the Schwarzschild radius for the neutron star with mass $M_\text{star}$. In general, we will utilize lower-case letters to refer to distances measured in gravitational units, with upper-case letters used to refer to distances measured in CGS units.

\subsection{Flux Measured by Distant Observer}
\label{sec:DistFlux}

In order to calculate pulse profiles and phase-averaged spectra in the frame of a distant observer, we will need to transform the radiation spectrum from the {\tt RM88} frame, where the radiation is generated, to the frame of the observer, where the radiation is detected. The energy flux spectrum measured by an observer located at radius $r$ in the general coordinate system, denoted by  $F_{\epsilon,\text{obs}}$, is computed using the fundamental integral \citep{RiffertandMeszaros1988}
\begin{equation}
    F_{\epsilon,\text{obs}}(r,\epsilon) = \iint_{4\pi}I_{\epsilon,\text{obs}}(r,\mu,\phi,\epsilon) \, \mu \, d\mu \, d\phi \ ,
    \label{fluxEqstarFrame}
\end{equation}
where the intensity $I_{\epsilon,\text{obs}}(r,\mu,\phi,\epsilon)$ and the variables $\epsilon$, $\mu$, and $\phi$ are measured in the frame of the observer, with $\mu = \cos\theta$. In our model, the input radiation intensity is specified in the local {\tt RM88} frame, and therefore we must transform the variables of integration in Equation~(\ref{fluxEqstarFrame}) into the {\tt RM88} frame. The formal transformation of Equation~(\ref{fluxEqstarFrame}) from the variables $(r,\epsilon,\mu,\phi)$ to the variables $(r_0,\epsilon_0,\mu_0,\phi_0)$ was carried out by {\tt RM88}. The result is given by the integral
\begin{equation}
    F_{\epsilon,\text{obs}}(r,\epsilon) = \frac{\epsilon^3}{\epsilon_0^3}\iint_{4\pi}I_{\epsilon,\text{star}}(r_0,\mu_0,\phi_0,\epsilon_0) \, \mu D(\mu,\phi;\mu_0,\phi_0) \, d\mu_0 \, d\phi_0 \ ,
    \label{fluxEq}
\end{equation}
where $I_{\epsilon,\text{star}}(r_0,\mu_0,\phi_0,\epsilon_0)$ is the intensity measured in the local {\tt RM88} frame, and $D(\mu,\phi;\mu_0,\phi_0)$ is the Jacobian for the coordinate transformation, with $\mu_0 = \cos\theta_0$. The relationship between the photon energy measured in the {\tt RM88} frame, $\epsilon_0$, appearing on the right-hand side of Equation~(\ref{fluxEq}), and the energy measured in the frame of the distant observer, $\epsilon$, depends on the emission radius, $r_0$, via the standard gravitational redshift formula \citep{Weinberg1972}
\begin{equation}
    \frac{\epsilon_0}{\epsilon} = 1+z\equiv
    \left(1-\frac{1}{r_0}\right)^{-1/2} \ .
    \label{eq:redshift1}
\end{equation}
The Jacobian, $D$, for the transformation between the frame of the distant observer and the {\tt RM88} frame appearing in Equation~(\ref{fluxEq}) is given by the formal expression
\begin{equation}
    D(\mu,\phi;\mu_0,\phi_0)=\abs{\pdv{\mu}{\mu_0}\pdv{\phi}{\phi_0}-\pdv{\mu}{\phi_0}\pdv{\phi}{\mu_0}} \ .
    \label{jacobianDef}
\end{equation}
In practice, the formal expression for the Jacobian given by Equation~(\ref{jacobianDef}) takes on different specific forms depending on whether the emission originates from a point on the column wall or the column top. We will briefly review the detailed results that were obtained by {\tt RM88} for these two cases in Sections~\ref{sec:WallJac} and \ref{sec:TopJac}.

\subsection{Column Wall Jacobian}
\label{sec:WallJac}

The expression {\tt RM88} derived for the Jacobian transformation governing the emission from the column wall, $D_\text{wall}$, at dimensionless emission radius $r_0$, is given by
\begin{equation}
    \mu D_\text{wall}(\mu,\phi;\mu_0,\phi_0) = \frac{1}{r^2}\frac{r_0^2}{r_0-3/2}\frac{1}{Y(r_0,\mu_0)}\frac{\sin\thetamax\cos\phi_0}{(\sin^2\Theta-\sin^2\thetamax\sin^2\phi_0)^{1/2}} \ ,
    \label{walljacobian}
\end{equation}
where
\begin{equation}
    Y(r_0,\mu_0) = \int_{r_0}^r q(x,b)\,dx+\begin{cases}
        0, & \mu_0\geq 0 \ , \\
        2\int_{r_m}^{r_0}
q(x,b)\,dx, & \mu_0<0 \ , \end{cases}
\label{yeq}
\end{equation}
and 
\begin{equation}
    q(x,b)=\frac{x-3/4}{x(x-3/2)^2}\left[x^2-b^2\left(1-\frac{1}{x}\right)\right]^{-1/2} \ .
\end{equation}
Here, $b$ represents the impact parameter, which is a constant of the motion, computed using
\begin{equation}
    b\equiv r_0\sqrt{\frac{1-\mu_0^2}{1-1/r_0}}=r\sqrt{\frac{1-\mu^2}{1-1/r}} \ .
    \label{eq:impact}
\end{equation}
The quantity $\thetamax$ appearing in Equation~(\ref{walljacobian}) represents the opening angle of the conical column, which is the maximum value of the variable angle $\Theta_0$ (see Figure~\ref{fig:rm88diagram}), and $r_0$ is the radius measured from the center of the neutron star to the emission point, in the general coordinate system. Finally, the parameter $r_m$ in Equation~(\ref{yeq}) represents the ``turning radius'' (the radius of closest approach to the central mass), which applies only to cases with $\mu_0 < 0$. The value of $r_m$ is the physically acceptable root of the equation
\begin{equation}
r_m^3 - b^2 (r_m-1) = 0 \ ,
\label{eq1.11c}
\end{equation}
which can be expressed in the analytical form
\begin{equation}
r_m = \frac{2 b}{\sqrt{3}} \cos\left\{\frac{1}{3}\left[\pi - \tan^{-1}\left(\frac{\sqrt{4 b^2-27}}{3 \sqrt{3}}\right)\right]\right\} \ .
\qquad
\mu_0 < 0 \ .
\label{eq1.11d}
\end{equation}

\subsection{Column Top Jacobian}
\label{sec:TopJac}

The Jacobian transformation for the photons emitted from the top of the accretion column, $D_\text{top}$, at dimensionless emission radius $r_0 = r_\text{top}$, is computed using
\begin{align}
    \mu D_\text{top}(\mu,\phi;\mu_0,\phi_0)&=\left(\frac{r_\text{top}}{r}\right)^2\frac{\mu_0}{1-1/r_\text{top}}\frac{\sin\Theta_0}{\cos\phi\sin\Theta}\nonumber \\
&\times\left(\frac{\cos\Theta_0\sin K+\cos\phi_0\cos K\sin\Theta_0}{\sin\Theta_0\cos K+\cos\phi_0\sin K\cos\Theta_0}\right) \ ,
    \label{topjacobian}
\end{align}
where $\Theta$ denotes the angle between the magnetic field axis and the line of sight to the observer, and $\Theta_0$ represents the angle between the magnetic field axis and the emission point on the upper surface of the accretion column, which must satisfy the relation $0 \le \Theta_0 \le \thetamax$, with $\thetamax$ denoting the opening angle of the conical column (see Figure~\ref{fig:rm88diagram}). The quantity $K$ in Equation~(\ref{topjacobian}) represents the angle between the emission point on the column top and the line of sight to the observer, which is further discussed in Section~\ref{sec:relchar}.

\subsection{Relativistic Characteristics}
\label{sec:relchar}

In order to fully define the geometry of the spinning neutron star and its magnetic field relative to a distant observer, it is necessary to introduce several fundamental angles. First, we define the inclination angle, $\Psi$, as the angle between the line of sight to the distant observer and the spin axis of the neutron star. This angle remains constant as the star spins, but it may precess over much longer timescales \citep[see][]{Katz1973,GerendBoynton1976,Staubert2009,Leahy2025}. Another important angle is the rotational latitude of the accretion column, denoted by $\varphi$, which represents the angle between the spin axis and the magnetic field axis of the accretion column. The value of $\varphi$ also remains constant as the star rotates. The angle between the axis of the accretion column and the line of sight to the distant observer is denoted by $\Theta$, which we refer to as the ``observation angle.'' The value of $\Theta$ varies periodically due to the rotation of the neutron star, which is represented by the rotational phase angle in the frame of the star, denoted by $\beta$. The phase angle is calibrated so that $\beta = 0$ corresponds to the minimum possible value of $\Theta$ that occurs during the stellar rotation. Finally, we will use $\Theta_0$ to denote the angle between the magnetic field axis and the emission point, located on either the wall or the top of the accretion column.

Using spherical trigonometry, it can be shown that the periodic variation of $\Theta$ during the rotation of the neutron star is related to the values of $\Psi$, $\varphi$, and $\beta$ via
\begin{equation}
    \cos\Theta = \cos\Psi\cos\varphi+\sin\Psi\sin\varphi\cos\beta \ .
    \label{thetaandbetaeq}
\end{equation}
It therefore follows that $\Theta = \Psi-\varphi$ when $\beta = 0$. A point of clarification must be made regarding our notation for $\Theta_0$. For emission occurring along the wall of the accretion column, we have $\Theta_0 = \thetamax$, where $\thetamax$ is the opening angle of the conical column (see Figure~\ref{fig:intensitydiagram}). However, for emission occurring on the top of the column, the value of $\Theta_0$ falls in the range $0 \le \Theta_0 \le \thetamax$ (see Figure~\ref{fig:rm88diagram}).

The formalism developed by {\tt RM88} is based on analysis of the characteristic equations describing the propagation of radiation along null geodesics in the Schwarzschild metric. They define the parameter $K$ as the angle between the emission location and the direction to the distant observer. In flat spacetime, for an observer located at infinity, straightforward geometric relations yield $K = \sin^{-1}(b/r_0)$, where $r_0$ is the emission radius and $b$ is the impact parameter given by Equation~(\ref{eq:impact}). However, in general relativity, this relation is modified by the curvature of spacetime around the neutron star, and for an observer located at finite distance $r$, we obtain the general expression
\begin{equation}
    K(r,r_0,\mu_0) = b\int_{r_0}^r\left[x^2-b^2\left(1-\frac{1}{x}\right)\right]^{-\frac{1}{2}}\frac{dx}{x}+Q(r_0,\mu_0) \ ,
    \label{keq}
\end{equation}
where
\begin{equation}
    Q(r_0,\mu_0) = \begin{cases} 
      0 \ , & \mu_0 \geq 0 \ , \\
      2b\int_{r_m}^{r_0}\left[x^2-b^2\left(1-\frac{1}{x}\right)\right]^{-\frac{1}{2}}\frac{dx}{x} \ , & \mu_0 \leq 0 \ ,
      \label{qeq}
   \end{cases}
\end{equation}
where $r_m$ is given by Equation~(\ref{eq1.11d}). By integrating one of the characteristic equations obtained from the metric, {\tt RM88} also obtain the important angular relation
\begin{equation}
    \cos\Theta = \cos\Theta_0\cos K-\cos\phi_0\sin\Theta_0\sin K \ ,
    \label{sphericaltrig}
\end{equation}
which reduces to $\Theta = \Theta_0 + K$ for the special case $\phi_0 = 0$. We also note that Equation~(\ref{sphericaltrig}) can also be obtained by considering spherical trigonometric relations on the unit sphere. In the following sections we will discuss the procedure for determining the emission location on the wall or top of the accretion column such that photons emitted from that location in a specified direction will reach a distant observer with a selected value of $\Theta$.

\subsubsection{Column Wall Geometry Correspondence}
\label{sec:colwallgeo}

In order to connect the intensity specified in the local {\tt RM88} frame with the observed flux, we must develop the ray-tracing procedure required to determine the emission radius $r_0$ for photons emitted at point $Q$ in a specified direction on the column wall such that they travel along null geodesics to reach a distant observer located at polar angle $\Theta$ relative to the axis of the accretion column, as depicted in Figure~\ref{fig:rm88diagram}. For given values of the rotational inclination angle, $\Psi$, and the accretion column rotational latitude, $\varphi$, Equation~(\ref{thetaandbetaeq}) allows us to compute the variation of the observation angle $\Theta$ between the magnetic field axis (which is also the central axis of the accretion column) and the line of sight to the distant observer, as the rotational phase angle $\beta$ varies through the domain $0\degree \le \beta \le 360\degree$. At a single moment in time, it is sufficient to consider a ``snapshot'' of the stellar rotation, which yields a single value for $\Theta$, as indicated in Figure~\ref{fig:rm88diagram}. Combining this value of $\Theta$ with a set of emission angles $(\theta_0,\phi_0)$ measured in the local {\tt RM88} frame, we now seek to determine the value of the emission radius, $r_0$, such that photons propagating from an emission point located on the column wall will reach the distant observer at angle $\Theta$. The value of $r_0$ can be derived by
setting $\Theta_0 = \thetamax$ in Equation~(\ref{sphericaltrig}), since the emission point is located on the column wall, and then solving for $K$. The result obtained is
\begin{align}
    \nonumber\cos K(r,r_0,\mu_0)&=\frac{\cos\Theta\cos \thetamax}{\cos^2 \thetamax+\cos^2\phi_0\sin^2 \thetamax}\\
    &\pm\frac{\cos\phi_0\sin \thetamax\sqrt{\cos^2\phi_0\sin^2 \thetamax+\cos^2\thetamax-\cos^2\Theta}}{\cos^2 \thetamax+\cos^2\phi_0\sin^2 \thetamax} \ ,
    \label{kanalytic}
\end{align}
where $\mu_0 = \cos\theta_0$.
Note that in general, Equation~(\ref{kanalytic}) may yield zero, one, or two physically acceptable values for $K(r,r_0,\mu_0)$. We can combine this result with Equation~(\ref{keq}), which links the value of $K(r,r_0,\mu_0)$ with the value of $r_0$, since $\mu_0$ is specified, and we assume that the distance to the observer, $r = R/R_\text{S}$, has a known value. Hence we conclude that in principle, a single emission vector $(\theta_0,\phi_0)$ in the {\tt RM88} frame may yield up to two acceptable values for the emission radius $r_0$ along the column wall, such that the photons reach the distant observer at observation angle $\Theta$.
Acceptable values for $r_0$ lie in the domain $r_\text{star} \le r_0 \le r_\text{top}$, where $r_\text{star} = R_\text{star}/R_\text{S}$ and $r_\text{top} = R_\text{top} / R_\text{S}$ denote the dimensionless stellar radius and the dimensionless column top radius, respectively.

In order to determine the variation of the emission radius $r_0$ as the star rotates, for a given set of emission angles $(\theta_0,\phi_0)$, we proceed as follows. First, we must select values for the column opening angle $\thetamax$, the inclination angle of the neutron star's rotation axis, $\Psi$, the rotational latitude of the accretion column, $\varphi$, and the rotational phase angle, $\beta$. Next, we use Equation~(\ref{thetaandbetaeq}) to determine the associated value of the observation angle $\Theta$. We can then combine Equations~(\ref{keq}) and (\ref{kanalytic}) to determine the resulting roots for the emission radius $r_0$. If, for a given set of emission angles $(\theta_0,\phi_0)$, the solution process yields zero acceptable values for the emission radius $r_0$, then we must conclude that no radiation will reach the observer for that value of $\beta$. On the other hand, if we obtain one acceptable root for $r_0$, then we must include emission from that point on the column wall in our calculation of the observed spectrum for that value of $\beta$. In certain situations, we may obtain two acceptable values for $r_0$, which implies that two separate null geodesics will reach the observer. In this case, we must include radiation generated from the emission points located at both values of $r_0$.

\subsubsection{Column Top Geometry Correspondence}
\label{sec:coltopgeo}

Next we consider the problem of determining the location of the emission point $Q$ on the column top such that photons emitted in a specified direction will reach a distant observer located at angle $\Theta$ relative to the central axis of the accretion column, as indicated in Figure~\ref{fig:rm88diagram}. Hence our present goal is to determine the required value of the emission angle $\Theta_0$, which must lie somewhere in the domain $0 \le \Theta_0 \le \thetamax$. We note that this is distinct from the case of the column wall emission, treated in Section~\ref{sec:colwallgeo}, where the goal was to determine the value of the emission radius $r_0$. In the case of the column top emission, the value of the emission radius is frozen at $r_0 = r_\text{top}$, and once the values of the emission angles $(\theta_0,\phi_0)$ are selected, we can compute the value of $K(r,r_0,\mu_0)$ using Equation~(\ref{keq}). With the value of $K$ determined, we can now solve Equation~(\ref{sphericaltrig}) to compute the associated value of $\Theta_0$. The result obtained is
\begin{align}
    \nonumber\cos {\Theta}_0&=\frac{\cos\Theta\cos K}{\cos^2 K+\cos^2\phi_0\sin^2 K}\\
    &\pm\frac{\cos\phi_0\sin K\sqrt{\cos^2\phi_0\sin^2 K+\cos^2 K-\cos^2\Theta}}{\cos^2 K+\cos^2\phi_0\sin^2 K}
    \label{theta0analytic} \ .
\end{align}
In analogy with Equation~(\ref{kanalytic}), we note that Equation~(\ref{theta0analytic}) may yield zero, one, or two acceptable roots for $\Theta_0$ for a given emission vector $(\theta_0,\phi_0)$. Note that here, $K(r,r_0,\mu_0)$ depends only on the value of $\mu_0 = \cos\theta_0$, since $r_0 = r_\text{top}$ for all points on the column top, and the distance to the observer, $r$, is assumed to be a known quantity. In the following section, we will discuss in detail our unitary emission model, in which the intensity distribution measured in the local {\tt RM88} frame is represented using a group of ``laser-like'' emission functions in a specified set of directions.

\section{UNITARY EMISSION MODEL}
\label{sec:UniMod}

One of the central unsolved problems related to the formation of X-ray pulsar spectra is the nature of the angular distribution of the radiation emitted from the surface of the accretion column. Detailed knowledge of this angular distribution is required in order to link the X-ray continuum spectrum emitted from the surface of the column with a relativistic formalism that can trace the paths of the photons along null geodesics to the distant observer. This problem has been approached in a variety of ways by previous authors. For example, \citet{Leahy_1991} assumed a simple ad hoc form for the angular distribution of the emitted radiation in developing his formalism for the computation of simulated pulse profiles that can be compared with the data for specific sources. In particular, \citet{Leahy_1991} assumed that the radiation distribution at the surface of an X-ray pulsar can be modeled using rings of emission surrounding the magnetic poles of the neutron star. He used this approach to compute pulse profiles based on the work of \citet{MeszarosandNagel1985a}, who suggested that the angular distribution of the surface intensity, $I$, can be described by either $I\propto\cos^2\theta_0$ or $I\propto\cos^4\theta_0$, where $\theta_0$ is the propagation angle relative to the local magnetic field direction.

In this paper, we will assume that the energy distribution of the X-ray continuum radiation emitted from the surface of the accretion column is given by the {\tt BW22} formalism. However, this formalism does not specify the angular distribution of the emitted radiation. Rather than adopting any ad hoc assumptions about the angular distribution of the emitted radiation, we will assume that we have no detailed a priori knowledge about this distribution, other than a few general constraints on the behavior. Instead, we will solve for the angular distribution of the radiation field in the local {\tt RM88} frame (also called the beaming pattern) as part of our procedure for fitting the pulse profile data for a given source. In order to accomplish this, we will adopt a novel approach in which the radiation field is specified using an expansion of ``laser-like'' emission vectors, each oriented in a specific unitary direction as observed in the {\tt RM88} frame. For a given value of the emission radius, $r_0$, the energy distribution of the intensity associated with each emission direction will be assigned using the {\tt BW22} continuum model, evaluated at that value of $r_0$. Hence each emission direction connected with a specific emission radius will have the same energy distribution. However, the values of the angular weight coefficients for each emission direction will be different, and will be determined as part of the fitting procedure. In some ways, our approach is analogous to that taken by \citet{PoutanenBeloborodov2006}, who developed an expansion method based on Fourier components that was used to fit observational pulse profile data. Similarly, \citet{Kraus_etal_1995} and \citet{Blum_Kraus_2000} formulated an expansion method in which the beaming pattern is computed by fitting the pulse profile data using their theoretical model, which is based on emission from rings and hot spots on the stellar surface.

\subsection{Emission Vectors in {\tt RM88} Frame}

In the case of the column top emission, the broadest possible angular range for the {\tt RM88} polar angle $\theta_0$ corresponds to the domain $0\degree < \theta_0 < 90\degree$, because emission at angles $\theta_0 > 90\degree$ would be directed into the accretion column, and would therefore not be connected with any null geodesics that can reach the distant observer (see Figure~\ref{fig:coordinatesys}). Furthermore, the emission from the top of the accretion column at radius $r = r_\text{top}$ satisfies the free-streaming boundary condition given by Equation~(\ref{ndottopfromgreens}). This boundary condition is generally expected to result in the partial collimation of the emitted radiation in the outward radial direction \citep{Paletou2018,Nagel1981a}. Motivated by this behavior, we therefore make the further restriction that the column top emission is confined to the domain $0\degree < \theta_0 < 45\degree$. For the column top emission, the {\tt RM88} azimuthal angle $\phi_0$ is unrestricted, so that $0\degree < \phi_0 < 360\degree$. In the case of the column wall emission, the domain for the {\tt RM88} polar angle $\theta_0$ is $0\degree < \theta_0 < 180\degree$, and the domain for the corresponding {\tt RM88} azimuthal angle $\phi_0$ is $-90\degree < \phi_0 < 90\degree$. These constraints are required in order to ensure that the emitted photons travel away from the column, rather than through the column wall into the interior region.

In our computational applications, we will model the sets of {\tt RM88} emission vectors for the radiation originating on the column wall or top using a discrete grid of $(\theta_{0i},\phi_{0j})$ values, where the indices $i$ and $j$ indicate the emission direction. Specifically, in our model for the column wall emission, we adopt uniformly-spaced grids for the values of $\theta_0$ and $\phi_0$, with
\begin{align}
    \theta_{0i}^\text{wall} &\in \left\{0\degree,1\degree,2\degree,\ldots,180\degree\right\} \ ,\nonumber\\
    \phi_{0j}^\text{wall} &\in \left\{-90\degree,-67.5\degree,-45\degree,\ldots,90\degree\right\} \ .
    \label{anglegridwall}
\end{align}
For the column top emission, the angular grids adopted are
\begin{align}
    \theta_{0i}^\text{top} &\in \left\{0\degree,1\degree,2\degree,\ldots,45\degree\right\} \ ,\nonumber\\
    \phi_{0j}^\text{top} &\in \left\{0\degree,40\degree,80\degree,\ldots,320\degree\right\} \ .
    \label{anglegridtop}
\end{align}
Our angular grids therefore comprise $N_\text{wall} = 180$ values for $\theta_{0i}^\text{wall}$, $N_\text{top} = 45$ values for $\theta_{0i}^\text{top}$, $M_\text{wall} = 9$ values for $\phi_{0j}^\text{wall}$, and $M_\text{top} = 9$ values for $\phi_{0j}^\text{top}$. We also note that each of the $(\theta_{0i},\phi_{0j})$ grids has a corresponding $(\mu_{0i},\phi_{0j})$ grid via the relation $\mu_{0i} = \cos\theta_{0i}$. 
The primary benefit of utilizing a discrete system of emission directions in the {\tt RM88} frame is that it provides us with the ability to build ``basis functions'' for the simulation of pulse profiles, with each basis function representing the discrete contribution to the observed pulse profile due to emission in a given direction $(\theta_{0i},\phi_{0j})$ originating on either the column wall or the column top.

The linearity of the fundamental transport equation (Equation~(\ref{eq:transEq})) implies that a linear combination of the basis functions can be used to simulate the pulse profile for a given XRP, which can then be compared with the observational data in order to determine the values of the angular weight coefficients for that specific source. We will refer to the basis functions as ``sub-profiles,'' with each one corresponding to a specific emission direction. Because we will be considering emission from a rotating neutron star with two accretion columns in our detailed applications, we will actually utilize four separate sets of sub-profiles, corresponding to emission originating on either the wall or top of column~1 or column~2.

\subsection{Azimuthal Isotropy}
\label{sec:AziIso}

The emission direction in the local {\tt RM88} frame in our model is specified using the polar angle $\theta_{0}$ and the azimuthal angle $\phi_{0}$, as indicated in Figure~\ref{fig:coordinatesys}. Based on the detailed results for the energy and angular dependence of the electron scattering cross section in a strong magnetic field obtained by \citet{Meszaros_Ventura_1979}, we expect that the scattering cross section for the photons in the accretion column depends only on the polar angle $\theta_0$, and that it does not depend on the azimuthal angle $\phi_0$. We therefore make the assumption here that the intensity amplitudes at a given location inside the accretion column will be distributed isotropically with respect to $\phi_0$. Within the context of our unitary emission model, this implies that for a given value of $\theta_{0i}$ in our polar grid, the intensity will be the same for all values of $\phi_{0j}$ in the azimuthal grid.

\subsection{Angular Expansion for Column Wall Emission}
\label{sec:AngExpWall}

For the case of the column wall emission, we will implement our unitary emission model by representing the energy and angular distribution of the radiation field escaping through the column wall as measured in the local {\tt RM88} frame using the expansion
\begin{equation}
    I_{\epsilon,\text{star}}^\text{wall}(R_0,\mu_0,\phi_0,\epsilon_0) = \sum_{i=1}^{N_\text{wall}} \sum_{j=1}^{M_\text{wall}} W_{i}^\text{wall} \, F_{\epsilon,{\rm cont}}^\text{wall}(R_0,\epsilon_0) \, \delta(\mu_{0}-\mu_{0i})\delta(\phi_{0}-\phi_{0j}) \ ,
    \label{intensitysumwallNEW2}
\end{equation}
where $N_\text{wall}$ and $M_\text{wall}$ denote the number of $\theta_{0i}$ and $\phi_{0j}$ values in the angular grids given by Equations~(\ref{anglegridwall}) and (\ref{anglegridtop}), respectively, $W_{i}^\text{wall}$ is the linear weight coefficient for the radiation emitted with polar angle $\theta_{0i} = \cos^{-1}\mu_{0i}$ in the local {\tt RM88} frame, and $F_{\epsilon,{\rm cont}}^\text{wall}$ represents the continuum wall flux computed using the {\tt BW22} model, defined in Equation~(\ref{fluxdefwall}). We note that the index ``$i$'' of the weight coefficient $W_{i}^\text{wall}$ corresponds to the polar angle $\theta_{0i} = \cos^{-1}\mu_{0i}$ in the local {\tt RM88} frame. We note that the weight coefficient $W_{i}^\text{wall}$ has only a single index because there is no dependence on the value of $\phi_{0j}$ due to the form of the electron scattering cross section in a strong magnetic field, as discussed in Section~\ref{sec:AziIso}.

The appearance of the angular $\delta$-functions in Equation~(\ref{intensitysumwallNEW2}) implies that we are expanding the emission over a set of precise ``laser-like'' emission directions. Each of these precise directions yields a contribution to the observed pulse profile that is weighted by the corresponding expansion coefficient. This is a novel approach that is in some ways analogous to the pulse-profile expansion method developed by \citet{Kraus_etal_1995} and \citet{Blum_Kraus_2000}. Combining Equations~(\ref{fluxdefwall}) and (\ref{intensitysumwallNEW2}) yields
\begin{equation}
    I_{\epsilon,\text{star}}^\text{wall}(R_0,\mu_0,\phi_0,\epsilon_0) = \sum_{i=1}^{N_\text{wall}} \sum_{j=1}^{M_\text{wall}} \frac{W_{i}^\text{wall} \, \epsilon_0\Dot{N}^\text{tot}_\epsilon(R_0,\epsilon_0)}{2\pi R_{0}\sin\thetamax} \, \delta(\mu_{0}-\mu_{0i})\delta(\phi_{0}-\phi_{0j}) \ ,
    \label{intensitysumwallNEW}
\end{equation}
where $\dot N^\text{tot}_\epsilon(R_0,\epsilon_0)$ is computed using Equation~(\ref{totalwallspec}). This expression provides the fundamental description of the energy and angular dependence of the radiation emitted through the column wall in the unitary emission model utilized here. Note that the weight coefficient $W_i^\text{wall}$ gives the amplitude of the intensity measured in the local {\tt RM88} frame due to emission in direction $\mu_{0i}$ relative to the {\tt BW22} continuum value $F_{\epsilon,{\rm cont}}^\text{wall}$, computed using Equation~(\ref{fluxdefwall}).

It is important to emphasize a point regarding the expansion coefficients, $W_{i}^\text{wall}$, appearing in the intensity expansion in Equation~(\ref{intensitysumwallNEW}). These weight coefficients are determined by fitting the model to the observed X-ray pulse profile data. However, since Equation~(\ref{intensitysumwallNEW}) gives the intensity in the {\tt RM88} frame, we must combine this expression with Equation~(\ref{fluxEq}) in order to compute the intensity distribution measured in the frame of a distant observer. We make no a priori assumptions about the shape of the angular intensity distribution (the beaming pattern), represented by the expansion coefficients, $W_{i}^\text{wall}$, and therefore the intensity distribution is an emergent property of the fit to the observed data. While this expansion method can produce excellent fits to the pulse profile data due to the quasi-orthogonal nature of the sub-profile basis functions, it is important to address the possible degeneracy of the parameter space, which can result in multiple sets of parameters yielding equivalently good fits to the data. We have addressed the possible degeneracy of the parameter space by developing a combined statistical metric that includes both the shape of the beaming pattern and the fit to the pulse profile data in order to determine the best overall set of theoretical parameters. This procedure is discussed in detail in Section~\ref{sec:parameterspacesearch}.

\subsection{Angular Expansion for Column Top Emission}

Proceeding in an analogous manner to the previous section, we now treat the emission escaping from the accretion column through the column top. In this case, the expansion representing the energy and angular distribution of the radiation field measured in the local {\tt RM88} frame is given by
\begin{equation}
    I_{\epsilon,\text{star}}^\text{top}(\mu_0,\phi_0,\epsilon_0) = \sum_{i=1}^{N_\text{top}} \sum_{j=1}^{M_\text{top}} W^\text{top}_{i} \, F_{\epsilon,{\rm cont}}^\text{top}(\epsilon_0) \, \delta(\mu_0-\mu_{0i})\delta(\phi_0-\phi_{0j}) \ ,
    \label{totalintensitydeftopNEW}
\end{equation}
where $N_\text{top}$ and $M_\text{top}$ represent the number of $\theta_{0i}$ and $\phi_{0j}$ values in the angular grids given by Equations~(\ref{anglegridwall}) and (\ref{anglegridtop}), respectively, $F_{\epsilon,{\rm cont}}^\text{top}$ denotes the {\tt BW22} continuum flux for the column top emission, defined in Equation~(\ref{fluxdeftop}), and $W^\text{top}_{i}$ represents the linear expansion coefficient for the column top emission generated with polar angle $\theta_{0i} = \cos^{-1}\mu_{0i}$. Once again, the angular $\delta$-functions in Equation~(\ref{totalintensitydeftopNEW}) express the ``laser-like'' emission directions, each of which yields a distinct contribution to the observed pulse profile that is weighted by the associated expansion coefficient, in a manner that is similar to the pulse-profile expansion method developed by \citet{Kraus_etal_1995} and \citet{Blum_Kraus_2000}. We can now combine Equations~(\ref{fluxdeftop}) and (\ref{totalintensitydeftopNEW}) to obtain the equivalent result
\begin{equation}
    I_{\epsilon,\text{star}}^\text{top}(\mu_0,\phi_0,\epsilon_0) = \sum_{i=1}^{N_\text{top}} \sum_{j=1}^{M_\text{top}} \frac{W^\text{top}_{i}\epsilon_0\,\Dot{\mathcal{N}}^\text{tot}_\epsilon(\epsilon_0)}{\Omega R_\text{top}^2} \, \delta(\mu_0-\mu_{0i})\delta(\phi_0-\phi_{0j}) \ ,
    \label{totalintensitydeftopNEW2}
\end{equation}
where $\dot{\mathcal{N}}^\text{tot}_\epsilon(\epsilon_0)$ is computed using Equation~(\ref{totaltopspec}). This relation describes the energy and angular dependence of the radiation escaping through the column top in the unitary emission model, where the weight coefficient $W_i^\text{top}$ expresses the intensity measured in the local {\tt RM88} frame due to emission in direction $\mu_{0i}$ relative to the {\tt BW22} continuum value $F_{\epsilon,{\rm cont}}^\text{top}$. Equation~(\ref{totalintensitydeftopNEW2}) provides the intensity in the {\tt RM88} frame, and therefore this expression must be combined with Equation~(\ref{fluxEq}) to compute the intensity distribution measured in the frame of a distant observer.

The expansion formalism defined above in Equations~(\ref{intensitysumwallNEW}) and (\ref{totalintensitydeftopNEW2}) provides a powerful and flexible methodology for expressing the energy and angular distribution of the radiation emitted through the walls and top of the conical accretion column, respectively, as measured in the {\tt RM88} frame. The associated sets of expansion coefficients $\{W^\text{wall}_{i}\}$ and $\{W^\text{top}_{i}\}$ will be determined by fitting the theoretical predictions to the pulse profile data for a given source, as discussed in Section~\ref{sec:spectrum}. In addition to fitting the pulse profile data, another important constraint that the expansion coefficients must satisfy is the requirement of local conservation, as discussed below.

\subsection{Local Energy Conservation Condition}
\label{sec:columnwallintensity}

\begin{figure}
    \centering
    \includegraphics[width=0.7\linewidth,]{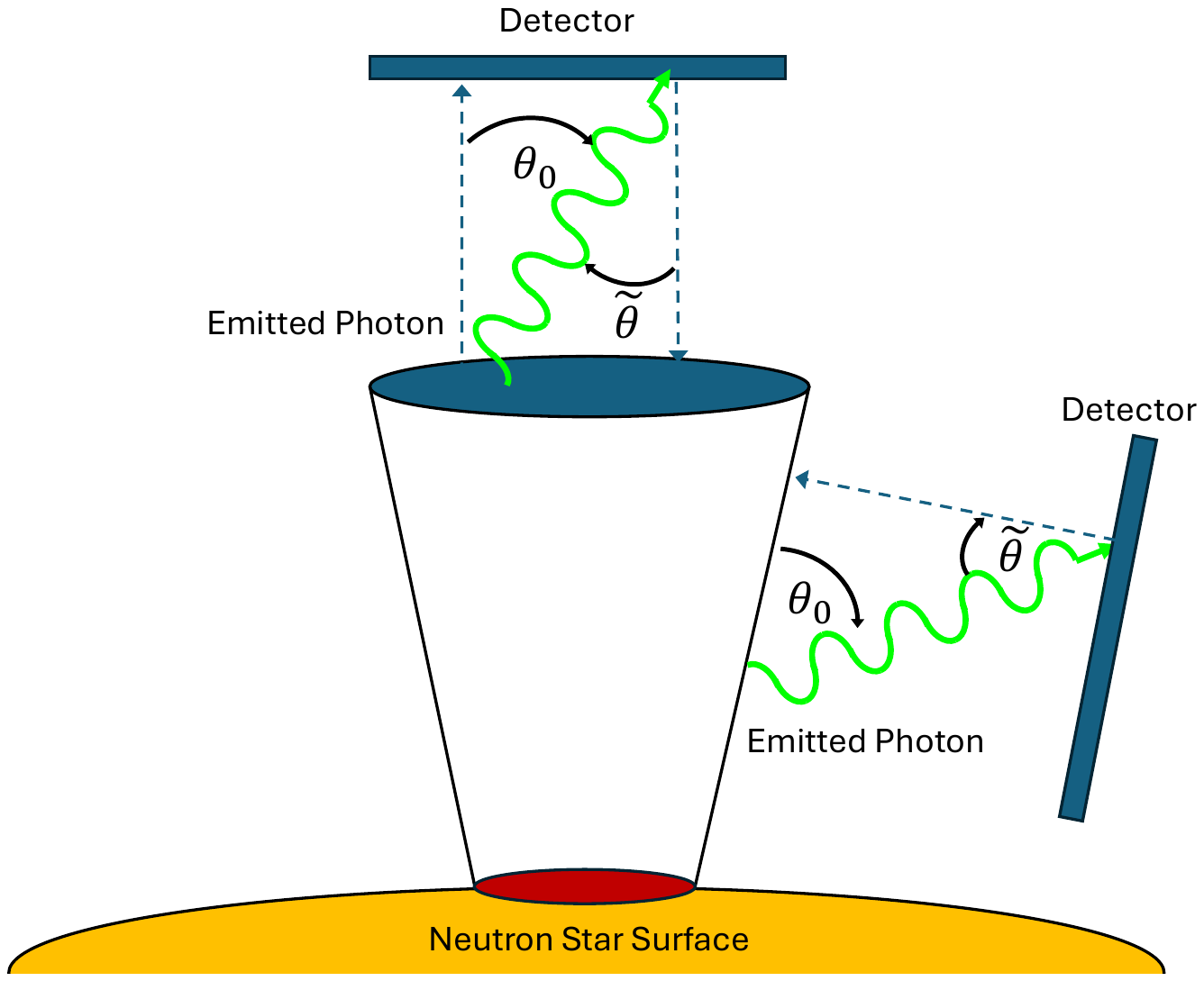}
    \caption{Representation of the coordinate transformation between the {\tt RM88} frame and the local detector frame. This transformation is required in order to calculate the energy flux in the local detector frame, which ensures local energy conservation. }
    \label{fig:detector}
\end{figure}

In order to ensure that energy is conserved in the frame of a local observer, who is stationary with respect to the neutron star, we must guarantee that the total energy flux, summed over all of the unitary emission directions, is equal to the continuum flux calculated using the {\tt BW22} model, which is given by Equation~(\ref{fluxdefwall}) for the column wall emission, or by Equation~(\ref{fluxdeftop}) for the column top emission. The first step is to consider how the energy flux measured by the local observer's detector, denoted by $F_{\epsilon,\text{star}}$, is related to the local intensity distribution. In general, the fundamental expression for the energy flux spectrum is given by \citep{RybickiandLightman1979}
\begin{equation}
    F_{\epsilon,\text{star}}(\epsilon_0) = \iint_{4\pi} I_{\epsilon,\text{star}}(\epsilon_0) \, \sin\tilde\theta\cos\tilde\theta \, d\tilde\theta \, d\tilde\phi \ ,
    \label{fluxintwoframes}
\end{equation}
where the detector-frame polar angle $\tilde\theta$ is measured with respect to the normal to the detector plane, as illustrated in Figure~\ref{fig:detector}. The detector-frame azimuthal angle $\tilde\phi$ corresponds to the rotation around the normal direction. The relation between the detector-frame angles $(\tilde\theta,\tilde\phi)$ and the corresponding {\tt RM88}-frame angles $(\theta_0,\phi_0)$ depends on whether the emission is escaping through the wall or top of the accretion column. We will consider each of these possibilities separately.

\subsubsection{Column Wall Energy Conservation}
\label{sec:energyconswall}

First we calculate the local energy flux spectrum for the radiation emitted through the column walls using Equation~(\ref{fluxintwoframes}). In this case, the {\tt RM88} coordinates $(\theta_0,\phi_0)$ are related to the local detector coordinates $(\tilde\theta,\tilde\phi)$ via a simple $90\degree$ rotation because the normal to the local detector plane is perpendicular to the local radial vector, as indicated in Figure~\ref{fig:detector}. In Appendix~\ref{sec:appendixA}, we demonstrate that under this coordinate transformation, we obtain (see Equation~(\ref{eq:DetFlux}))
\begin{equation}
    F_{\epsilon,\text{star}}^\text{wall}(\epsilon_0) = \iint_{4\pi} I_{\epsilon,\text{star}}^\text{wall}(\epsilon_0) \, (\sin\theta_0\cos\phi_0)\sin\theta_0 \, d\theta_0 \, d\phi_0 \ ,
    \label{eq:wallDet1}
\end{equation}
where the factor $(\sin\theta_0\cos\phi_0)$ represents the dot product between the emission vector and the normal to the local detector surface. Using our standard substitution $\mu_0 = \cos\theta_0$, Equation~(\ref{eq:wallDet1}) can be rewritten as
\begin{equation}
    F_{\epsilon,\text{star}}^\text{wall}(\epsilon_0) = \iint_{4\pi} I_{\epsilon,\text{star}}^\text{wall}(\epsilon_0) \sqrt{1-\mu_0^2}\cos\phi_0 \, d\mu_0 \, d\phi_0 \ .
    \label{fluxintegration}
\end{equation}
This relation can be used to derive an energy conservation constraint for the column wall expansion coefficients $W_{i}^\text{wall}$ appearing in Equation~(\ref{intensitysumwallNEW}) as follows. By substituting Equation~(\ref{intensitysumwallNEW}) into Equation~(\ref{fluxintegration}) and reversing the order of integration and summation, we obtain
\begin{align}
    F_{\epsilon,\rm{star}}^\text{wall}(R_0,\epsilon_0) = \sum_{i=1}^{N_\text{wall}} \sum_{j=1}^{M_\text{wall}} &\int_0^{2\pi}\int_{-1}^{1}\frac{W_{i}^\text{wall}\epsilon_0\Dot{N}^\text{tot}_\epsilon(R_0,\epsilon_0)}{2\pi R_{0}\sin\thetamax} \nonumber\\
    &\times \delta(\mu_{0}-\mu_{0i})\delta(\phi_{0}-{\phi}_{0j})\sqrt{1-{\mu_0}^2}\cos{\phi}_0 \, d{\mu}_0 \, d{\phi}_0 \ .
    \label{eq:Fwall1}
\end{align}
The evaluation of the integrals is trivial, and we find that Equation~(\ref{eq:Fwall1}) reduces to
\begin{equation}
    F_{\epsilon,\rm{star}}^\text{wall}(R_0,\epsilon_0) = \sum_{i=1}^{N_\text{wall}} \sum_{j=1}^{M_\text{wall}} \frac{W_{i}^\text{wall}\epsilon_0\Dot{N}^\text{tot}_\epsilon(R_0,\epsilon_0)}{2\pi R_{0}\sin\thetamax}\sqrt{1-{\mu_{0i}}^2}\cos{\phi}_{0j} \ .
    \label{starfluxuni}
\end{equation}
Equation~(\ref{starfluxuni}) gives the energy flux measured in the local {\tt RM88} frame as an expansion of contributions to the flux from photons escaping in each emission direction.

In order to ensure local energy conservation, the flux computed using Equation~(\ref{starfluxuni})  must agree with the energy flux computed using the {\tt BW22} continuum model, $F_{\epsilon,{\rm cont}}^\text{wall}(R_0,\epsilon_0)$, given by Equation~(\ref{fluxdefwall}). In other words, we require that
\begin{equation}
    F_{\epsilon,\rm{star}}^\text{wall}(R_0,\epsilon_0) = F_{\epsilon,{\rm cont}}^\text{wall}(R_0,\epsilon_0) \ .
    \label{eq:WallEnCon1}
\end{equation}
Combining Equations~(\ref{fluxdefwall}), (\ref{starfluxuni}), and (\ref{eq:WallEnCon1}) yields the energy conservation constraint
\begin{equation}
    \frac{\epsilon_0\Dot{N}^\text{tot}_\epsilon(R_0,\epsilon_0)}{2\pi R_0\sin\thetamax}
    = \sum_{i=1}^{N_\text{wall}} \sum_{j=1}^{M_\text{wall}} \frac{W_{i}^\text{wall}\epsilon_0\Dot{N}^\text{tot}_\epsilon(R_0,\epsilon_0)}{2\pi R_{0}\sin\thetamax}\sqrt{1-{\mu_{0i}}^2}\cos{\phi}_{0j} \ .
    \label{starfluxcont} 
\end{equation}
At this juncture, it is important to emphasize that the energy conservation calculation is carried out in the frame of a stationary observer located at radius $R_0$, so that in this application, the value of $R_0$ is independent of the emission direction $(\mu_{0i},\phi_{0j})$. Hence, the factors of $\dot N^\text{tot}_\epsilon(R_0,\epsilon_0)$ on both sides of Equation~(\ref{starfluxcont}) cancel, and upon simplification, we obtain
\begin{equation}
    e^\text{wall}\equiv\sum_{i=1}^{N_\text{wall}} \sum_{j=1}^{M_\text{wall}} W_{i}^\text{wall}\sqrt{1-\mu_{0i}^2}\cos\phi_{0j} = 1 \ ,
    \label{energyconswall}
\end{equation}
where $e^\text{wall}$ denotes the energy conservation constant for the column wall emission.
In order to guarantee local energy conservation throughout the column, the set of weight coefficients $\{W_{i}^\text{wall}\}$ obtained by the fit to the observational data must satisfy Equation~(\ref{energyconswall}), with $e^\text{wall} = 1$. If a fit yields a set of coefficients $\{W_{i}^\text{wall}\}$ which do not obey Equation~(\ref{energyconswall}), then that fit is not considered to be physically acceptable.

\subsubsection{Column Top Energy Conservation}
\label{sec:energyconstop}

Next we consider the case of the emission escaping from the top of the accretion column. In this case, the local detector plane is oriented parallel to the upper surface of the accretion column, and therefore we have the simple relation $\tilde \theta = \theta_0$ according to Figure~\ref{fig:detector}. In addition, we have $\tilde\phi = \phi_0$ for the azimuthal angle. Hence the energy flux spectrum measured by the local detector for the column top emission, represented by $F_{\epsilon,\text{star}}^\text{top}$, is given by the integral
\begin{equation}
    F_{\epsilon,\text{star}}^\text{top}(\epsilon_0) = \iint_{4\pi} I_{\epsilon,\text{star}}^\text{top}(\epsilon_0) \, \sin\theta_0\cos\theta_0 \, d\theta_0 \, d\phi_0 \ ,
    \label{eq:topDet1}
\end{equation}
with no change of coordinates required to transform the fundamental flux integral (Equation~(\ref{fluxintwoframes})) from the detector frame into the {\tt RM88} frame. We can also use the relation $\mu_0 = \cos\theta_0$ to rewrite Equation~(\ref{eq:topDet1}) in the equivalent form
\begin{equation}
    F_{\epsilon,\text{star}}^\text{top}(\epsilon_0) = \iint_{4\pi} I_{\epsilon,\text{star}}^\text{top}(\epsilon_0) \, \mu_0 \, d\mu_0 \, d\phi_0 \ .
    \label{eq:topDet2}
\end{equation}
We can use this result to derive an energy conservation constraint for the column top expansion coefficients $W^\text{top}_{i}$ appearing in Equation~(\ref{totalintensitydeftopNEW2}). First we substitute into Equation~(\ref{eq:topDet2}) using Equation~(\ref{totalintensitydeftopNEW2}), which yields
\begin{equation}
    F_{\epsilon,\text{star}}^\text{top}(\epsilon_0) = \sum_{i=1}^{N_\text{top}} \sum_{j=1}^{M_\text{top}} \int_{0}^{2\pi}\int_{-1}^{1}\frac{W^\text{top}_{i}\epsilon_0\,\Dot{\mathcal{N}}^\text{tot}_\epsilon(\epsilon_0)}{\Omega R_\text{top}^2} \, \delta(\mu_0-\mu_{0i}) \delta(\phi_0-\phi_{0j})\mu_0d\mu_0d\phi_0 \ ,
    \label{fluxfromcoltopsum}
\end{equation}
where we have reversed the order of integration and summation. The integrals are trivial to perform, and Equation~(\ref{fluxfromcoltopsum}) reduces to
\begin{equation}
     F_{\epsilon,\text{star}}^\text{top}(\epsilon_0) = \sum_{i=1}^{N_\text{top}} \sum_{j=1}^{M_\text{top}} \frac{W^\text{top}_{i}\epsilon_0\,\Dot{\mathcal{N}}^\text{tot}_\epsilon(\epsilon_0)}{\Omega R_\text{top}^2} \ \mu_{0i} \ .
     \label{eq:TopEnCon1}
\end{equation}

The requirement of local energy conservation imposes the condition that the total flux computed using Equation~(\ref{eq:TopEnCon1}) must equal the flux generated by the {\tt BW22} continuum model, $F_{\epsilon,{\rm cont}}^\text{top}$, given by Equation~(\ref{fluxdeftop}). We therefore require that
\begin{equation}
    F_{\epsilon,\rm{star}}^\text{top}(\epsilon_0) = F_{\epsilon,{\rm cont}}^\text{top}(\epsilon_0) \ .
    \label{eq:TopEnCon2}
\end{equation}
Combining Equations~(\ref{fluxdeftop}), (\ref{eq:TopEnCon1}), and (\ref{eq:TopEnCon2}) yields
\begin{equation}
     \frac{\epsilon_0\,\Dot{\mathcal{N}}^\text{tot}_\epsilon(\epsilon_0)}{\Omega R_\text{top}^2} = \sum_{i=1}^{N_\text{top}} \sum_{j=1}^{M_\text{top}} \frac{W^\text{top}_{i}\epsilon_0\,\Dot{\mathcal{N}}^\text{tot}_\epsilon(\epsilon_0)}{\Omega R_\text{top}^2} \ \mu_{0i} \ ,
\end{equation}
which simplifies to yield
\begin{equation}
    e^\text{top}\equiv\sum_{i=1}^{N_\text{top}} \sum_{j=1}^{M_\text{top}} W_{i}^\text{top} \, \mu_{0i} = 1 \ ,
    \label{energyconstop}
\end{equation}
where $e^\text{top}$ is defined as the energy conservation constant for the column top emission. A distribution of expansion coefficients $W_{i}^\text{top}$ satisfying Equation~(\ref{energyconstop}) with $e^\text{top} = 1$ will guarantee that local energy conservation is maintained at every emission point on the column top.

In Section~\ref{sec:spectrum}, we develop the key procedure that will be used to establish the values of the sets of expansion coefficients $\{W_{i}^\text{wall}\}$ and $\{W_{i}^\text{top}\}$ by fitting a given set of observational pulse profile data. Once the fit is obtained, we must also ensure that Equations~(\ref{energyconswall}) and (\ref{energyconstop}) are satisfied, since otherwise the solution does not conserve energy, and is therefore not physically acceptable. There is also an additional subtlety that becomes important when we start examining in detail the null geodesics reaching a specific observer located at radius $R$ and with angle $\Psi$ relative to the neutron star spin axis, for a given set of emission angles $(\theta_0,\phi_0)$ in the {\tt RM88} frame. The question is whether that selected set of emission angles is ever visible to the observer during the rotation of the neutron star, or if the associated null geodesics never reach the observer for any value of the rotational phase angle $\beta$, in which case it is entirely invisible. If it is never visible, then the associated expansion coefficient must be computed using an alternative method that satisfies the energy conservation requirements $e^\text{wall} = 1$ and $e^\text{top} = 1$.

\subsection{Flux Measured by a Distant Observer}
\label{sec:spectrum}

In the preceding sections, our primary focus has been on the computation of the energy flux in the local {\tt RM88} frame using the {\tt BW22} continuum model, with the goal of ensuring local energy conservation. However, in order to fit to observational pulse profile data for a given source, we must also be able to compute the spectrum measured in the frame of a distant observer located at radius $R$ and with polar angle $\Theta$ with respect to the central axis of the accretion column. Hence our next task is to connect the calculation of the local intensity distribution using the {\tt BW22} continuum model with the relativistic formalism that links the local {\tt RM88} frame with that of a distant observer. The procedure we will develop will allow us to solve for the sets of intensity expansion coefficients $\{W_{i}^\text{wall}\}$ and $\{W_{i}^\text{top}\}$ by fitting observational pulse profile data.

The fundamental integral used to calculate the energy flux spectrum measured in the frame of a distant observer, denoted by $F_{\epsilon,{\rm obs}}$, is given by Equation~(\ref{fluxEq}). This equation is based on a coordinate transformation between the frame of the observer and the local {\tt RM88} frame, which is where the intensity is computed using the {\tt BW22} continuum model. The fundamental integral in Equation~(\ref{fluxEq}) takes on two different forms depending on whether the emission originates on the column wall or the column top. We distinguish between these two possibilities by writing
\begin{equation}
    F_{\epsilon,{\rm obs}}^\text{wall}(\epsilon,\beta) = \frac{\epsilon^3}{\epsilon_0^3}\iint_{4\pi}I_{\epsilon,\text{star}}^\text{wall}(\mu_0,\phi_0,\epsilon_0) \, \mu D_\text{wall}(\mu,\phi;\mu_0,\phi_0) \, d\mu_0 \, d\phi_0 \ ,
    \label{eq:WallFlux}
\end{equation}
for the phase-dependent energy flux spectrum resulting from the column wall emission, and
\begin{equation}
    F_{\epsilon,{\rm obs}}^\text{top}(\epsilon,\beta) = \frac{\epsilon^3}{\epsilon_0^3}\iint_{4\pi}I_{\epsilon,\text{star}}^\text{top}(\mu_0,\phi_0,\epsilon_0) \, \mu D_\text{top}(\mu,\phi;\mu_0,\phi_0) \, d\mu_0 \, d\phi_0 \ ,
    \label{eq:TopFlux}
\end{equation}
for the phase-dependent energy flux spectrum resulting from the column top emission, where the Jacobians $D_\text{wall}$ and $D_\text{top}$ are defined in Equations~(\ref{walljacobian}) and (\ref{topjacobian}), respectively. We note that the value of the {\tt RM88} frame energy, $\epsilon_0$, appearing on the right-hand side in Equations~(\ref{eq:WallFlux}) and (\ref{eq:TopFlux}) is related to the observer-frame energy, $\epsilon$, on the left-hand side via Equation~(\ref{eq:redshift1}). The next step is to combine Equation~(\ref{eq:WallFlux}) and (\ref{eq:TopFlux}) with the expansions for the wall and top emission given by Equations~(\ref{intensitysumwallNEW}) and (\ref{totalintensitydeftopNEW2}), respectively. The detailed procedure is carried out in Appendix~\ref{AppendixC}, and the results are summarized below.

\subsection{Total Energy Flux Spectrum}
\label{sec:gFunc}

The total phase-dependent energy flux spectrum measured by a distant observer, \break $F_{\epsilon,\text{obs}}^\text{tot}(\epsilon,\beta)$, is obtained by adding together all four emission components, taking into account the radiation escaping through the walls and tops of both accretion columns. We can therefore express the total phase-dependent energy flux spectrum using
\begin{equation}
    F_{\epsilon,\text{obs}}^\text{tot}(\epsilon,\beta) = F_{\epsilon,\text{obs}}^\text{wall,1}(\epsilon,\beta) + F_{\epsilon,\text{obs}}^\text{wall,2}(\epsilon,\beta) + F_{\epsilon,\text{obs}}^\text{top,1}(\epsilon,\beta) + F_{\epsilon,\text{obs}}^\text{top,2}(\epsilon,\beta) \ ,
    \label{totalfluxdefinition1}
\end{equation}
or, equivalently,
\begin{align}
    F_{\epsilon,\text{obs}}^\text{tot}(\epsilon,\beta)&=M_\text{wall} \sum_{i=1}^{N_\text{wall}} W_{i}^\text{wall,1} \, g_i^\text{wall,1}(\epsilon,\beta) + M_\text{wall} \sum_{i=1}^{N_\text{wall}} W_{i}^\text{wall,2} \, g_i^\text{wall,2}(\epsilon,\beta)\nonumber\\
    &+ M_\text{top} \sum_{i=1}^{N_\text{top}} W_{i}^\text{top,1} \, g_i^\text{top,1}(\epsilon,\beta) + M_\text{top} \sum_{i=1}^{N_\text{top}} W_{i}^\text{top,2} \, g_i^\text{top,2}(\epsilon,\beta) \ ,
    \label{totalfluxdefinition}
\end{align}
where the sets of angular expansion coefficients for the walls and tops of the two columns are denoted by $\{W_{i}^\text{wall,1}, W_{i}^\text{wall,2}, W_{i}^\text{top,1}, W_{i}^\text{top,2}\}$, and the corresponding sets of azimuthally-averaged subs-spectra, $\{g_i^\text{wall,1}(\epsilon,\beta)$, $g_i^\text{wall,2}(\epsilon,\beta)$, $g_i^\text{top,1}(\epsilon,\beta)$, $g_i^\text{top,2}(\epsilon,\beta)\}$ are calculated using either Equation~(\ref{gFunAvgWallPreCyc}) or (\ref{gFunAvgTopPreCyc}), as discussed in Appendix~\ref{AppendixC}. The expansion coefficients represent the angular distribution (beaming pattern) of the intensity in the {\tt RM88} frame, and therefore Equations~(\ref{totalfluxdefinition1}) and (\ref{totalfluxdefinition}) are somewhat analogous to the angular expansions used by \citet{Kraus_etal_1995} and \citet{Blum_Kraus_2000}. The parameters $N_\text{wall}$ and $N_\text{top}$ appearing in Equation~(\ref{totalfluxdefinition}) represent the total number of $\theta_0$ values in the angular grids for the column wall and the column top emission, respectively (see Equations~(\ref{anglegridwall}) and (\ref{anglegridtop})).

The total flux given by Equation~(\ref{totalfluxdefinition}) provides us with a complete description of the spectrum measured by a distant observer. We note that the flux $F_{\epsilon,\text{obs}}^\text{tot}(\epsilon,\beta)$ is also an implicit function of the neutron star's rotational phase angle, $\beta$, because this angle plays a central role in the ray-tracing procedure developed in Section~\ref{sec:RelForm}. In particular, the variation of $\beta$ as the star spins determines the variation of observation angle $\Theta$ between the line of sight to the observer and the magnetic field axis, via Equation~(\ref{thetaandbetaeq}). By performing a term-by-term integration of Equation~(\ref{totalfluxdefinition}) with respect to the observer-frame photon energy, $\epsilon$, we can generate an expression for the pulse profile, that can be compared with the observational pulse profile data in order to determine the values of the expansion coefficients. This procedure is carried out in Section~\ref{sec:pulseProfile}. Once the expansion coefficients are determined, we can then use Equation~(\ref{totalfluxdefinition}) to compute the theoretical prediction for the phase-averaged X-ray spectrum of the source. The comparison between the predicted spectrum and the observational data will provide a powerful test of the validity of our model.

\section{PULSE PROFILES AND PHASE-AVERAGED SPECTRA}

\label{sec:pulseProfile}

The formalism we have developed in the preceding sections provides us with a powerful tool for fitting pulse profiles, which we can utilize to determine source parameters such as the inclination angle, $\Psi$, and the rotational latitudes of the two accretion columns, $\varphi_1$ and $\varphi_2$. We can also use the new formalism to constrain the values of the physical input parameters for the {\tt BW22} continuum model, such as the accretion rate, $\dot M$, the electron temperature, $T_e$, the column top radius, $R_\text{top}$, the conical opening angles $\thetamax$ and $\thetamin$, the angle-averaged electron scattering cross section, $\bar\sigma$, and the parameters $\sigma_{||}$ and $\sigma_\perp$, which denote the scattering cross sections for radiation propagating parallel or perpendicular to the magnetic field, respectively. We can also constrain the values of the dimensionless parameters $\alpha$, $\xi$, and $\psi$, defined in Equations~(\ref{eq:alpha}), (\ref{eq:xi}), and (\ref{eq:psi}), respectively. In this section, we will develop the methodology required to utilize our unitary emission model to compute simulated pulse profiles.

\subsection{Calculation of the Pulse Profile}

The detailed physics of cyclotron resonant absorption is highly complex, and depends on the energy and angular dependences of the electron scattering cross section in a strong magnetic field \citep{Ventura1979,Meszaros_Ventura_1979}, and therefore a comprehensive treatment of this process is beyond the scope of this paper. We will therefore adopt the formalism used by \citet{BeckerandWolff2007} and {\tt BW22}, who introduced an overall energy-dependent multiplicative function to approximate the effect of cyclotron absorption. Computationally, we will implement the cyclotron absorption process by multiplying the basis functions by an attenuation function that follows a Gaussian distribution in energy, centered at the cyclotron energy. This attenuation factor is given by
\begin{equation}
    d_\text{cyc}(\epsilon)  = 1 - \frac{A_0}{\sigma_\text{c}\sqrt{2\pi}}e^{-(\epsilon-\epsilon_\text{c})/(2\sigma_\text{c})^2} \ ,
    \label{dcyc}
\end{equation}
where $\epsilon_c$ is the cyclotron energy, $\sigma_c$ is the standard deviation or width parameter for the attenuation, and the parameter $A_0$ represents the relative strength of the absorption feature.

Observational pulse profiles are obtained by integrating the observed phase-dependent photon number spectrum, denoted by $F_\#(\epsilon,\beta)$, with respect to time. The units of the photon spectrum are $F_\#(\epsilon,\beta) \propto {\rm photons\ s}^{-1}\ {\rm cm}^{-2}$. Pulse profiles are usually plotted in units of counts s$^{-1}$, rather than flux units, because the counts are measured in the detector plane, and they therefore represent a convolution of the impinging photon number spectrum, $F_\#(\epsilon,\beta)$, multiplied by the detector response function, $\mathscr{D}(\epsilon)$, which gives the effective area of the detector.  In order to calculate the photon number spectrum, $F_\#(\epsilon,\beta)$, emitted from the pulsar based on any given set of geometry angles, we begin with the total flux, $F_{\epsilon,\text{obs}}^\text{tot}(\epsilon,\beta)$, computed using Equation~(\ref{totalfluxdefinition}). In general, the phase-dependent photon number flux spectrum is related to the flux $F_{\epsilon,\text{obs}}^\text{tot}(\epsilon,\beta)$ via
\begin{equation}
    F_\#(\epsilon,\beta) = \frac{d_\text{cyc}(\epsilon)}{\epsilon} \, F_{\epsilon,\text{obs}}^\text{tot}(\epsilon,\beta) \ ,
    \label{numberspecdef}
\end{equation}
where the cyclotron absorption function, $d_\text{cyc}(\epsilon)$, is defined in Equation~(\ref{dcyc}).
We can obtain an explicit expression for $F_\#(\epsilon,\beta)$ by combining Equations~(\ref{totalfluxdefinition}) and (\ref{numberspecdef}), which yields for the phase-dependent photon number spectrum
\begin{align}
    F_\#(\epsilon,\beta)&=\frac{d_\text{cyc}(\epsilon)}{\epsilon} \, M_\text{wall} \sum_{i=1}^{N_\text{wall}} \left[W_{i}^\text{wall,1} \, g_i^\text{wall,1}(\epsilon,\beta) +  W_{i}^\text{wall,2} \, g_i^\text{wall,2}(\epsilon,\beta)\right]\nonumber\\
    &+ \frac{d_\text{cyc}(\epsilon)}{\epsilon}
     \, M_\text{top} \sum_{i=1}^{N_\text{top}} \left[W_{i}^\text{top,1} \, g_i^\text{top,1}(\epsilon,\beta) + W_{i}^\text{top,2} \, g_i^\text{top,2}(\epsilon,\beta)\right] \ .
    \label{eq:PhaseDepSpec}
\end{align}
The theoretical pulse profile function, in units of counts\ s$^{-1}$, denoted by $S(\beta)$, is computed from the phase-dependent photon number spectrum, $F_\#(\epsilon,\beta)$, using
\begin{equation}
    S(\beta) = \int_{\epsilon_{\text{min}}}^{\epsilon_{\text{max}}} \, \mathscr{D}(\epsilon) \, F_\#(\epsilon,\beta) \, d\epsilon
    \label{pulseProfileEq} \ ,
\end{equation}
where $\beta$ is the rotational phase angle of the spinning neutron star, the photon number spectrum $F_\#(\epsilon,\beta)$ is computed using Equation~(\ref{eq:PhaseDepSpec}), the function $\mathscr{D}(\epsilon)$ represents the effective area curve for the detector, and $\epsilon_\text{min}$ and $\epsilon_\text{max}$ denote the minimum and maximum energies for the observation, respectively.
The rotational phase angle $\beta$ varies in the range $0\degree \le \beta \le 360\degree$ over the course of one rotation period. We note that $F_\#(\epsilon,\beta)$ also has an implicit dependence on $\beta$, via the ray-tracing procedure discussed in Section~\ref{sec:RelForm}, because of the associated variation of the observation angle $\Theta$ as the star spins, as expressed by Equation~(\ref{thetaandbetaeq}). Since the inclination angle, $\Psi$, and the rotational latitude of the accretion column, $\varphi$, are fixed model parameters, there is generally a one-to-one correspondence between $\beta$ and $\Theta$.

We are now prepared to define the theoretical expansion used to model the observed pulse profile. Recall from Section~\ref{sec:spectrum} that the flux emitted from the walls and tops of the two accretion columns is modeled using four sets of azimuthally-averaged, energy- and phase-dependent ``sub spectra,'' denoted by $\{g_i^\text{wall,1}(\epsilon,\beta)\}$, $\{g_i^\text{wall,2}(\epsilon,\beta)\}$, $\{g_i^\text{top,1}(\epsilon,\beta)\}$, and $\{g_i^\text{top,2}(\epsilon,\beta)\}$, which are defined by Equations~(\ref{gFunAvgWallPreCyc}) and (\ref{gFunAvgTopPreCyc}) for the column wall and column top emission, respectively. Each of these basis functions corresponds to the contribution to the observed spectrum resulting from emission with polar angle $\theta_{0i}$ as viewed in the {\tt RM88} frame, and they are averaged over all of the values of the azimuthal emission angle, $\phi_{0j}$. Using Equation~(\ref{totalfluxdefinition}), the azimuthally-averaged basis functions are added together and weighted by the set of coefficients $\{W_{i}^\text{wall,1}, W_{i}^\text{wall,2}, W_{i}^\text{top,1}, W_{i}^\text{top,2}\}$ to generate the total observed number flux spectrum, $F_\#(\epsilon,\beta)$.

The theoretical pulse profile count-rate function, $S(\beta)$, is obtained by integrating the total observed phase-dependent photon number flux spectrum, $F_\#(\epsilon,\beta)$, with respect to energy using Equation~(\ref{pulseProfileEq}). Substituting Equation~(\ref{eq:PhaseDepSpec}) into Equation~(\ref{pulseProfileEq}), we obtain
\begin{align}
     S(\beta) =  M_\text{wall} \sum_{i=1}^{N_\text{wall}}\left[W_i^\text{wall,1}\, h_i^\text{wall,1}(\beta) + W_i^\text{wall,2}\, h_i^\text{wall,2}(\beta)\right]
     \nonumber\\
     + M_\text{top} \sum_{i=1}^{N_\text{top}}\left[W_i^\text{top,1}\, h_i^\text{top,1}(\beta) + W_i^\text{top,2}\, h_i^\text{top,2}(\beta)\right] \ ,
     \label{pulseprofilefrombasis}
 \end{align}
where the phase-dependent ``sub-profiles,'' $h_{i}^\text{wall,1}(\beta)$, $h_{i}^\text{wall,2}(\beta)$, $h_{i}^\text{top,1}(\beta)$, and $h_{i}^\text{top,2}(\beta)$ are obtained by integrating the respective sub-spectra with respect to energy according to the definitions
\begin{align}
    h_{i}^\text{wall,1}(\beta)\equiv\int_{\epsilon_\text{min}}^{\epsilon_\text{max}} d_\text{cyc}(\epsilon) \, g_{i}^\text{wall,1}(\epsilon,\beta) \, \mathscr{D}(\epsilon) \, \frac{d\epsilon}{\epsilon} \ , \nonumber\\
    h_{i}^\text{wall,2}(\beta)\equiv\int_{\epsilon_\text{min}}^{\epsilon_\text{max}} d_\text{cyc}(\epsilon) \, g_{i}^\text{wall,2}(\epsilon,\beta) \, \mathscr{D}(\epsilon) \, \frac{d\epsilon}{\epsilon} \ , \nonumber\\
    h_{i}^\text{top,1}(\beta)\equiv\int_{\epsilon_\text{min}}^{\epsilon_\text{max}} d_\text{cyc}(\epsilon) \, g_{i}^\text{top,1}(\epsilon,\beta) \, \mathscr{D}(\epsilon) \, \frac{d\epsilon}{\epsilon} \ , \nonumber\\
    h_{i}^\text{top,2}(\beta)\equiv\int_{\epsilon_\text{min}}^{\epsilon_\text{max}} d_\text{cyc}(\epsilon) \, g_{i}^\text{top,2}(\epsilon,\beta) \, \mathscr{D}(\epsilon) \, \frac{d\epsilon}{\epsilon} \ .
    \label{subprofiles}
\end{align}
The sub-profiles $h_{i}^\text{wall,1}(\beta)$, $h_{i}^\text{wall,2}(\beta)$, $h_{i}^\text{top,1}(\beta)$, and $h_{i}^\text{top,2}(\beta)$ play a central role in our formalism for fitting observational pulse profile data using our unitary emission model. Each of these functions represents the contribution to the theoretical pulse profile due to radiation emitted in a specific set of directions in the {\tt RM88} frame, via escape through either the wall or the top of column~1 or column~2. Hence the sub-profiles are effectively basis functions, in units of counts per second, which can be added together in a linear fashion using the sets of angular weight coefficients $\{W_{i}^\text{wall,1}, W_{i}^\text{wall,2}, W_{i}^\text{top,1}, W_{i}^\text{top,2}\}$ to simulate the variation of the pulse profile with the rotational phase angle $\beta$. In this work, the fitting is accomplished using the \textit{FindFit} command in the \textit{Mathematica} environment, which is a very powerful nonlinear fitting routine. We find that in practice, the resulting pulse profile fits are essentially unique, because the families of sub-profiles form quasi-orthogonal sets.

\subsection{Calculation of Phase-averaged Spectrum}
\label{sec:phaseAvgSpec}

Another key highlight of our model is the ability to compute the true phase-averaged X-ray spectrum for a given XRP, using the same expansion coefficients that were obtained by fitting the pulse profile data for the source. The resulting phase-averaged X-ray spectrum is therefore fully consistent with the associated pulse profile. In this sense, there is no new ``fitting'' associated with the computation of the phase-averaged spectrum, and instead, the spectrum represents a separate theoretical prediction made by the model. The associated X-ray photon number spectrum, $F_\#(\epsilon,\beta)$, is computed using the energy-dependent sub-spectra represented by the functions $g_i^\text{wall,1}$, $g_i^\text{wall,2}$, $g_i^\text{top,1}$, and $g_i^\text{top,2}$ introduced in Equations~(\ref{gFunAvgWallPreCyc}) and (\ref{gFunAvgTopPreCyc}), which are combined together in Equation~(\ref{eq:PhaseDepSpec}). The X-ray spectrum is then integrated with respect to the rotational phase angle, $\beta$, to obtain the phase-averaged X-ray spectrum, which is essentially an independent theoretical prediction of our model, that can be compared with the data for the same source that was analyzed using the pulse-profile fitting methodology. The comparison of the theoretical prediction for the phase-averaged spectrum with the spectral data for the same source provides a powerful, independent test of the validity of the model, which represents one of the most significant advantages of our theoretical approach.

The phase-averaged theoretical photon number spectrum, $\langle F_\#(\epsilon) \rangle$, is computed using the integral expression
\begin{equation}
    \langle F_\#(\epsilon) \rangle \equiv \frac{1}{2\pi} \int_{0}^{2\pi} F_\#(\epsilon,\beta) \, d\beta
    \label{eq:PhaseAveSpec} \ ,
\end{equation}
where $F_\#(\epsilon,\beta)$ is the phase-dependent photon number spectrum, computed using Equation~(\ref{eq:PhaseDepSpec}). In practice, Equation~(\ref{eq:PhaseAveSpec}) is evaluated using an explicit average over a grid of discrete values for $\beta$ defined by
\begin{equation}
    \beta\in\{0\degree,5.625\degree,11.25\degree,\ldots,360\degree\}
\end{equation}
We note that the phase-averaged spectrum, $\langle F_\#(\epsilon)\rangle$, computed using Equation~(\ref{eq:PhaseAveSpec}) not only includes the geometry of the rotating neutron star and the locations and properties of the two accretion columns, but it also includes all of the effects of GR, which are non-negligible for emission from the vicinity of the neutron star surface, since the gravitational redshift from the surface of the star is $z_\text{star} \approx 0.3$ (see Equation~(\ref{eq:starZ})). This means that any photon energies used in the {\tt BW22} model itself (which is used to generate the spectrum in the {\tt RM88} frame) will correspond to lower values in the frame of the distant observer. Furthermore, the gravitational redshift will also lead to adjustments in some of the fundamental theory parameters in our work, such as the electron temperature, $T_e$, and the accretion rate, $\dot M$, which implies that in our model for Her X-1, for example, we will use input parameters that are somewhat different from the values adopted by {\tt BW22} in their model for the same source.

\subsection{Energy Conservation and Non-visible Geodesics}
\label{sec:VisInvis}

One important subtlety of the formalism we have developed here is the distinction between visible and non-visible geodesics. We discuss this in detail here because it bears on the process of determining the sets of angular expansion coefficients $\{W_{i}^\text{wall,1}, W_{i}^\text{wall,2}, W_{i}^\text{top,1}, W_{i}^\text{top,2}\}$ in such a way that energy conservation is satisfied via Equations~(\ref{energyconswall}) and (\ref{energyconstop}), which specify that we must have $e^\text{wall} = 1$ and $e^\text{top} = 1$ in order to fully conserve energy in the local {\tt RM88} frame.

First we remind the reader that the pulse profile we are computing is for a specific observer, located at radius $r$ from the neutron star, and with inclination angle $\Psi$ relative to the neutron star's spin axis. As discussed in Section~\ref{sec:relchar}, for a specific {\tt RM88} emission direction $(\theta_{0i},\phi_{0j})$ chosen from the angular grids defined in Equations~(\ref{anglegridwall}) and (\ref{anglegridtop}), and for a given value of the neutron star's rotational phase angle $\beta$, there will in general be either zero, one, or two emission points on the accretion column such that the geodesics of the photons emitted with that {\tt RM88} emission vector will reach the distant observer. In addition, these emission locations will vary in a periodic manner as the star spins, and the angle $\Theta$ between the observer and the accretion column axis varies according to Equation~(\ref{thetaandbetaeq}). Hence the possibility exists that a particular emission direction $(\theta_{0i},\phi_{0j})$ may {\sl never} be visible to a distant observer at a specified location {\sl at any point} during the spin of the neutron star. When this occurs, then the corresponding sub-profile $h_i(\beta)$ as defined in Equations~(\ref{subprofiles}) will yield $h_i(\beta) = 0$ for all values of the rotational phase angle $\beta$, and we therefore refer to the associated emission direction $(\theta_{0i},\phi_{0j})$ as ``invisible'' to the specified distant observer. Although unobservable, these ``invisible'' emission directions are nonetheless significant in our formalism, because they represent directions in which power is emitted in the {\tt RM88} frame. We must therefore consider the power emitted in these directions when calculating the energy conservation constants $e^\text{wall} $ and $e^\text{top}$ using Equations~(\ref{energyconswall}) and (\ref{energyconstop}), respectively. We emphasize that this unobservable emission power has no effect on the predicted pulse profile or the associated phase-averaged X-ray spectrum. This in turn gives us a powerful mechanism to ensure that local energy conservation is satisfied, as explained below.

Based on the above discussion of ``visible'' and ``invisible'' emission directions, it is convenient to break the energy-conservation sums given in Equations~(\ref{energyconswall}) and (\ref{energyconstop}) into two parts. The first part expresses the energy conservation sum for the ``visible'' emission directions, with associated null geodesics that reach the distant observer for some range of values of the rotational phase angle $\beta$, and the second part corresponds to the ``invisible'' energy conservation sum, associated with emission directions that are never visible to the observer. We therefore split the energy conservation constants $e^\text{wall}$ and $e^\text{top}$ into two parts by writing
\begin{equation}
    e^\text{wall} = e^\text{wall}_\text{vis} + e^\text{wall}_\text{invis} = 1 \ ,
    \label{visplusinviswall}
\end{equation}
and
\begin{equation}
    e^\text{top} = e^\text{top}_\text{vis} + e^\text{top}_\text{invis} = 1 \ ,
    \label{visplusinvistop}
\end{equation}
where $e^\text{wall}_\text{vis}$ and $e^\text{top}_\text{vis}$ represent the energy conservation fractions corresponding to the visible emission directions, and $e^\text{wall}_\text{invis}$ and $e^\text{top}_\text{invis}$ represent the energy conservation fractions corresponding to the invisible emission directions.

We have established that excess power can be directed into invisible emission directions in order to satisfy local energy conservation without affecting the theoretical predictions for the pulse profile or the phase-averaged spectrum. However, this does not completely settle the question of how to distribute this power over the grid of polar emission angles $\theta_{0i}$ in the {\tt RM88} frame. In order to answer this question, we need to think about the angular distribution of the radiation escaping from the accretion column after experiencing its last scattering. We anticipate that the angular dependence of the scattered radiation field will resemble that of the scattering cross section. This problem was considered by \citet{herold_et_al_1979} and \citet{Loudas_et_al_2021}, who demonstrated that the differential cyclotron scattering cross section has the behavior
\begin{equation}
    \frac{d\sigma}{d\Omega}\propto(1+\cos^2\theta_0)
    \ ,
    \label{eq:CrossScat}
\end{equation}
where $\theta_0$ is the angle of the scattered photon with respect to the magnetic field direction.

It is impossible to have complete certainty about the angular distribution of the radiation escaping from the X-ray pulsar accretion column in the ``invisible'' emission directions, because this emission is, by definition, unobservable, and therefore unconstrained by the data. In the absence of any further information, we will therefore approximate the angular distribution of this radiation component using Equation~(\ref{eq:CrossScat}), which provides a means for ``filling in'' the missing power in order to guarantee local energy conservation. We will therefore set the values of the expansion coefficients for the invisible emission directions using
\begin{equation}
    W_{i}^\text{wall} = P^\text{wall}_0 \, (1+\cos^2\theta_{0i}) \ ,
    \label{invisibleWall}
\end{equation}
for the wall emission, and \citep{Leahy2004a,Leahy2004b}
\begin{equation}
    W_{i}^\text{top} = P^\text{top}_0 \, \frac{1}{\sigma \sqrt{2 \pi}} \, e^{-\theta_{0i}^2/(2 \sigma^2)} \ ,
    \label{invisibleTop}
\end{equation}
for the top emission, where the values of ``$i$'' correspond to invisible emission directions only, and the constants $P^\text{wall}_0$ and $P^\text{top}_0$ are determined by enforcing Equations~(\ref{visplusinviswall}) and~(\ref{visplusinvistop}). The Gaussian form in Equation~(\ref{invisibleTop}) is physically motivated by the application of the Eddington-Barbier relation at the column top, which implies that the emission escaping through the upper surface is strongly collimated in the radial direction due to the low value for the parallel scattering cross section, $\bar\sigma$, in the highly magnetized plasma \citep{Paletou2018,Nagel1981a}. The variance $\sigma$ is set using $\sigma=16$ in our numerical applications.

\section{ASTROPHYSICAL APPLICATIONS}
\label{sec:results}

\begin{figure}[htbp]
    \centering
    \includegraphics[width=0.7\linewidth]{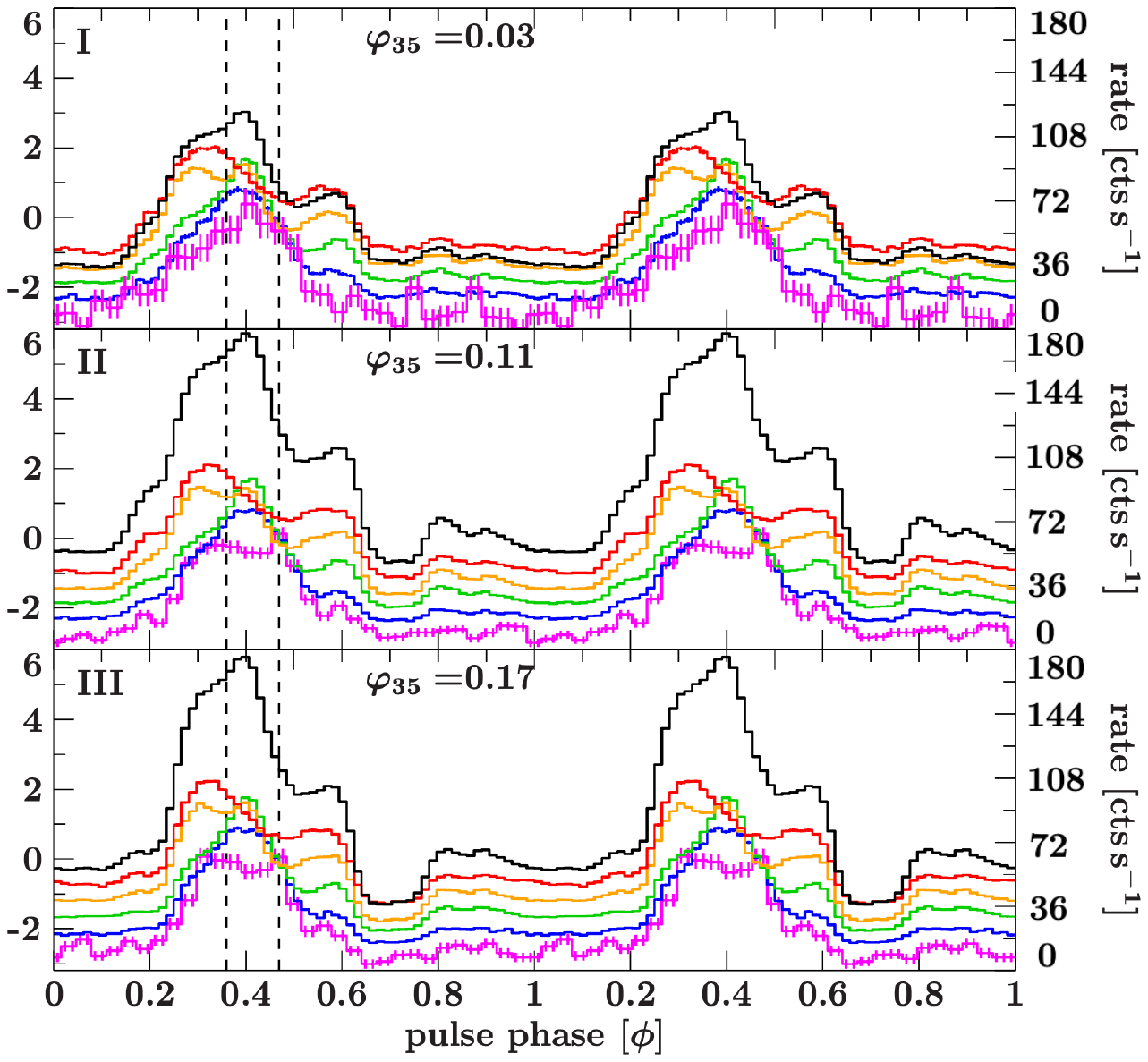}
    \caption{Pulse profile data for Her X-1 presented by \citet{Furst_et_al_2013} for three different phases during the 35 day precession cycle. The black curves represent the total count rate in the 3-79\,keV energy range, and the various colored curves correspond to a variety of energy ranges, as discussed in the text. We focus here on analysis of the total count rate data for Observation~II. Note that the scale on the right-hand side was adjusted to correct a normalization error in the original plot.}
    \label{fig:furstetal}
\end{figure}

One of the primary goals of this paper is to utilize our new relativistic model to fit the pulse profile data for Her X-1 corresponding to Observation~II from \citet{Furst_et_al_2013}, plotted in Figure~\ref{fig:furstetal}. Her X-1 is one of the most widely-studied pulsating X-ray sources, due to a variety of factors. First, Her X-1 is quite bright due to its relatively nearby location in the galaxy, with a distance of $\sim 6.6\,$kpc from Earth \citep{Bailer-JonesEtal2021,ReynoldsEtal1997}. The rotational period of the neutron star in Her X-1 is 1.24\,s, and the energy-dependent pulse profile and its evolution have been widely studied \citep{Staubert_etal2019,Staubert_etal2020}. Due to its high luminosity $(\xlum\sim10^{37}>L_\text{crit})$, Her X-1 also provides an interesting test for the radiation-dominated shock dynamical model incorporated into the {\tt BW22} continuum model, which forms the basis for our work here. Our selection of Her X-1 as the first application for our new model is further motivated by the large body of existing work analyzing its emission characteristics, including the orientation of the neutron star spin axis and the rotational latitudes of the emission regions. For example, \citet{Leahy_1991} and \citet{Blum_Kraus_2000} analyzed the Her X-1 pulse profile to obtain solutions for the various geometry parameters based on the assumption that the emission regions comprise hot spots or rings on the stellar surface. The availability of these alternative interpretations of the pulse profile data for Her X-1 provides a valuable point of comparison with the geometric results we obtain using our model.

Her X-1 exhibits a 35-day activity cycle due to the precession of the accretion disk, which causes the observed flux to dip twice per cycle, resulting from obscuration when the accretion disk blocks our line of sight. The observed pulse profile for Her X-1 in multiple X-ray energy channels obtained using {\sl NuSTAR} was presented by \citet{Furst_et_al_2013}. They discuss three observations, all made during the unobscured or ``main-on'' phase. Observation~I focuses on the turn-on, when the flux rises as the columns become unobscured, Observation~II was obtained during the brightest part of the cycle when the main-on phase is at its peak, and Observation~III focuses on the turn-off when the flux begins to dip. The pulse profile data for all three observations are plotted in Figure~\ref{fig:furstetal}, in which the various colors correspond to the count rates measured in different energy ranges. The red, orange, green, blue, and magenta curves report the count rates in the 3-6\,keV, 6-10\,keV, 10-20\,keV, 20-40\,keV, and 40-79\,keV energy ranges, respectively, and the black curves represent the total count rate from 3-79\,keV. We note that the scale on the right-hand side of Figure~\ref{fig:furstetal} was adjusted to correct a normalization error in the original plot. Here we will focus on the analysis of the 3-79\,keV count rate data obtained during Observation II. We note that the calibration of the scale on the right side of Figure~\ref{fig:furstetal} has been adjusted relative to Figure~3 from \citet{Furst_et_al_2013} in order to correct an error that rendered the normalization of their pulse profile inconsistent with the normalization of the spectral data plotted in their Figure~6.

The phase-averaged X-ray spectrum of Her X-1 is plotted in Figure~\ref{fig:BW22HerX1_spect}a. The main features of the spectrum include the characteristic power-law slope in the photon energy range $10\,{\rm keV} \lapprox \epsilon \lapprox 30\,{\rm keV}$, and the roughly exponential cutoff for photon energies $\epsilon \gtrsim 40\,$keV. Another important feature in the spectrum of Her X-1 is the dip in the observed spectrum centered around photon energy $\epsilon = 37.44\pm 0.07\,$keV \citet{Staubert_etal2020}, which is clearly visible in Figure~\ref{fig:BW22HerX1_spect}a. This feature is widely interpreted as the result of the resonant absorption of photons at the cyclotron energy $\epsilon_c$ \citep[see][]{Truemperetal}. According to the work of \citet{BeckerandWolff2007} and {\tt BW22}, the phase-averaged spectrum of Her X-1 is dominated by Comptonized bremsstrahlung radiation due to its high accretion rate. However, the presence of cyclotron and blackbody seed photons also contributes non-negligibly, and therefore we will consider all three sources of  seed radiation when applying our new model to this source.

The flow velocity profile describing the dynamical structure of the accretion column in Her X-1 was also obtained by {\tt BW22}, and is plotted in Figure~\ref{fig:BW22HerX1_spect}b. The velocity starts at the local Newtonian free-fall value far above the neutron star, reaches a maximum value at radius $R \sim 1.5\,R_\text{star}$, and then subsequently decreases due to the strong radiation pressure gradient. Ultimately, the gas comes to rest at the stellar surface, where the accreting matter merges with the crust of the neutron star.

\subsection{Computational Procedure}
\label{sec:CompProc}

In order to use our formalism to analyze a specific astrophysical source, we must first select values for the physical parameters required to evaluate the {\tt BW22} continuum model using the expressions given in Section~\ref{numspecfromsource}. The input parameters for the {\tt BW22} model comprise the electron temperature, $T_e$, the accretion rate, $\dot M$, the magnetic field strength, $B$, the radius at the column top, $R_\text{top}$, the dimensionless parameters $\alpha$, $\xi$, and $\psi$, and the outer and inner opening angles for the conical accretion column, denoted by $\thetamax$ and $\thetamin$, respectively. The parameters $\alpha$, $\xi$, and $\psi$ are defined in Equations~(\ref{eq:alpha}), (\ref{eq:xi}), and (\ref{eq:psi}), respectively. After the physical input parameters for the continuum model have been determined, the next step is to select the geometry parameters for the accreting, spinning neutron star. The pulse profile we are computing is measured by an observer located at radius $r$ from the neutron star, with inclination angle $\Psi$ with respect to the star's rotational axis. We must therefore select values for $\Psi$ and $r$, and we must also select values for the rotational latitudes of the two accretion columns, $\varphi_1$ and $\varphi_2$.

We must also allow for the possibility of phase shifts in the rotational phase angle, $\beta$, for each of the two accretion columns, because the $\beta = 0$ point in the physical stellar rotation has an arbitrary calibration relative to the observational plots of the pulse profile. These rotational phase shifts are denoted by $\Delta\beta_1$ for Column~1 and $\Delta\beta_2$ for Column~2, respectively. Hence the fundamental values of the rotational phase angle for the two columns, denoted by $\beta_1$ and $\beta_2$, are computed using
\begin{equation}
    \beta_1 = \beta + \Delta\beta_1 \ , \qquad
    \beta_2 = \beta + \Delta\beta_2 \ ,
    \label{eq:betaShift}
\end{equation}
where $\beta$ is the observational value.
Hence the complete magnetic and rotational geometry of the neutron star and the two accretion columns is specified by the set of five angles $(\Psi,\varphi_1,\varphi_2,\Delta\beta_1,\Delta\beta_2)$.

Based on relations from spherical trigonometry, we find that the angle between the central axes of the two accretion columns, denoted by $\Xi$, is given by
\begin{equation}
    \cos\Xi = \cos\varphi_1\cos\varphi_2 + \cos(\Delta\beta_1-\Delta\beta_2) \sin\varphi_1 \sin\varphi_2 \ .
    \label{eq:Xi}
\end{equation}
Once the geometry parameters are specified, we can then compute the sets of sub-profiles $h_{i}^\text{wall,1}(\beta)$, $h_{i}^\text{wall,2}(\beta)$, $h_{i}^\text{top,1}(\beta)$, and $h_{i}^\text{top,2}(\beta)$ that represent the basis functions for the theoretical pulse-profile expansion given in Equation~(\ref{pulseprofilefrombasis}). With the sets of sub-profiles determined, then we can use Equation~(\ref{pulseprofilefrombasis}) to fit the theoretical pulse-profile function $S(\beta)$ to a specified set of pulse profile data. In carrying out the fits presented in this section, we have utilized the cyclotron absorption function given by Equation~(\ref{dcyc}), with the associated parameters set using $A_0 = 14\,$keV, $\epsilon_\text{c}=37.7$keV, and $\sigma = 8\,$keV. The source distance for Her X-1 assumed here is $R = 6.6\,$kpc.

The fitting method we utilize employs the {\sl Mathematica} routine {\sl FindFit}, which is based on a powerful nonlinear least-squares algorithm \citep{Mathematica,FindFit}. The algorithm determines the best-fit solution for the theoretical pulse profile by simultaneously varying the four coupled sets of angular weight coefficients $\{W_{i}^\text{wall,1}, W_{i}^\text{wall,2}, W_{i}^\text{top,1}, W_{i}^\text{top,2}\}$ that describe the angular distributions of the radiation escaping from the surfaces of the two accretion columns, as viewed in the local {\tt RM88} frame. With the expansion coefficients determined, we can then compute the true phase-averaged X-ray spectrum using Equation~(\ref{eq:PhaseAveSpec}). Comparison of the resulting spectrum with the observational data provides an important validation test for our model. We note that the weight coefficients $\{W_{i}^\text{wall,1}, W_{i}^\text{wall,2}, W_{i}^\text{top,1}, W_{i}^\text{top,2}\}$ must also satisfy the energy-conservation sums given in Equations~(\ref{energyconswall}) and (\ref{energyconstop}), which also involves consideration of the distinction between ``visible'' and ``invisible'' geodesics, as discussed in Section~\ref{sec:VisInvis}. Furthermore, we also require a minimum level of emission in all directions as viewed in the local {\tt RM88} frame. In practice, we require that $W_{i}^\text{wall,1}$ and $W_{i}^\text{wall,2}$ always exceed the value $0.003/M_\text{wall}$, and we require that $W_{i}^\text{top,1}$ and $W_{i}^\text{top,2}$ always exceed the value $0.001/M_\text{top}$, where $M_\text{wall}$ and $M_\text{top}$ represent the number of $\phi_{0j}$ values in the column wall and column top angular grids, respectively (see Equations~(\ref{anglegridwall}) and (\ref{anglegridtop})).

In our computational application to Her X-1, we will explore two scenarios regarding the possible correlation of the emission patterns between the two accretion columns. For simplicity, and to reduce the number of free parameters, in both scenarios, we will assume that the physical structures of the two columns are identical, meaning that the values of the physical input parameters $T_e$, $\dot M$, $R_\text{top}$, $\thetamax$, $\thetamin$, $\alpha$, $\xi$, and $\psi$ (all compared with {\tt BW22} in Table~\ref{tbl-3new}), as well as the computed model parameters $\bar\sigma$, $\sigma_\perp$, and $\sigma_{||}$ (compared with {\tt BW22} in Table~\ref{tbl-4new}), are the same for the two columns. In particular, it is important to emphasize that the parameter $\dot M$ represents the accretion rate for each of the two accretion columns, and therefore the total accretion rate in the frame of the star, denoted by $\dot M_\text{tot}$, is given by $\dot M_\text{tot} = 2\,\dot M$. This is further discussed in Section~\ref{SecMdot}. The flow velocity profiles are also assumed to be the same, with $k_\infty = 1$ and $k_0 = 0$ for both columns. Once the values of the input parameters for the continuum model are specified, then we must also determine the geometry for the accreting, spinning neutron star by selecting values for the rotational inclination angle, $\Psi$, the rotational latitudes of the two accretion columns, $\varphi_1$ and $\varphi_2$, and the rotational phase shifts, $\Delta\beta_1$ and $\Delta\beta_2$.

The difference between Scenario~1 and Scenario~2 enters when we consider the fitting of the theoretical pulse-profile function, $S(\beta)$, given by Equation~(\ref{pulseprofilefrombasis}), to the observational pulse profile data for Her X-1. In Scenario~1, the wall intensity angular distribution for Column~1 is constrained to be identical to that for Column~2. Furthermore, the top intensity angular distribution for Column~1 is likewise constrained to be identical to that for Column~2. We can express this constraint using the conditions $\{W_{i}^\text{wall,1}\} = \{W_{i}^\text{wall,2}
\}$ and \{$W_{i}^\text{top,1}\} = \{W_{i}^\text{top,2}\}$. Physically, this scenario is motivated by the idea that the neutron star's global magnetic field is likely to be roughly dipolar, in which case it may be reasonable to assume that the magnetic topology within the two accretion columns should be similar, yielding similar scattering cross section distributions in the magnetized plasma.

While Scenario~1 may have some reasonable physical basis, it can nonetheless be viewed as an over-idealization of the actual situation. In reality, even if the global stellar magnetic field is nearly dipolar, there may still be subtle distortions in the structure of the neutron star magnetosphere that may lead to important differences in the magnetic field distributions inside the two accretion columns \citep{MushtukovEtal2025}. The distortions in the structure of the magnetosphere could arise as a result of complicated multipole geometries, in which the central axes of the two accretion columns are separated by less than $180\degree$ \citep{HuangEtal2025}. Therefore, in Scenario~2, we will relax the assumption of identical intensity distributions in the two columns, and instead allow all four sets of angular expansion coefficients $\{W_{i}^\text{wall,1}, W_{i}^\text{wall,2}, W_{i}^\text{top,1}, W_{i}^\text{top,2}\}$ to vary independently in the fitting process. We note that Scenario~2 has twice the expansion coefficients available to accomplish the fit to the pulse profile data, compared with Scenario~1, and therefore we may reasonably expect to obtain better fits to the observational pulse profile data in the former case. We present the detailed results obtained by applying either Scenario~1 or Scenario~2 to the analysis of the Her X-1 data below.

\subsection{Parameter Space Search Strategy}
\label{sec:parameterspacesearch}

The calculation of the X-ray continuum spectrum in the local {\tt RM88} frame is carried out using the {\tt BW22} model with physical input parameters $T_e$, $\dot M$, $R_\text{top}$, $\thetamax$, $\thetamin$, $\alpha$, $\xi$, and $\psi$, which also imply values for the scattering cross sections $\bar\sigma$, $\sigma_\perp$, and $\sigma_{||}$. These parameters must be varied in order to obtain an acceptable fit to the phase-averaged X-ray spectrum for the selected source. Once the continuum parameters are determined, we explore the rotational and magnetic geometry of the neutron star and the accretion columns by varying the set of five angles $(\Psi,\varphi_1,\varphi_2,\Delta\beta_1,\Delta\beta_2)$. For a neutron star with accretion columns located approximately on opposite sides of the star, changes in these angles do not have a strong effect on the phase-averaged spectrum. Hence variation of the angles is mainly used to bring the simulated pulse profile into agreement with the data. The multidimensional nature of the theoretical parameter space, combined with the nonlinearity of our relativistic model, requires a systematic search in order to find sets of parameters that yield good agreement with the phase-averaged spectrum and the pulse profile data for a given source. A corollary of the search is the requirement that we investigate the uniqueness (or degeneracy) of each of the successful geometrical solutions identified.

The selection of the criteria for determining the ``successful'' geometrical solutions is a rather subtle issue, and some arbitrary decisions must be made in order to develop a quantitative approach. We have identified two crucial aspects of the candidate solution that must be taken in consideration, namely the fit of the theoretical pulse-profile model compared with the data, and the shapes of the angular distributions of the radiation emitted from the surfaces of the two accretion columns as viewed in the {\tt RM88} frame. We have developed a quantitative method for measuring and combining these two metrics that allows us to quantitatively evaluate the overall quality of a candidate solution, as explained in detail below.

Our first step is to establish the formula we will use to quantitatively evaluate the fit of the theoretical model to the observational pulse profile data. This is carried out using the reduced chi-squared statistic, denoted by $\chi_\text{red}^2$, computed using \citep[e.g.,][]{PEDERSEN1997171}
\begin{equation}
    \chi_\text{red}^2=\frac{1}{\tt{dof}}\sum_{n=1}^{N_\text{data}}\frac{\left(D_n\Delta t_n-T_n\Delta t_n\right)^2}{\sigma_n^2} \ ,
    \label{reducedchisquare}
\end{equation}
where $N_{\rm data}$ denotes the number of data points, $N_\text{free}$ represents the number of free parameters in the theoretical model, $\Delta t_n$ is the time interval in the $n$th time bin for the observational data, and the quantity ${\tt dof} = N_\text{data} - N_\text{free}$ denotes the number of degrees of freedom. The quantities ${D}_{n}$, ${T}_{n}$, and $\sigma_n$ appearing in Equation~(\ref{reducedchisquare}) denote the observational count rate, the computed theoretical count rate, and the observational error, respectively. Each of these three values is a function of the observational pulsar rotation phase angle, $\beta$, which has $N_\text{data}$ discrete values, represented by $\beta_n$. The values of $N_{\rm data}$, $N_\text{free}$, and ${\tt dof}$ for Scenarios~1 and 2 are listed in Table~\ref{tbl-1new}.

The observational error, $\sigma_n$, in Equation~(\ref{reducedchisquare}) is computed using Poisson statistics, and therefore $\sigma_n$ is equal to the square root of the mean number of counts in the $n$th phase bin. We can therefore write 
\citep{Leahy2004a,FerrignoEtal2023}
\begin{equation}
    \sigma_n = \sqrt{D_n\,\Delta t_n} \ .
    \label{sigmandef}
\end{equation}
Combining Equations~(\ref{reducedchisquare})  and (\ref{sigmandef}) yields
\begin{equation}
    \chi_\text{red}^2 = \frac{1}{\tt{dof}}\sum_{n=1}^{N_\text{data}}\frac{\left(D_n-T_n\right)^2\Delta t_n}{D_n} \ .\label{chiSquaredStat}
\end{equation}
The first criterion in our evaluation of a candidate pulsar geometry, represented by the angles $(\Psi,\varphi_1,\varphi_2,\Delta\beta_1,\Delta\beta_2)$, is the calculation of the reduced chi-squared statistic, $\chi_\text{red}^2$, using Equation~(\ref{chiSquaredStat}).
The second criterion we consider is based on the shape of the angular distribution (the beaming pattern) of the intensity radiated from the walls and tops of the two accretion columns, as viewed in the local {\tt RM88} frame. This is discussed below.

The second step is the development of a method for quantitatively evaluating the shapes of the angular distributions (the beaming patterns) for the radiation emitted from the walls and tops of the two accretion columns. The electron scattering cross section in a strong magnetic field can have a strong angular dependence, but this tends to get smoothed out once averaged over energy \citep{Ventura1979,Meszaros_Ventura_1979}. Since the {\tt BW22} model utilizes energy-averaged scattering cross sections, we expect that the beaming pattern obtained here should be relatively smooth, without excessive peaks or valleys with respect to variation of the {\tt RM88} emission angle $\theta_0$, which we refer to as ``spikiness.'' In general, the emission from the column tops tend to be collimated around the outward radial direction as a consequence of the free-streaming boundary condition imposed there. This radial collimation naturally results in beaming patterns for the column tops that are relatively smooth, and therefore physically reasonable. However, the emission from the column walls is not as strongly influenced by the boundary conditions, and therefore it is important to measure the smoothness of the beaming patterns for the wall emission, represented by the sets of angular weight coefficients $\{W_{i}^\text{wall,1}\}$ and $\{W_{i}^\text{wall,2}\}$. We evaluate the ``spikiness'' of the beaming patterns for the wall emission from the two accretion columns using a new statistical metric, $\zeta$, that quantifies the angle-to-angle variation of the intensity, defined by
\begin{equation}
\zeta \;\equiv\; \sum_{i=1}^{N_\text{obs}-1}
\left(\dfrac{W_i - W_{i+1}}{\overline{W}}\right)^2 \ ,
\label{zetaStat}
\end{equation}
where
\begin{equation}
    \overline{W} \equiv \frac{1}{N_\text{obs}}\sum_{i=1}^{N_\text{obs}} W_i \ ,
    \label{eq:123}
\end{equation}
and $N_\text{obs}$ represents the number of values of $\theta_0$ corresponding to the ``visible'' geodesics, meaning those that reach the distant observer. Models with small values of $\zeta$ represent intensity patterns that do not display large variations with respect to the {\tt RM88} emission angle $\theta_0$, and are therefore relatively smooth. The sum in Equation~(\ref{zetaStat}) is carried out separately for the wall intensity distributions of each accretion column and the results are added together to obtain the combined wall statistic, denoted by $\zeta_\text{wall}$.

In order to evaluate the overall quality of a candidate model, we need to take into consideration both the smoothness of the angular distribution, measured using the $\zeta$ establish a single overall metric that can be used to compare the quality of the fits and intensity distributions for many different models using a single number, it is convenient to define a total statistic, denoted by $\Gamma$, which is based on a weighted combination the $\chi^2_{\rm red}$ and $\zeta$ metrics defined above. We therefore adopt the phenomenological definition
\begin{equation}
    \Gamma^2 = A \left(\frac{\chi_\text{red}^2}{\bar{\chi}_\text{red}^2}\right)^2
    + B\left(\frac{\zeta_\text{wall}}{\bar{\zeta}_\text{wall}}\right)^2 \ ,
    \label{GammaStat}
\end{equation}
where $A$ and $B$ represent the statistical weight parameters, describing the contribution of each metric to the value of $\Gamma$. The normalization parameters $\bar{\chi}^2_\text{red}$ and $\bar{\zeta}_\text{wall}$ appearing in Equation~(\ref{GammaStat}) represent the averages of each of the two metrics across the investigated parameter space. The assignment of values for the weight parameters $A$ and $B$ is of course rather arbitrary, and the detailed results will inevitably depend on the values selected for these parameters. However, in the absence of a complete multidimensional and angle-dependent theory, we must try to make a reasonable guess. Based on the fact that Equation~(\ref{GammaStat}) also includes the normalization factors $\bar{\chi}^2_\text{red}$ and $\bar{\zeta}_\text{wall}$, we conclude that a reasonable choice for the weight parameters is $A = 0.5$ and $B = 0.5$. Hence we will weight the $\chi^2_{\rm red}$ and $\zeta$ metrics equally when computing the composite statistic $\Gamma$ in our analysis of the observational data for Her X-1. We rank all of the computed models according to the value of $\Gamma$, with smaller values of $\Gamma$ characterizing intensity distributions that do not display excessive angular variations, while also providing excellent fits to the pulse profile data. The best overall model corresponds to the lowest value of $\Gamma$ obtained for a given scenario, corresponding to a specific selection of the set of rotational and magnetic angles $(\Psi,\varphi_1,\varphi_2,\Delta\beta_1,\Delta\beta_2)$.

The initial step in our exploration of the parameter space for each of the two scenarios considered here is a low-resolution search, conducted over the full range of all of the angular variables, in order to determine the specific zones where acceptable fits to the data are likely to be found. Once these promising zones are identified, we then perform follow-up, high-resolution searches in the specific domains, which differ somewhat for Scenarios~1 and 2. In the case of Scenario~1, with two identical accretion columns, the high-resolution search was conducted across the following parameter region:
\begin{equation}
\Psi\in\{30\degree,31\degree,32\degree,33\degree,34\degree\} \ ,
\end{equation}
\begin{equation}
    \varphi_1\in\{42\degree,43\degree,\cdots,67\degree,68\degree\} \ ,
\end{equation}
\begin{equation}
    \varphi_2\in\{108\degree,109\degree,\cdots,127\degree,128\degree\} \ .
\end{equation}
For Scenario~2, with two independent accretion columns, the high-resolution search was conducted in the following parameter region:
\begin{equation}
    \Psi\in\{72\degree,73\degree,74\degree,75\degree,76\degree,77\degree,78\degree\} \ ,
\end{equation}
\begin{equation}
    \varphi_1\in\{45\degree,46\degree,\cdots,64\degree,65\degree\} \ ,
\end{equation}
\begin{equation}
    \varphi_2\in\{110\degree,111\degree,\cdots,129\degree,130\degree\} \ .
\end{equation}
The results obtained using our statistical analysis in each of the two high-resolution searches are depicted graphically using contour plots of the combined statistical metric $\Gamma$ computed using Equation~(\ref{GammaStat}). We discuss these results below, and we also provide a summary in Table~\ref{tbl-1new}.

\begin{deluxetable}{cccccccc}
\tabletypesize{\scriptsize}
\tablecaption{Statistical Results for Best Fit Models
\label{tbl-1new}}
\tablewidth{0pt}
\tablehead{
\colhead{Scenario}
& \colhead{$\chi_{\mathrm{red}}^2$}
& \colhead{$\zeta_{\mathrm{wall}}$}
& \colhead{$\Gamma$}
& \colhead{$\log \Gamma$}
& \colhead{$N_\text{free}$}
& \colhead{$N_\text{data}$}
& \tt{dof}
}
\startdata
1
&2.199
&0.545
&0.531
&-0.275
&29
&65
&36
\\
2
&2.508
&0.254
&0.201
&-0.698
&58
&65
&7
\\
\enddata


\end{deluxetable}

\begin{figure}[htbp]
    \centering
    
    \begin{minipage}{0.32\textwidth}
        \centering
        \includegraphics[width=\linewidth]{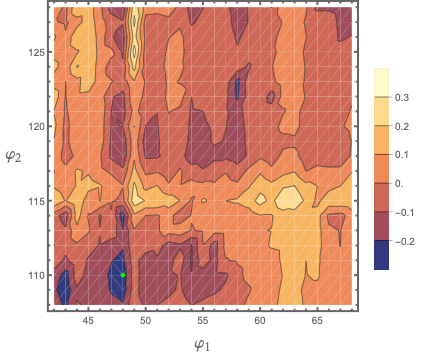}
        \caption*{}
    \end{minipage}
    \hfill
    \begin{minipage}{0.32\textwidth}
        \centering
        \includegraphics[width=\linewidth]{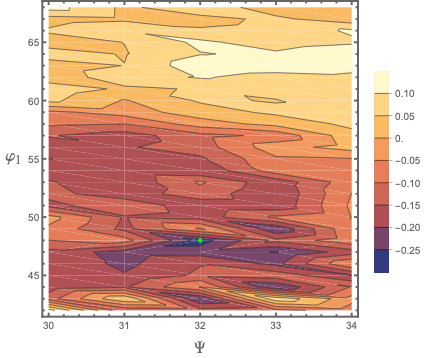}
        \caption*{}
    \end{minipage}
    \hfill
    \begin{minipage}{0.32\textwidth}
        \centering
        \includegraphics[width=\linewidth]{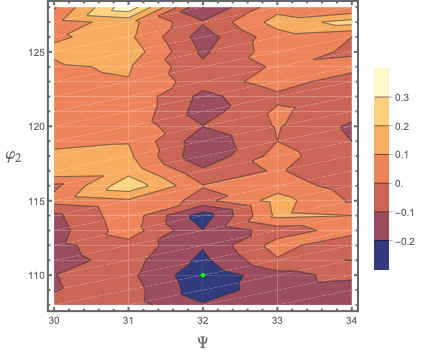}
        \caption*{}
    \end{minipage}
    
    \caption{Contour plots of $\log\Gamma$, computed using Equation~(\ref{GammaStat}), for the model fits conducted across the parameter space for Scenario~1, corresponding to the identical column model discussed in Section~\ref{sec:IdenticalColumns}. The best fit model corresponds to the minimum value of $\Gamma$, obtained for $(\Psi = 32\degree,\varphi_1 = 48\degree,\varphi_2 = 110\degree)$, indicated by the green dot in each plot. The plots, from left to right, represent contours of $\log\Gamma$ as a function of ($\varphi_1,\varphi_2$) holding $\Psi = 32\degree$, as a function of $(\Psi,\varphi_1)$ holding $\varphi_2 = 110\degree$, and as a function of $(\Psi,\varphi_2)$ holding $\varphi_1 = 48\degree$.}
    \label{fig:9}
\end{figure}

\begin{figure}[htbp]
    \centering
    
    \begin{minipage}{0.32\textwidth}
        \centering
        \includegraphics[width=\linewidth]{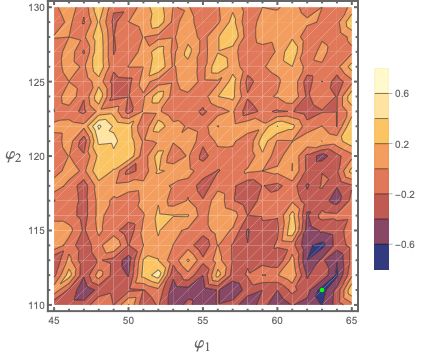}
        \caption*{}
    \end{minipage}
    \hfill
    \begin{minipage}{0.32\textwidth}
        \centering
        \includegraphics[width=\linewidth]{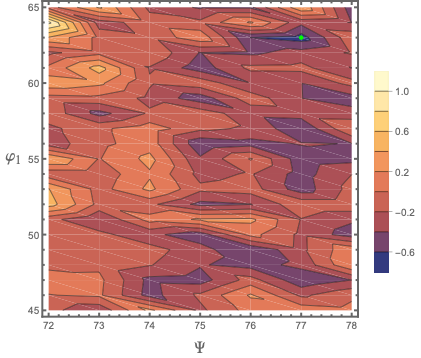}
        \caption*{}
    \end{minipage}
    \hfill
    \begin{minipage}{0.32\textwidth}
        \centering
        \includegraphics[width=\linewidth]{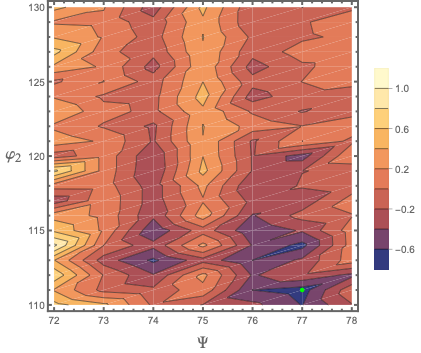}
        \caption*{}
    \end{minipage}
    
    \caption{Contour plots of $\log\Gamma$, computed using Equation~(\ref{GammaStat}), for the model fits conducted across the parameter space for Scenario~2, corresponding to the independent column model discussed in Section~\ref{sec:IndependentColumns}. The best fit model corresponds to the minimum value of $\Gamma$, obtained for $(\Psi = 77\degree,\varphi_1 = 63\degree,\varphi_2 = 111\degree)$, indicated by the green dot in each plot. The plots, from left to right, represent contours of $\log\Gamma$ as a function of ($\varphi_1,\varphi_2$) holding $\Psi = 77\degree$, as a function of $(\Psi,\varphi_1)$ holding $\varphi_2 = 111\degree$, and as a function of $(\Psi,\varphi_2)$ holding $\varphi_1 = 63\degree$.}
    \label{fig:10}
\end{figure}

The contour plots of the $\Gamma$ statistic corresponding to Scenarios~1 and 2 are presented in Figures~\ref{fig:9} and \ref{fig:10}, respectively. Each of these figures includes three separate contour plots in which one of the variables $(\Psi,\varphi_1,\varphi_2)$ is held constant using the best fit value while the other two parameters are varied. The best fit models for the identical and independent column scenarios are indicated by the minimum values of $\Gamma$ in the contour plots, which are labeled by the green dots in Figures~\ref{fig:9} and \ref{fig:10}. The systematic shapes of the contours around the minimum values of $\Gamma$ provides a statistical justification for our choice of the best fit models for each scenario. For both scenarios, we have fixed the shifts in the rotational phase angles for the two columns using the values $\Delta\beta_1 = 196\degree$ and $\Delta\beta_2 = 4\degree$ in order to align the peaks in the sub-profiles with the dominant and subdominant peaks seen in the observed pulse profile (see Figures~\ref{fig:furstetal},~\ref{fig:subprofilesidentical}, and~\ref{fig:subprofiles}). The best fit model for Scenario~1, with two identical accretion columns, has the values $(\Psi = 32\degree,\varphi_1 = 48\degree,\varphi_2=110\degree,\Delta\beta_1 = 196\degree,\Delta\beta_2 = 4\degree)$ and the best fit model for Scenario~2, with two independent columns, has the values $(\Psi = 77\degree,\varphi_1 = 63\degree,\varphi_2 = 111\degree,\Delta\beta_1 = 196\degree,\Delta\beta_2 = 4\degree)$. It is also interesting to compute the normalization parameters $\bar{\chi}^2_\text{red}$ and $\bar{\zeta}_\text{wall}$ appearing in Equation~(\ref{GammaStat}). For Scenario~1, we obtain $\bar{\chi}^2_\text{red} = 4.369$ and $\bar{\zeta}_\text{wall} = 0.976$, and for Scenario~2, we obtain $\bar{\chi}^2_\text{red} = 21.404$ and $\bar{\zeta}_\text{wall} = 0.985$. The detailed sets of results for each of these models are presented in Sections~\ref{sec:IdenticalColumns} and \ref{sec:IndependentColumns} for Scenarios~1 and 2, respectively, and the statistical results are summarized in Table~\ref{tbl-1new}, which includes the values for $N_\text{free}$, $N_\text{data}$, {\tt dof}$, \bar{\chi}^2_\text{red}$, $\bar{\zeta}_\text{wall}$, $\Gamma$, and $\log\Gamma$ obtained for the best-fit models under the Scenario~1 and 2 assumptions.

\subsection{Scenario 1: Identical Column Model for Her X-1}
\label{sec:IdenticalColumns}

Next we consider in detail the results obtained under the assumption of Scenario~1, in which the angular intensity distribution for the wall emission from Column~1 is constrained to be identical to that for Column~2, and the column top angular intensity distribution for Column~1 is likewise constrained to be identical to that for Column~2. In this case, the fitting of the observational pulse profile data using Equation~(\ref{pulseprofilefrombasis}) for the theoretical pulse-profile function, $S(\beta)$, involves obtaining the solutions for two coupled sets of weight coefficients. We therefore have
\begin{equation}
    \text{Scenario 1:}
    \begin{cases}
    W_{i}^\text{wall,1} &= W_{i}^\text{wall,2} \ ,\\
    W_{i}^\text{top,1} &= W_{i}^\text{top,2} \ .
\end{cases}
\end{equation}
We will subsequently present the detailed results obtained under the Scenario~1 constraint for the sets of sub-profiles, the theoretical pulse-profile fit, the weight coefficients (beaming pattern), the associated phase-averaged spectrum, and the time-averaged intensity measured in the co-moving frame.

\subsubsection{Pulse Profile and Weight Coefficients}
\label{sec:identicalcolassump}

\begin{figure}[htbp]
    \centering
    \begin{subfigure}[b]{0.4\linewidth}
        \centering
        \includegraphics[width=\linewidth]{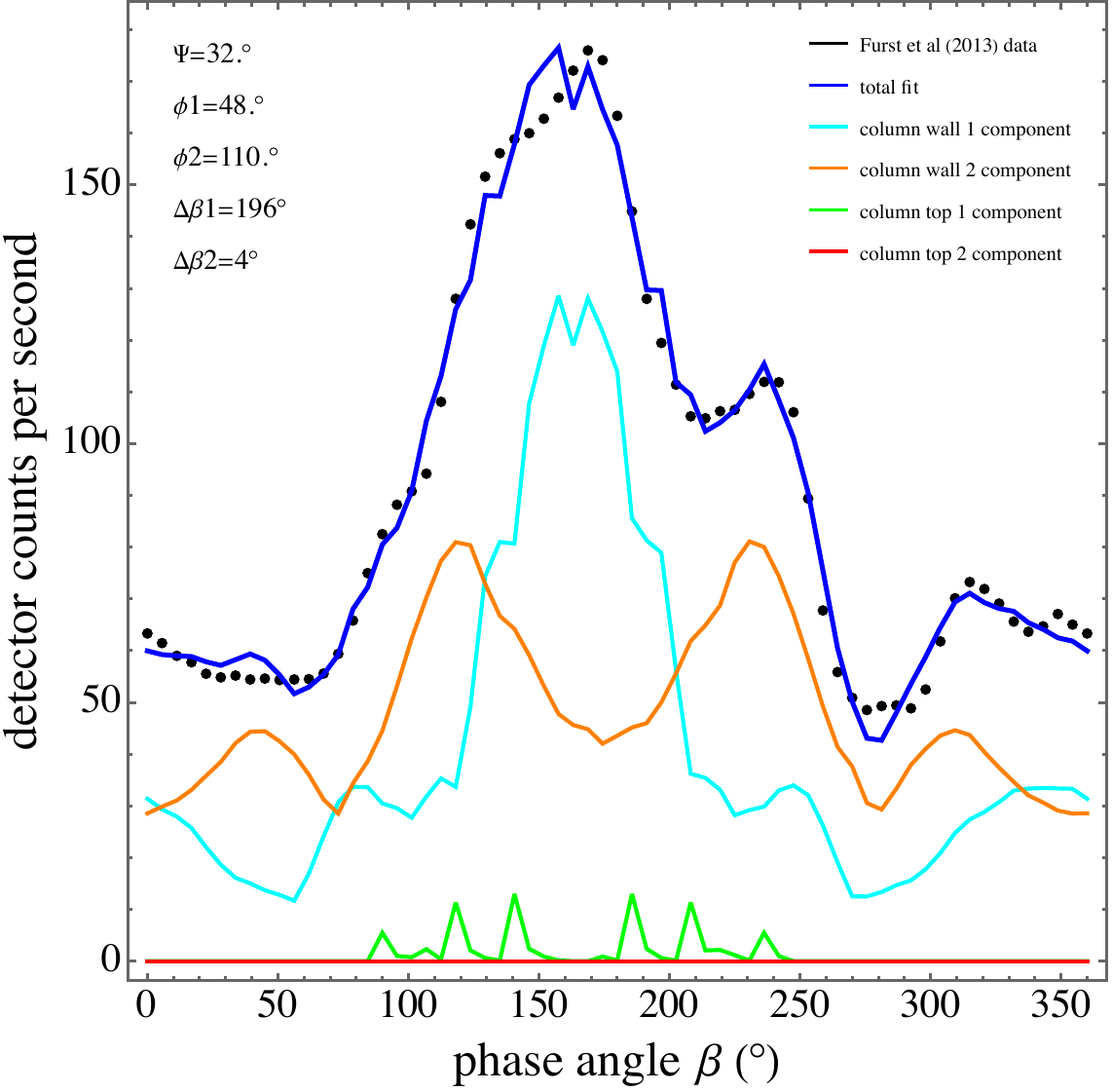}
        \caption{}
        \label{fig:11a}
    \end{subfigure}
    \hfill
    \begin{subfigure}[b]{0.4\linewidth}
        \centering
        \includegraphics[width=\linewidth]{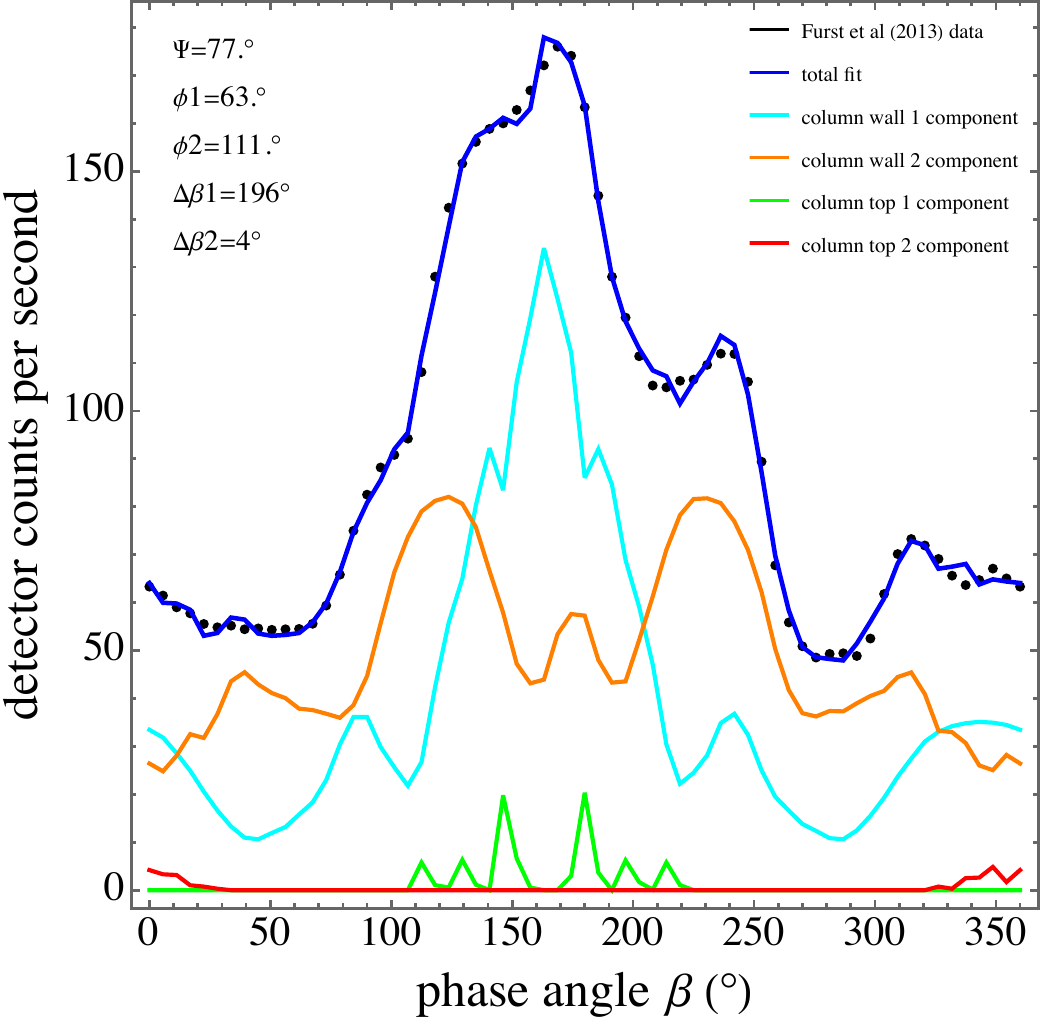}
        \caption{}
        \label{fig:11b}
    \end{subfigure}

    \caption{Theoretical pulse profiles, $S(\beta)$, for a rotating neutron star with two accretion columns, computed using Equation~(\ref{pulseprofilefrombasis}) and assuming either (a) the Scenario~1 constraint of identical angular intensity distributions for the two accretion columns, or (b) the Scenario~2 assumption of independent angular intensity distributions. The total pulse profile is indicated by the blue curve, and the individual components radiated from Wall~1, Wall~2, Top~1, and Top~2 are plotted in cyan, orange, green, and red, respectively. The black points represent the Her X-1 Observation~II data from \citet{Furst_et_al_2013}. See the discussion in the text.}
    \label{fig:11}
\end{figure}

\begin{figure}[htbp]
    \centering
    \begin{subfigure}[b]{0.48\linewidth}
        \centering
        \includegraphics[width=\linewidth]{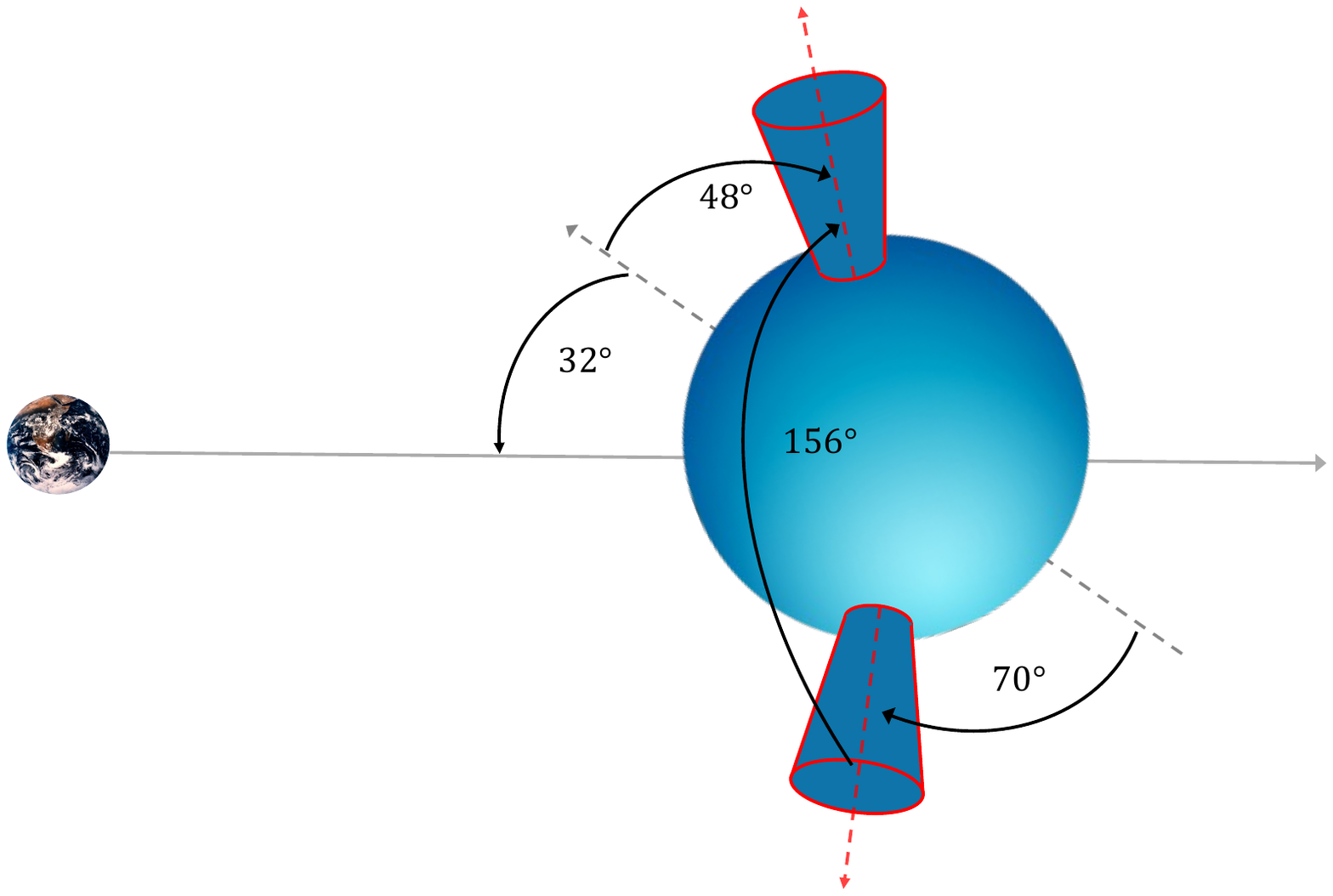}
        \caption{}
        \label{fig:12a}
    \end{subfigure}
    \hfill
    \begin{subfigure}[b]{0.48\linewidth}
        \centering
        \includegraphics[width=\linewidth]{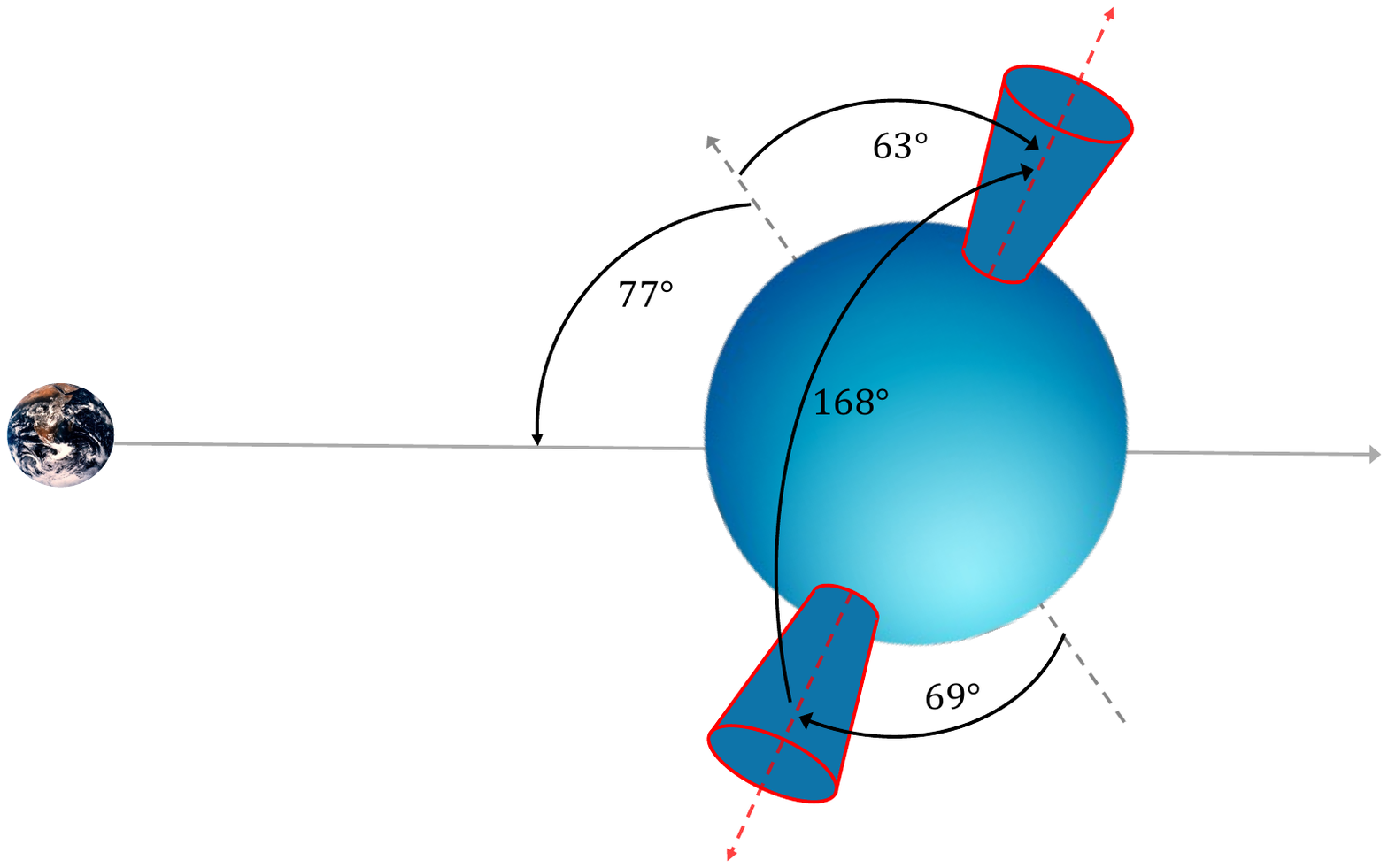}
        \caption{}
        \label{fig:12b}
    \end{subfigure}

    \caption{Visualizations of the rotational and magnetic geometry of Her X-1 obtained under (a) the Scenario~1 constraint of identical angular intensity distributions for the two accretion columns, or (b) the Scenario~2 assumption of independent angular intensity distributions.}
    \label{fig:HerX1OrientationComparison}
\end{figure}

In Scenario~1, we constrain Column~1 and Column~2 to radiate with identical angular intensity distributions, so that the two columns exhibit the same beaming patterns. The theoretical pulse-profile function, $S(\beta)$, computed using Equation~(\ref{pulseprofilefrombasis}) under the Scenario~1 constraint is plotted in Figure~\ref{fig:11a}, and compared with the pulse profile data corresponding to Observation~II from \citet{Furst_et_al_2013}. After an extensive survey of the multidimensional parameter space, we find that under the Scenario~1 constraint, the geometric orientation that yields the best qualitative fit to the \citet{Furst_et_al_2013} pulse profile data is given by
\begin{equation}
    \Psi=32\degree \ , \ \varphi_1=48\degree \ , \ \varphi_2=110\degree \ , \ \Delta\beta_1=196\degree \ , \ \Delta\beta_2=4\degree \ ,
\end{equation}
as illustrated in Figure~\ref{fig:12a}. The black points in Figure~\ref{fig:11a} represent the observational data in the $3-79\,$keV energy range (see Figure~\ref{fig:furstetal}), and the fit obtained using our unitary emission model is plotted in dark blue. Additionally, the cyan, orange, green, and red curves in Figure~\ref{fig:11a} represent the individual contributions made to the total pulse profile due to the emission from the wall of Column~1, the wall of Column~2, the top of Column~1, and the top of Column~2, respectively.
We note that the angle between the central axes of the two accretion columns in this case is $\Xi = 155.77\degree$ (see Equation~(\ref{eq:Xi})), and the corresponding offset of the second column from the dipolar position relative to the first column, denoted by $\delta$, is given by $\delta = 180\degree - \Xi = 24.23\degree$. For the fit in Figure~\ref{fig:11a}, the reduced chi-squared value calculated using Equation~(\ref{chiSquaredStat}) is $\chi^2_\text{red} = 2.199$.

\begin{figure}[htbp]
    \centering
    \includegraphics[width=0.8\linewidth]{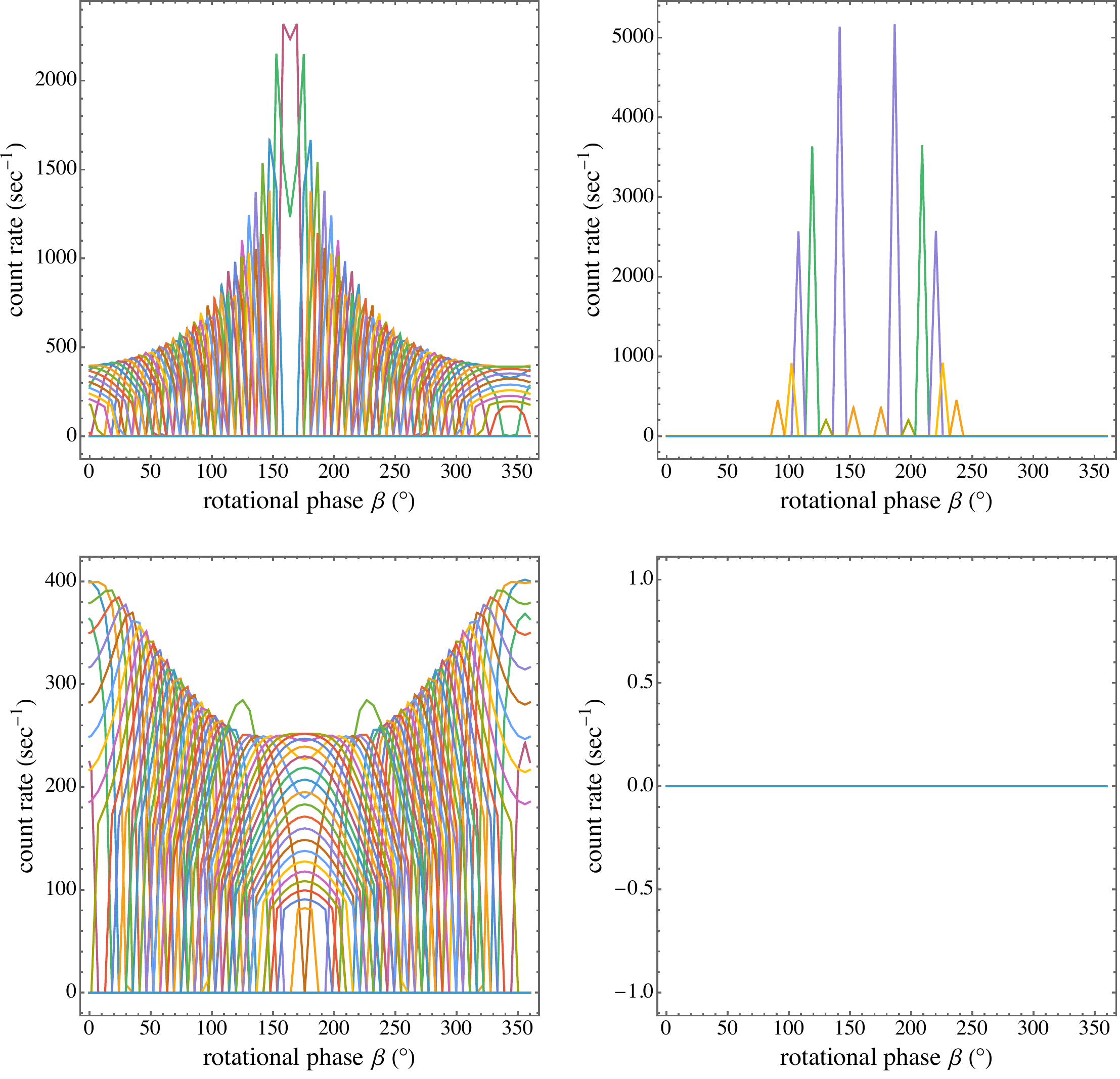}
    \caption{Sub-profiles representing emission out of the walls and tops of both accretion columns are plotted as functions of the rotational phase angle $\beta$ for Scenario~1, in which the two columns are constrained to have identical angular intensity distributions. The parameters used to compute the sub-profiles are $\Psi=32\degree$, $\varphi_1=48\degree$, $\varphi_2=110\degree$, $\Delta\beta_1=196\degree$, and $\Delta\beta_2=4\degree$. Note that the emission from the top of Column~2 is not visible during any portion of the neutron star's rotation.}
    \label{fig:subprofilesidentical}
\end{figure}

The sub-profiles $h_{i}^\text{wall,1}(\beta)$, $h_{i}^\text{wall,2}(\beta)$, $h_{i}^\text{top,1}(\beta)$, and $h_{i}^\text{top,2}(\beta)$ defined in Equations~(\ref{subprofiles}) obtained in the Scenario~1 calculation performed here are plotted as functions of the rotational phase angle $\beta$ in Figure~\ref{fig:subprofilesidentical}. These sub-profiles indicate the contribution to the observed pulse profile made by emission generated in each direction $(\theta_{0i},\phi_{0j})$ in the angular grids defined in Equations~(\ref{anglegridwall}) and (\ref{anglegridtop}). The two wall distributions are identical in keeping with the Scenario~1 constraint, as are the two top distributions. It is interesting to note that for the particular geometric configuration considered here, the emission from the top of Column~2 is never visible to the observer, for any value of the phase angle $\beta$, as indicated by the red line with zero amplitude in Figure~\ref{fig:11a}. We can also see this effect in Figure~\ref{fig:subprofilesidentical}, where it is apparent that all of the sub-profiles for the Column~2 top emission are zero. This is a result of the restriction $\theta_{0i} \le 45\degree$ that we imposed on the angular distribution for the radiation emitted from the column tops (see Equation~(\ref{anglegridtop})), combined with the value $\Psi = 32\degree$ for the observer's inclination angle relative to the neutron star's spin axis, and the value $\varphi_2 = 110\degree$ for the rotational latitude of Column~2. A visualization of this geometry is provided in Figure~\ref{fig:12a}.

\begin{figure}[htbp]
\centering
\includegraphics[width=0.7\linewidth]{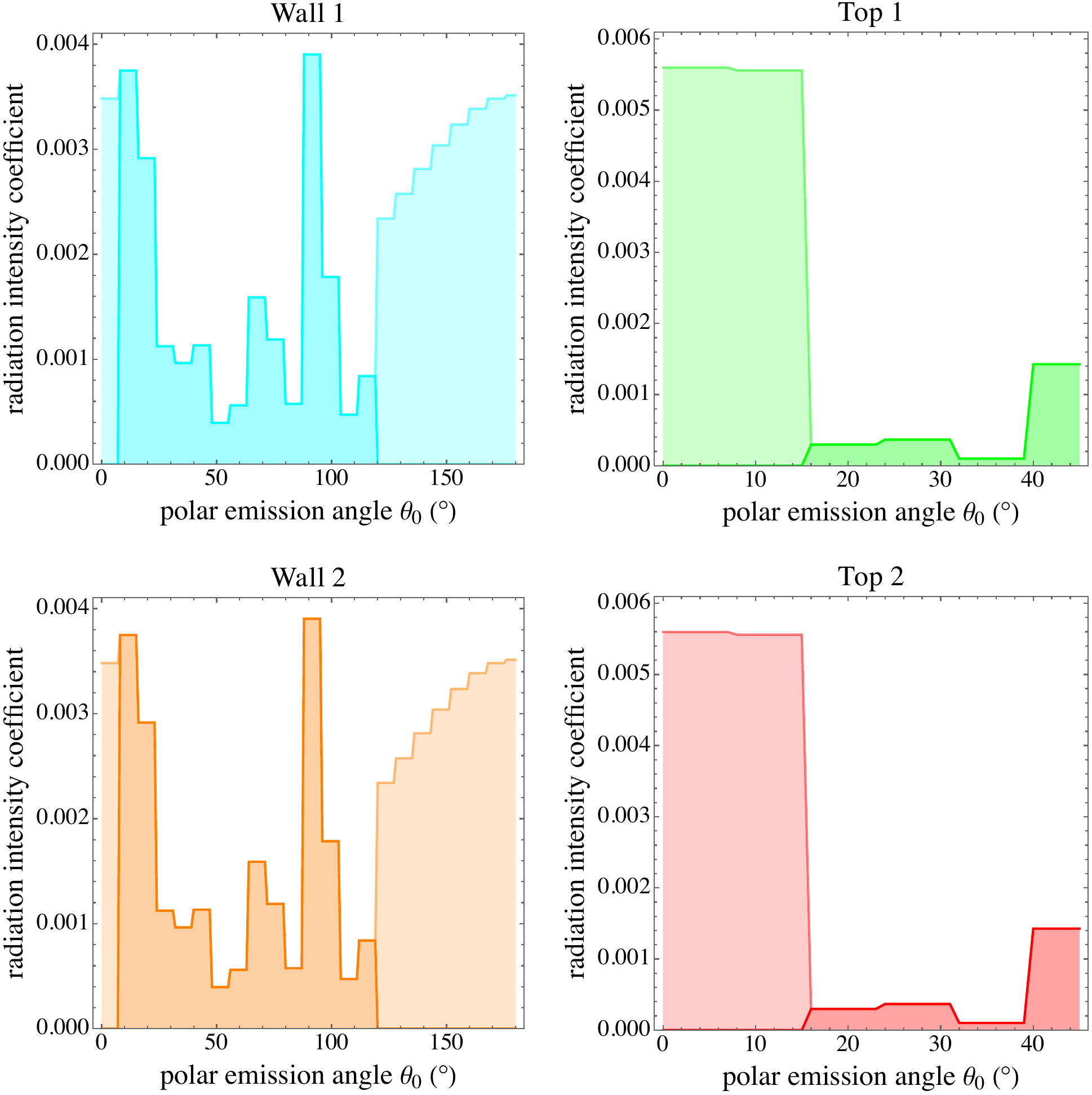}       \caption{Weight coefficients $W$ representing the angular distribution (beaming pattern) of the intensity in the local {\tt RM88} frame for the wall and top emission components, under the Scenario~1 constraint that the two columns have identical emission patterns. These coefficients yield the theoretical pulse-profile fit plotted in Figure~\ref{fig:11a}. Within the plots, the lighter-shaded regions represent emission generated in directions that never reach the observer during the star's rotation, in which case the associated amplitudes were computed using Equations~(\ref{invisibleWall}) and (\ref{invisibleTop}).}
\label{fig:herx1IdenticalColumnWs}
\end{figure}

The sets of weight coefficients $\{W_{i}^\text{wall,1}, W_{i}^\text{wall,2}, W_{i}^\text{top,1}, W_{i}^\text{top,2}\}$ obtained by fitting Equation~(\ref{pulseprofilefrombasis}) for the theoretical pulse profile function, $S(\beta)$, to the observational pulse profile data based on the Scenario~1 constraint are plotted in Figure~\ref{fig:herx1IdenticalColumnWs}. This essentially indicates the beaming pattern in the {\tt RM88} frame. In order to reduce the number of free parameters in the model, we group together eight consecutive values of $\theta_0$, separated by 1\degree, into ``bundles,'' which are assigned the same values for the angular weight coefficient. This is carried out for both the wall and top intensity distributions. These coefficients represent the angular distribution of the intensity measured in the local {\tt RM88} frame for the wall and top emission components, under the constraint that the two columns have identical angular intensity distributions. These sets of coefficients were used to compute the theoretical pulse profile plotted in Figure~\ref{fig:11a}. As discussed in Section~\ref{fig:subprofilesidentical}, some of the emission directions $(\theta_{0i},\phi_{0j})$ are unobservable during any part of the neutron star's spin for the specific observer considered here, at rotational inclination angle $\Psi = 32\degree$. This effect can also be seen in the zero-amplitude sub-profiles plotted in Figure~\ref{fig:subprofilesidentical}. For the non-visible emission directions, the weight coefficients were calculated using Equations~(\ref{invisibleWall}) and (\ref{invisibleTop}). These coefficients are indicated by the lighter shading in Figure~\ref{fig:herx1IdenticalColumnWs}.

\begin{figure}[htbp]
    \centering
    \begin{subfigure}[b]{0.45\linewidth}
        \centering
        \includegraphics[width=\linewidth]{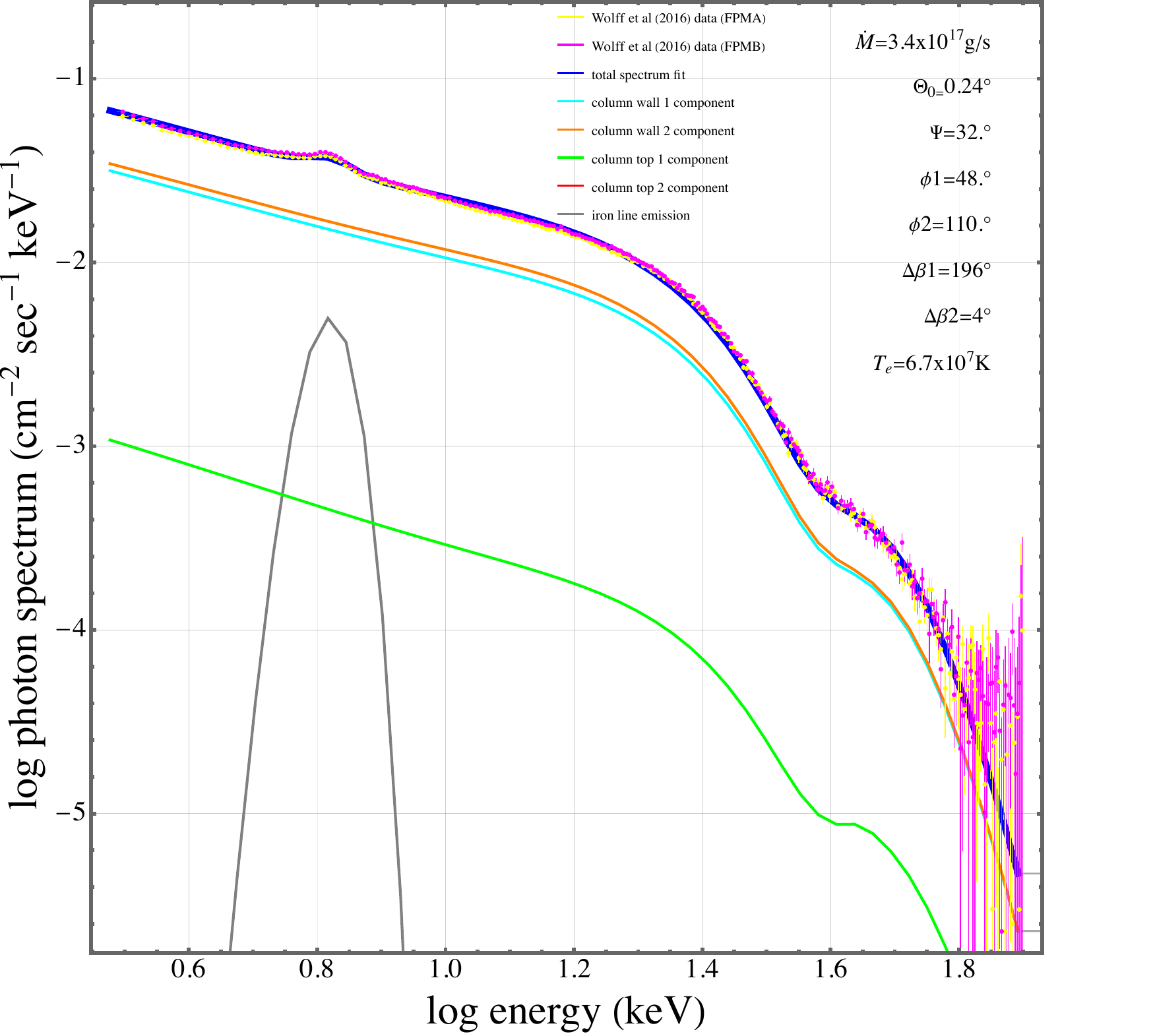}
        \caption{}
        \label{fig:herx1IdenticalColumnSpec}
    \end{subfigure}
    \hfill
    \begin{subfigure}[b]{0.45\linewidth}
        \centering
        \includegraphics[width=\linewidth]{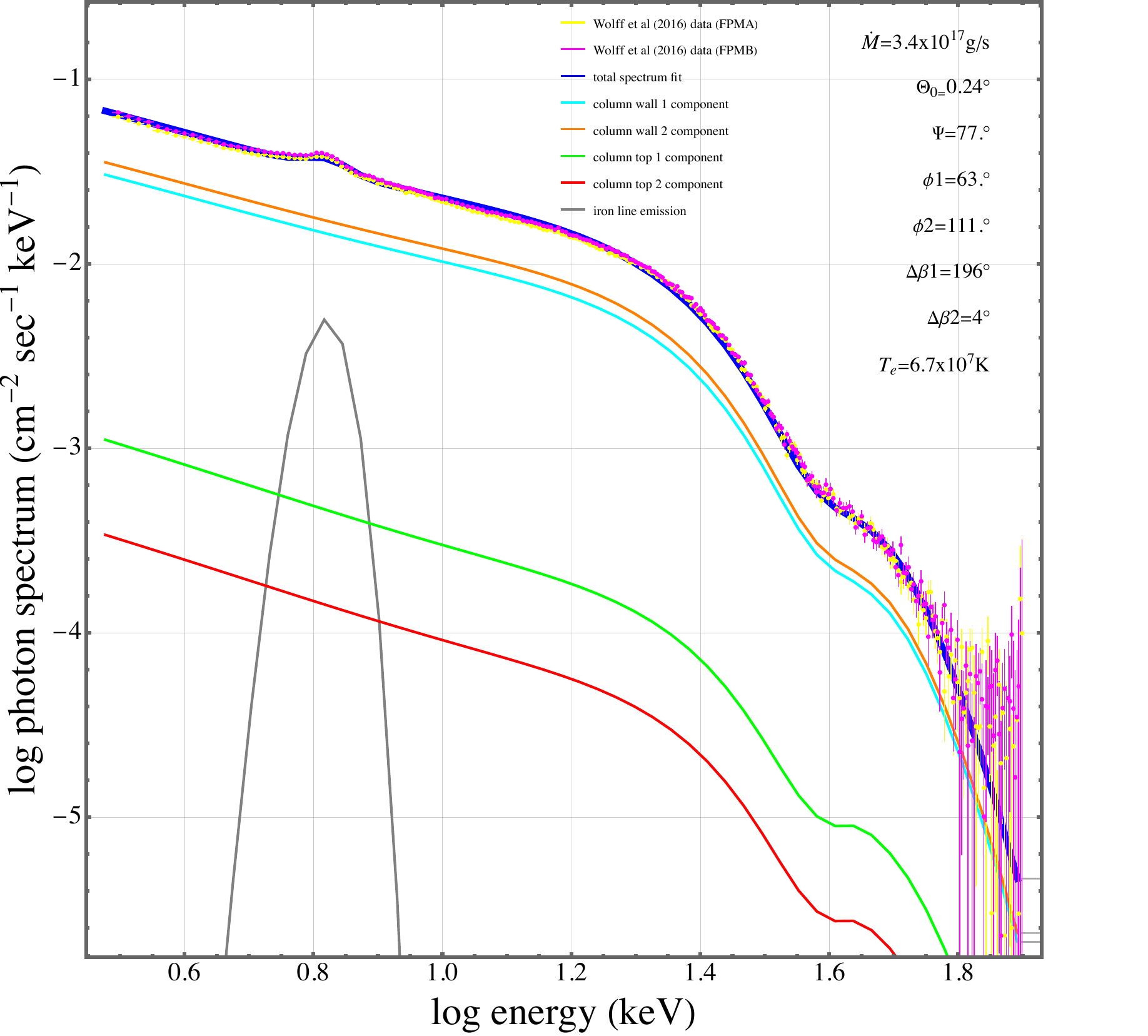}
        \caption{}
        \label{fig:specplot}
    \end{subfigure}

    \caption{Comparison of the phase-averaged theoretical photon number spectra for Her X-1 computed using Equation~(\ref{eq:PhaseAveSpec}), indicated by the blue curves, compared with the {\sl NuSTAR} data from \citet{Wolff_etal_2016}, indicted by the cyan and magenta points. Panel (a) corresponds to Scenario~1, in which the two accretion columns are constrained to have identical angular intensity distributions, and panel (b) corresponds to Scenario~2, in which the angular distributions of the two columns are treated as independent. The individual emission components from the walls and tops of both columns are plotted in cyan, orange, green, and red according to the included legend.}
    \label{fig:herx1SpecComparison}
\end{figure}

\begin{figure}[htbp]
    \centering
    \includegraphics[width=\linewidth]{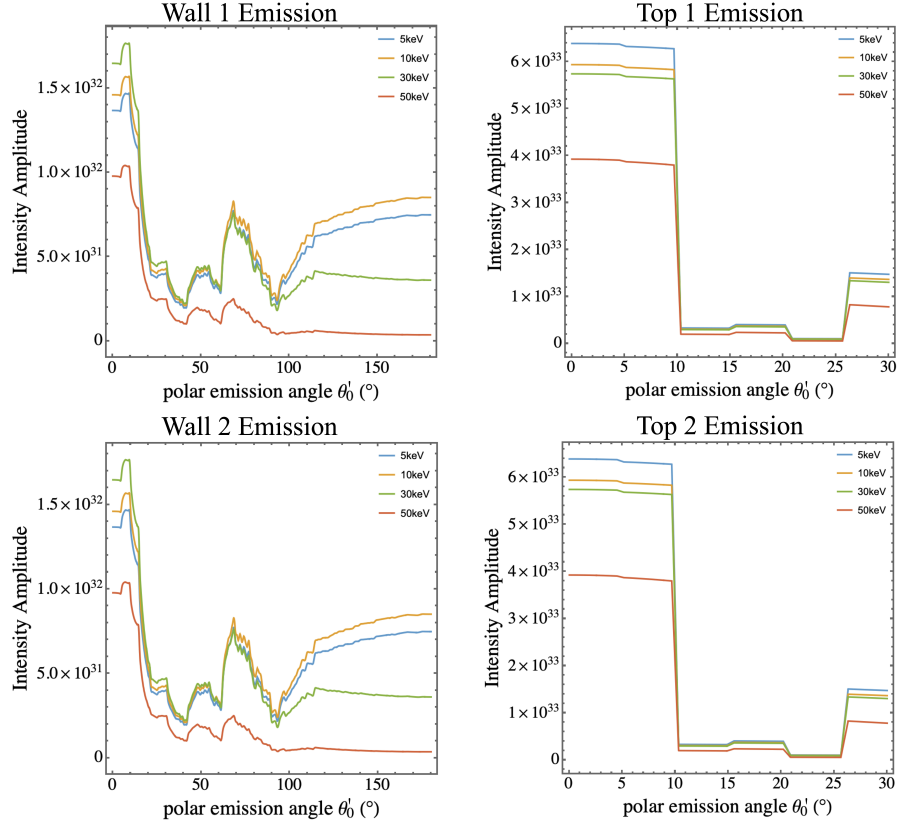}
    \caption{Co-moving frame intensity amplitude functions $Q^\text{wall}(\mu_{0}',\epsilon_0')$ and $Q_i^\text{top}(\mu_{0}',\epsilon_0')$, computed using Equations~(\ref{eq:B27newTEXT}) and (\ref{eq:132new}), respectively, under the Scenario~1 assumption of two identically emitting accretion columns. The distributions are plotted as functions of the co-moving polar angle $\theta_0'$, where $\mu_0'=\cos\theta_0'$. The values of the co-moving photon energy are $\epsilon_0' = 5\,$keV, $10\,$keV, $30\,$keV, and $50\,$keV, as indicated by the color legend.}
    \label{fig:IdenticalIntensity}
\end{figure}

\subsubsection{Phase-averaged Spectrum}
\label{sec:IdentSpect}

Once the sets of weight coefficients $\{W_{i}^\text{wall,1}, W_{i}^\text{wall,2}, W_{i}^\text{top,1}, W_{i}^\text{top,2}\}$ have been determined by fitting the observational pulse profile data using the expansion in basis functions given by Equation~(\ref{subprofiles}), we can then use the same sets of basis functions to compute the phase-averaged X-ray spectrum based on Equations~(\ref{eq:PhaseDepSpec}) and (\ref{eq:PhaseAveSpec}). We note that in Scenario~1 considered here, the intensity angular distributions for the two columns are constrained to be identical, and therefore the weight coefficients satisfy the relations $\{W_{i}^\text{wall,1}\} = \{W_{i}^\text{wall,2}
\}$ and \{$W_{i}^\text{top,1}\} = \{W_{i}^\text{top,2}\}$. Utilizing the sets of weight coefficients plotted in Figure~\ref{fig:herx1IdenticalColumnWs}, the theoretical result we obtain for the phase-averaged spectrum of Her X-1 under Scenario~1 is presented in Figure~\ref{fig:herx1IdenticalColumnSpec}, which also includes the phase-averaged Her X-1 spectrum from \citet{Wolff_etal_2016} for comparison. The spectrum reported by \citet{Wolff_etal_2016} was obtained using \textit{NuSTAR}, corresponding to ObsID 30002006005, which is the same observation that \cite{Furst_et_al_2013} analyzed to obtain the pulse profile plotted in Figure~\ref{fig:furstetal}.  The observation occurred on 2012 September 22 from 04:20:32 to 18:35:00 UTC, and comprises data captured by the FPMA and FPMB modules. For a detailed overview of the observational methods utilized by the \textit{NuSTAR} mission, see \citet{harrison_etal_2013}. It is important to emphasize that the theoretical spectrum plotted in Figure~\ref{fig:herx1SpecComparison}a is not a fit to the data, but rather it represents the X-ray spectrum corresponding to the pulse-profile fit presented in Figure~\ref{fig:11a}. Hence this is an independent calculation, and therefore the agreement between the theoretical X-ray spectrum and the spectral data represents a rigorous test of both the self-consistency and the physical validity of the model. Following the analysis presented in \citet{Wolff_etal_2016}, the theoretical spectrum in Figure~\ref{fig:herx1SpecComparison}a also includes a Gaussian iron line centered at photon energy $6.4\,$keV.

\subsubsection{Time-averaged Intensity Distribution in Co-moving Frame}
\label{sec:distofintensity}

The Scenario~1 weight coefficients $\{W_{i}^\text{wall,1}, W_{i}^\text{wall,2}, W_{i}^\text{top,1}, W_{i}^\text{top,2}\}$ plotted in Figure~{\ref{fig:herx1IdenticalColumnWs}} have been obtained by fitting the pulse profile function, $S(\beta)$, given by Equation~(\ref{pulseprofilefrombasis}), using the pulse profile data for Her X-1 plotted in Figure~{\ref{fig:furstetal}}. It is important to recall that the weight coefficients describe the angular distribution of the radiation escaping from the wall and top surfaces of the accretion column, relative to the {\tt BW22} continuum value, as measured in the local {\tt RM88} frame. The corresponding phase-averaged photon number spectrum is plotted in Figure~\ref{fig:herx1SpecComparison}a. Since the {\tt RM88} frame is stationary with respect to the star, it is also interesting to compute the intensity distribution as measured in the frame of an observer who is co-moving with the plasma, as this provides insight into the angular and energy distribution of the radiation in the reference frame in which it it is actually generated. We will therefore calculate the time-average of the co-moving frame intensity distribution as a function of the co-moving polar angle $\theta_0'$ and the co-moving photon energy $\epsilon_0'$. The detailed analysis is carried out in Appendix~\ref{sec:appendixB} and we refer the reader to that section of the paper for the complete derivation.

We note that in Scenario~1 considered here, the intensity angular distributions for the two columns are constrained to be identical, and therefore the weight coefficients satisfy the relations $\{W_{i}^\text{wall,1}\} = \{W_{i}^\text{wall,2}
\}$ and \{$W_{i}^\text{top,1}\} = \{W_{i}^\text{top,2}\}$. Referring to Equation~(\ref{eq:B25new}), we can evaluate the time average of the co-moving frame wall intensity, $\mathscr{I}^\text{tot}_{\epsilon,\text{wall}}(\mu_0',\phi_0',\epsilon_0')$, using the expression
\begin{equation}
    \mathscr{I}^\text{tot}_{\epsilon,\text{wall}}(\mu_0',\phi_0',\epsilon_0') =  Q^\text{wall}(\mu_0',\epsilon_0')\sum_{j=1}^{M_\text{wall}} \, \delta(\phi_{0}'-\phi_{0j}') \ ,
    \label{eq:B25newTEXT}
\end{equation}
where the amplitude function, $Q^\text{wall}(\mu_0',\epsilon_0')$, is defined by
\begin{align}
    Q^\text{wall}(\mu_0',\epsilon_0') &\equiv \sum_{i=1}^{N_\text{wall}}\frac{1}{t_\text{tot}'} \,  \, \frac{W_i^\text{wall}\epsilon_0\Dot{N}_\epsilon^\text{tot}(R_0^*,\epsilon_0)}{2\pi R_0^*\sin\thetamax} \frac{(1+\mu_{0i}\abs{\vel(R_0^*)/c})^3}{(1-\mu_{0i}^2)\abs{\vel'(R_0^*)/c}} \, \frac{1}{\abs{\vel(R_0^*)}} \ .
    \label{eq:B27newTEXT}
\end{align}
in which $R_0^*$ is a function of $\mu_0'$ and $\mu_{0i}$ via Equation~(\ref{eq:Rroot3}), and the value of the {\tt RM88}-frame photon energy, $\epsilon_0$, is given by (see Equation~(\ref{eq:EnLor2}))
\begin{equation}
    \epsilon_0 = \epsilon_0' \, \gamma(R_0^*) \, (1-\mu_{0}' \, \abs{\vel(R_0^*)/c}) \ .
    \label{eq:EnLor2text}
\end{equation}
The co-moving time-average column-wall intensity amplitude function $Q^\text{wall}(\mu_0',\epsilon_0')$ is plotted in Figure~\ref{fig:IdenticalIntensity} under Scenario~1, in which the two columns are constrained to emit the same intensity angular distributions. The intensity curves are plotted as functions of the co-moving polar angle, $\theta_0' = \cos^{-1} \mu_0'$, for four values of the co-moving frame photon energy, $\epsilon_0'$, with $\epsilon_0' = 5\,$keV, $10\,$keV, $30\,$keV, and $50\,$keV. We note that the wall intensity patterns have an angular distribution consistent with ordinary-mode emission in the regions $\theta_0' \sim 0\degree$ and $\theta_0' \sim 180\degree$, and extraordinary-mode emission in the region $\theta_0' \sim 90\degree$, as expected based on the angular dependence of the cyclotron scattering cross section \citep{Nagel1981a,Ventura1979}.

The intensity distribution measured by a co-moving observer located at the column top, with radius $R_0 = R_\text{top}$, is computed using Equation~(\ref{eq:B41new}), which can be rewritten as
\begin{equation}
    \mathscr{I}_{\epsilon,\text{top}}^\text{tot}(\mu_{0}',\phi_0',\epsilon_0') = \sum_{i=1}^{N_\text{top}} Q_i^\text{top}(\mu_{0}',\epsilon_0') \, \delta(\mu_0'-\mu_{0i}') \, \sum_{j=1}^{M_\text{top}}\delta(\phi_{0}'-\phi_{0j}') \ ,
    \label{eq:131new}
\end{equation}
where the amplitude function for the column-top emission, $Q_i^\text{top}(\mu_{0}',\epsilon_0')$, is defined by
\begin{equation}
    Q_i^\text{top}(\mu_{0}',\epsilon_0') \equiv \frac{1}{\gamma(R_\text{top})} \, \frac{1}{1-\mu_{0}' \, \abs{\vel(R_\text{top})/c}} \, \frac{W^\text{top}_{i}\epsilon_0\,\Dot{\mathcal{N}}^\text{tot}_\epsilon(\epsilon_0)}{\Omega R_\text{top}^2} \ .
    \label{eq:132new}
\end{equation}
Here, $\gamma(R_\text{top})$ represents the bulk Lorentz factor at the column top, computed using Equation~(\ref{LorentzFactorRop}), and the {\tt RM88}-frame photon energy, $\epsilon_0$, is computed using (see Equation~(\ref{eq:B42new}))
\begin{equation}
    \epsilon_0 = \epsilon_0' \, \gamma(R_\text{top}) \, (1-\mu_{0}' \, \abs{\vel(R_\text{top})/c}) \ .
    \label{eq:133}
\end{equation}
We plot the co-moving column-top intensity amplitude function $Q_i^\text{top}(\mu_{0}',\epsilon_0')$ in Figure~\ref{fig:IdenticalIntensity} as a function of the co-moving polar angle, $\theta_0' = \cos^{-1} \mu_0'$, for four values of the co-moving frame photon energy, $\epsilon_0'$, with $\epsilon_0' = 5\,$keV, $10\,$keV, $30\,$keV, and $50\,$keV under Scenario~1, in which the two columns emit the same intensity angular distributions. Note that in Figure~\ref{fig:IndepIntensity}, we are evaluating $Q_i^\text{top}(\mu_{0}',\epsilon_0')$ at the discrete values $\mu_0' = \mu_{0i}'$ computed from the {\tt RM88} grid values $\mu_{0i}$ using the Lorentz transformation
\begin{equation}
    \mu_{0i}' = \frac{\mu_{0i}+\abs{\vel(R_\text{top})/c}}{1+\mu_{0i} \, \abs{\vel(R_\text{top})/c}} \ .
    \label{eq:134new}
\end{equation}
The plot of $Q_i^\text{top}(\mu_{0}',\epsilon_0')$ in Figure~\ref{fig:IndepIntensity} exhibits a strong feature in the region $\theta_0' \sim 0\degree$ that is consistent with the beaming along the outward radial direction due to the free-streaming boundary condition imposed at the column top. The angular distribution of the column-top intensity is dominated by a strong feature in the region $0 < \theta_0' \lesssim 25\degree$ that is interpreted as a consequence of the radial beaming associated with the free-streaming boundary condition imposed at the upper surface of the column \citep{Paletou2018,Nagel1981a}.

\subsection{Scenario 2: Independent Column Model for Her X-1}
\label{sec:IndependentColumns}

Next we consider in detail the results obtained under the assumption of Scenario~2, in which the two accretion columns are allowed to have completely independent angular distributions of the intensity radiated from the column walls and tops. In this case, the fitting of the observational pulse profile data using Equation~(\ref{pulseprofilefrombasis}) for the theoretical pulse-profile function, $S(\beta)$, involves obtaining the solutions for four separate sets of weight coefficients, $\{W_{i}^\text{wall,1}, W_{i}^\text{wall,2}, W_{i}^\text{top,1}, W_{i}^\text{top,2}\}$, so that we have
\begin{equation}
    \text{Scenario 2:}
    \begin{cases}
    W_{i}^\text{wall,1} &= {\rm independent} \ ,\\
    W_{i}^\text{wall,2} &= {\rm independent} \ ,\\
    W_{i}^\text{top,1} &= {\rm independent} \ ,\\
    W_{i}^\text{top,2} &= {\rm independent} \ .
\end{cases}
\end{equation}
With the availability of twice as many weight coefficients as in Scenario~1, we expect that the fits obtained in Scenario~2 will be of higher quality than those obtained under the Scenario~1 identical-columns constraint (see Figure~\ref{fig:11a}). As discussed previously, due to the complex multipolar magnetic topology likely to be present in accretion-powered X-ray pulsars, there may be differences between the two accretion columns in the way that the magnetic field is distributed, which may produce subtle variations in the electron scattering cross sections. In the following sections, we will present the detailed results obtained under the Scenario~2 assumption for the weight coefficients, the sets of sub-profiles, the theoretical pulse-profile fit, the associated phase-averaged spectrum, and the time-averaged intensity measured in the co-moving frame.

\subsubsection{Pulse Profile and Weight Coefficients}

\begin{figure}[htbp]
    \centering
    \includegraphics[width=0.8\linewidth]{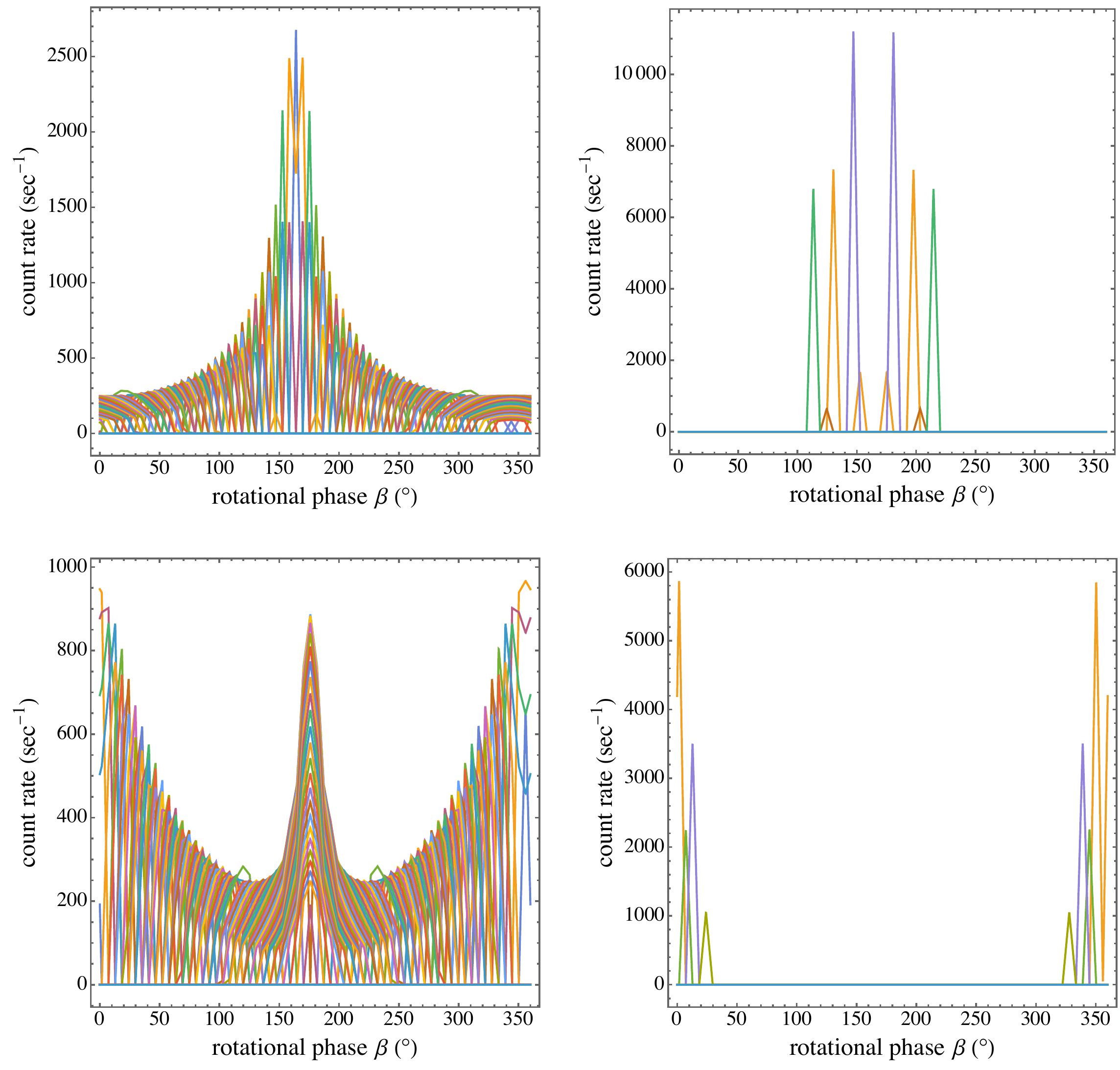}
    \caption{Sub-profiles computed for emission out of the walls and tops of both columns plotted as functions of the rotational phase angle $\beta$ for Scenario~2, in which the two columns have independent angular intensity distributions. The parameters used to generate the sub-profiles are $\Psi=77\degree$, $\varphi_1=63\degree$, $\varphi_2=111\degree$, $\Delta\beta_1=180\degree$, and $\Delta\beta_2=4\degree$.}
    \label{fig:subprofiles}
\end{figure}

In Scenario~2, we remove the constraint that the angular intensity distributions must be identical for the two accretion columns, and instead they are allowed to be independent. By removing this constraint, we have doubled the number of free parameters, allowing better fits to be generated. The theoretical pulse profile function, $S(\beta)$, obtained by fitting Equation~(\ref{pulseprofilefrombasis}) to the pulse profile data for Her X-1 under the independent column assumption of Scenario~2 is plotted in Figure~\ref{fig:11b}, where we also reproduce the Observation~II pulse profile data from \citet{Furst_et_al_2013} for comparison. Under the independent columns assumption, after a broad scan of the multidimensional parameter space, we find that the best qualitative fit to the pulse profile data is obtained by setting the geometry parameters using
\begin{equation}
    \Psi=77\degree \ , \ \varphi_1 = 63\degree \ , \ \varphi_2 = 111\degree \ , \ \Delta\beta_1 = 196\degree \ , \ \Delta\beta_2 = 4\degree \ ,
\end{equation}
as depicted in Figure~\ref{fig:12b}. The black points in Figure~\ref{fig:11b} correspond to the $3-79\,$keV energy range observational data plotted in Figure~\ref{fig:furstetal}, and the theoretical fit obtained using Equation~(\ref{pulseprofilefrombasis}) is indicated by the dark blue curve. The cyan, orange, green, and red curves in Figure~\ref{fig:11b} represent the contributions to the total pulse profile due to the emission from the wall of Column~1, the wall of Column~2, the top of Column~1, and the top of Column~2, respectively.
In Scenario~2, we find that the angle between the central axes of the two accretion columns, computed using Equation~(\ref{eq:Xi}), is $\Xi = 167.51\degree$, and the offset of the second column from the dipolar position relative to the first column is $\delta = 180\degree - \Xi = 12.49\degree$. For the Scenario~2 fit in Figure~\ref{fig:11b}, the reduced chi-squared value computed using Equation~(\ref{chiSquaredStat}) is $\chi^2_\text{red} = 2.508$. This value is actually slightly higher than the Scenario~1 value obtained from the fit in Figure~\ref{fig:11a}, which is $\chi^2_\text{red} = 2.199$, although the visual quality of the fit in Figure~\ref{fig:11b} is better than that in Figure~\ref{fig:11a}. This paradoxical result stems from the difference in the number of degrees of freedom, {\tt dof}, for the two scenarios considered here (see Table~\ref{tbl-1new}).

The sub-profiles $h_{i}^\text{wall,1}(\beta)$, $h_{i}^\text{wall,2}(\beta)$, $h_{i}^\text{top,1}(\beta)$, and $h_{i}^\text{top,2}(\beta)$ computed using Equations~(\ref{subprofiles}) under the Scenario~2 assumption are plotted as functions of the phase angle $\beta$ in Figure~\ref{fig:subprofiles}. These sub-profiles depict the pulse profile contribution made by emission in each direction in the angular grids $(\theta_{0i},\phi_{0j})$ defined in Equations~(\ref{anglegridwall}) and (\ref{anglegridtop}). Note that in Scenario~2, explored here, the angular distributions for the emission from the two walls are distinct from each other, as are the two angular distributions for the column top emission. It this case, there is emission visible from the top of Column~2 for certain emission directions. This rotational configuration is illustrated in Figure~\ref{fig:12b}.

\begin{figure}[htbp]
    \centering
    \includegraphics[width=0.7\linewidth]{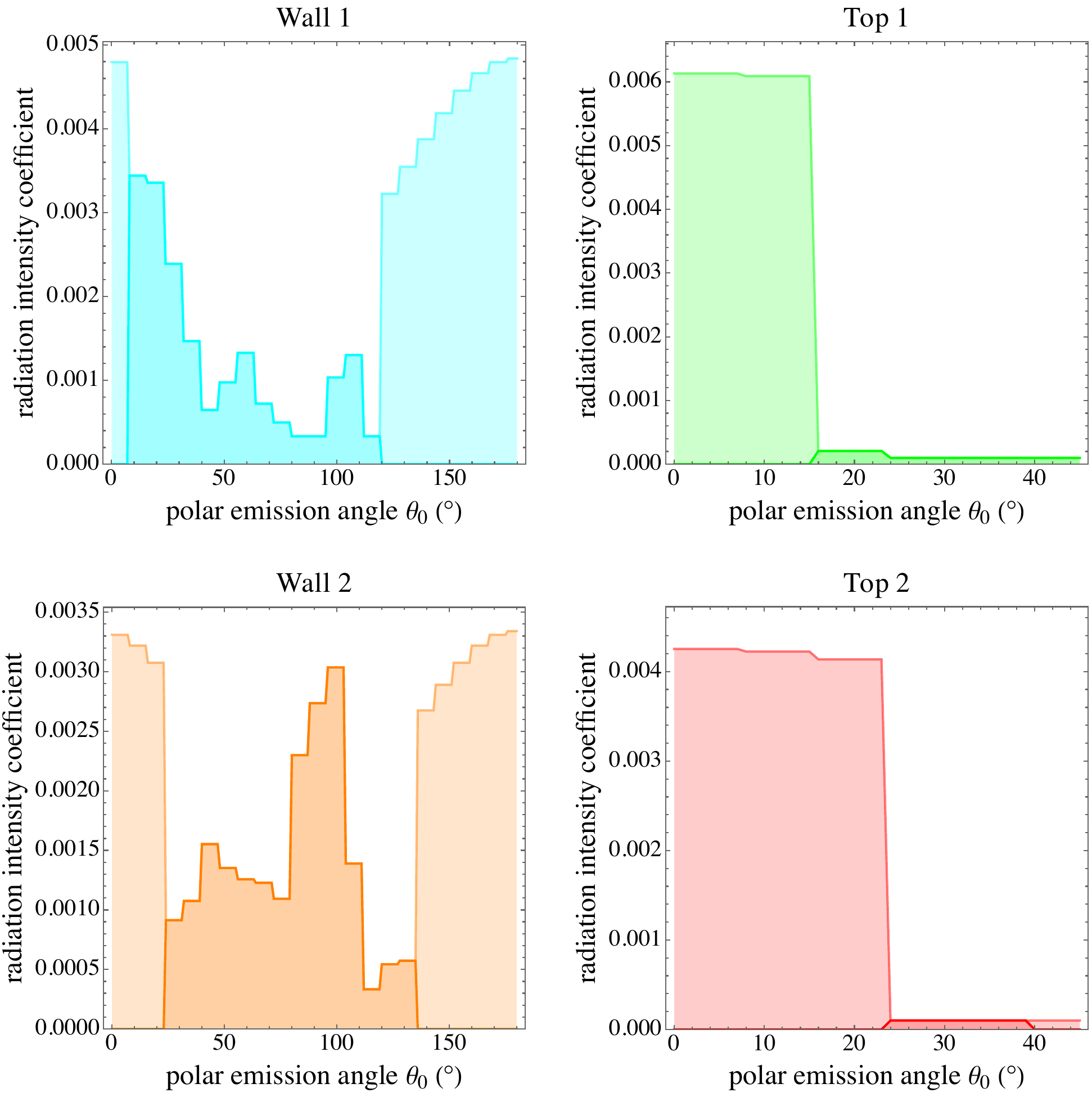}
    \caption{Weight coefficients $W$ describing the angular distribution (beaming pattern) of the intensity in the local {\tt RM88} frame for the wall and top emission components, under the Scenario~2 assumption that the two columns have independent emission patterns. These are the coefficients used to compute the theoretical pulse-profile fit plotted in Figure~\ref{fig:11b}. The lighter-shaded regions within the plots represent emission generated in directions that never reach the observer during the star's rotation, with associated amplitudes computed using Equations~(\ref{invisibleWall}) and (\ref{invisibleTop}).}
    \label{fig:herx1Ws}
\end{figure}

The sets of weight coefficients $\{W_{i}^\text{wall,1}, W_{i}^\text{wall,2}, W_{i}^\text{top,1}, W_{i}^\text{top,2}\}$ obtained by fitting Equation~(\ref{pulseprofilefrombasis}) to the observational pulse profile data under the Scenario~2 assumption are plotted in Figure~\ref{fig:herx1Ws}. These plots depict the beaming pattern in the {\tt RM88} frame. The number of free parameters in the model in the model is reduced by grouping together eight consecutive values of $\theta_0$, separated by 1\degree, into ``bundles,'' which are assigned the same values for the angular weight coefficient. The same procedure is used for both the wall and top intensity distributions. These coefficients describe the angular distribution of the intensity measured in the local {\tt RM88} frame for the wall and top emission, under the assumption that the columns have independent angular intensity distributions. These coefficients also produced the theoretical pulse profile depicted in Figure~\ref{fig:11b}. The unobservable emission directions $(\theta_{0i},\phi_{0j})$ are indicated by the zero-amplitude sub-profiles plotted in Figure~\ref{fig:subprofilesidentical}. For the non-visible emission directions, the weight coefficients were calculated using Equations~(\ref{invisibleWall}) and (\ref{invisibleTop}), and are indicated by the lighter shading in Figure~\ref{fig:herx1Ws}.

\subsubsection{Phase-averaged Spectrum}

We can now use the sets of basis functions to compute the phase-averaged X-ray spectrum based on Equations~(\ref{eq:PhaseDepSpec}) and (\ref{eq:PhaseAveSpec}). In Scenario~2 explored here, the intensity angular distributions represented by the weight coefficients $\{W_{i}^\text{wall,1}, W_{i}^\text{wall,2}, W_{i}^\text{top,1}, W_{i}^\text{top,2}\}$ for the two columns are allowed to be independent. Based on the sets of weight coefficients plotted in Figure~\ref{fig:herx1Ws}, the theoretical result we obtain for the phase-averaged spectrum of Her X-1 under Scenario~2 is depicted in Figure~\ref{fig:specplot}, along with the phase-averaged Her X-1 spectrum from \citet{Wolff_etal_2016}. It should be emphasized that the theoretical spectrum plotted in Figure~\ref{fig:specplot} is an independent prediction of the theoretical model, that was computed using weight coefficients obtained from the fit to the pulse profile data, rather than resulting from a fit to the spectral data. Hence the good agreement between theory and data strongly supports the validity of the model. The theoretical spectrum plotted in Figure~\ref{fig:specplot} also includes a Gaussian iron line at photon energy $6.4\,$keV \citep[see][]{Wolff_etal_2016}.

\subsubsection{Intensity Distribution in Plasma Frame}

\begin{figure}[htbp]
    \centering
    \includegraphics[width=\linewidth]{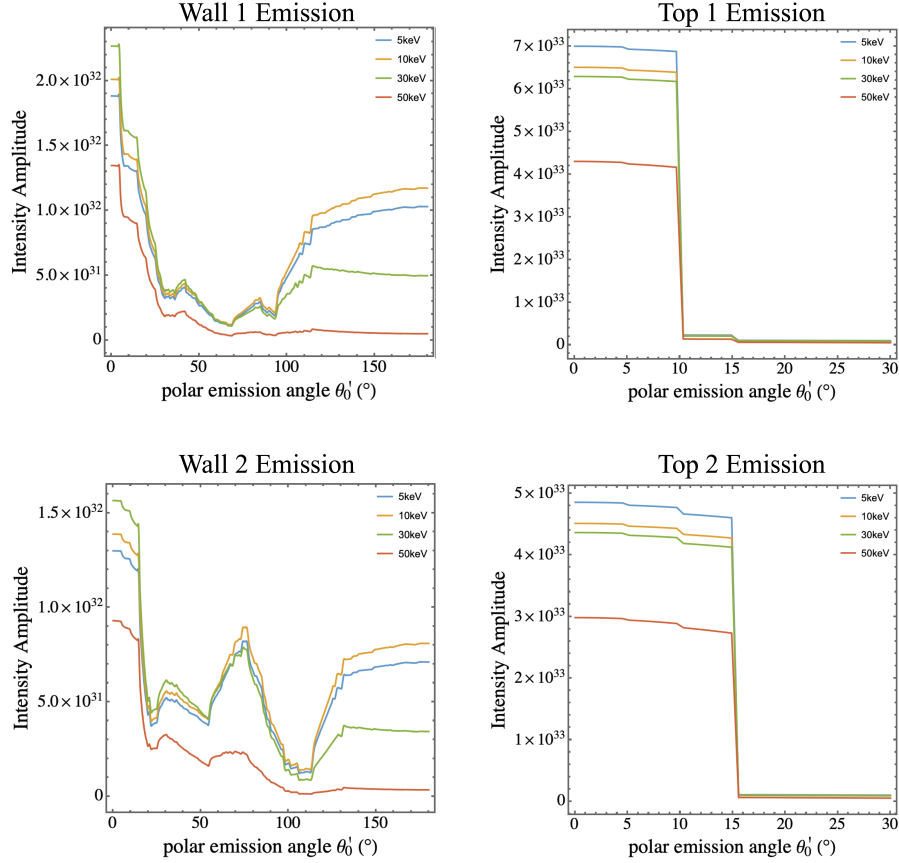}
    \caption{Co-moving frame intensity amplitude functions $Q^\text{wall}(\mu_{0}',\epsilon_0')$ and $Q_i^\text{top}(\mu_{0}',\epsilon_0')$, computed using Equations~(\ref{eq:B27newTEXT}) and (\ref{eq:132new}), respectively, based on the Scenario~2 assumption, in which the two columns have independent emission patterns. The distributions are plotted as functions of $\theta_0'$, where $\mu_0'=\cos\theta_0'$. The values of the photon energy in the co-moving frame are $\epsilon_0'=5\,$keV, 10\,keV, 30\,keV, and 50\,keV, as indicated by the color legend.}
    \label{fig:IndepIntensity}
\end{figure}

In Scenario~2 explored here, the intensity angular distributions for the two columns are independent, and consequently the four sets of weight coefficients $\{W_{i}^\text{wall,1}\}$, $\{W_{i}^\text{wall,2}
\}$, $\{W_{i}^\text{top,1}\}$, and $\{W_{i}^\text{top,2}\}$ are all separately computed during the fitting process. In Figure~\ref{fig:IndepIntensity}, we plot the amplitude function $Q^\text{wall}(\mu_0',\epsilon_0')$ representing the time-average of the wall emission intensity measured in the co-moving frame computed using Equation~(\ref{eq:B27newTEXT}). The co-moving frame photon energy is set using $\epsilon_0' = 5\,$keV, $10\,$keV, $30\,$keV, and $50\,$keV, and the intensity is plotted as a function of the co-moving polar angle, $\theta_0' = \cos^{-1} \mu_0'$. The wall intensity patterns display a combination of ordinary-mode emission for $\theta_0' \sim 0\degree$ and $\theta_0' \sim 180\degree$, and extraordinary-mode emission for $\theta_0' \sim 90\degree$ \citep{Nagel1981a,Meszaros_Ventura_1979,BeckerandWolff2022}.

The column-top intensity amplitude functions, $Q_i^\text{top}(\mu_0',\epsilon_0')$, obtained under Scenario~2 and computed using Equation~(\ref{eq:132new}), are also plotted in Figure~\ref{fig:IndepIntensity}. The energy values selected in the co-moving frame are $\epsilon_0' = 5\,$keV, $10\,$keV, $30\,$keV, and $50\,$keV, and the amplitudes are plotted as functions of the co-moving polar angle, $\theta_0' = \cos^{-1} \mu_0'$. The emission from the column top is dominated by a strong feature in the angular domain $0 < \theta_0' \lesssim 25\degree$ that is interpreted as the result of the radial beaming associated with the free-streaming boundary condition \citep{Paletou2018,Nagel1981a}.

\subsection{Quantitative Evaluation of Pulse Profile Fits}

The theoretical results computed using the Scenario~1 and Scenario~2 assumptions, presented in Sections~\ref{sec:IdenticalColumns} and \ref{sec:IndependentColumns}, respectively, both provide excellent fits to the pulse profile data for Her X-1, corresponding to Observation~II from \citet{Furst_et_al_2013}. However, the independent-column assumption of Scenario~2 provides a noticeably better overall fit to the data for the pulse profile than does the identical-column assumption of Scenario~1. This is expected, because the relaxation of the Scenario~1 constraint that the two accretion columns must have the same wall weight coefficients and also the same top weight coefficients essentially doubles the number of free parameters that can be varied in Scenario~2, which significantly enhances the quality of the resulting fit.

It is interesting to further assess our results by comparing the co-moving, time-averaged intensity distributions computed using Scenarios~1 and 2, plotted in Figures \ref{fig:IdenticalIntensity} and \ref{fig:IndepIntensity}, respectively. One may note that the angular distribution of intensity depicted in Figure~\ref{fig:IdenticalIntensity} for Scenario~1 is spikier, with more broad peaks, as compared to the Scenario~2 distribution plotted in Figure~\ref{fig:IndepIntensity}, which appears smoother, with only one broad central peak. We argue that the Scenario~2 angular distribution represents a more reasonable physical behavior, because we would expect peaks in the intensity distribution in the wings ($\theta_0' \sim 0 \degree$ or $\theta_0' \sim 180\degree$), corresponding to ordinary-mode emission, with one broad peak in the normal direction ($\theta_0' \sim 90\degree$), corresponding to extraordinary-mode emission, since this behavior reflects the imprinting of the angular dependence of the cyclotron scattering cross section on the emergent radiation field \citep{Nagel1981a}. Given that the independent column case exhibits an intensity distribution that is closer to our expectation based on the angular dependence of the scattering cross section, this provides further support Scenario~2 assumption.

\section{CONCLUSION AND DISCUSSION}
\label{sec:conc}

We have developed a new, unified model for the production of the X-ray emission from an accretion-powered X-ray pulsar with two accretion columns, that includes a detailed continuum calculation, as well as general relativistic effects, and the rotational and magnetic configuration of the neutron star. The components of the continuum X-ray spectrum emitted from the wall and top surfaces of the two accretion columns are calculated using the rigorous analytical model formulated by {\tt BW22}. The input physical parameters for the calculation of the continuum spectrum include the electron temperature, $T_e$, the accretion rate, $\dot M$, and the magnetic field strength, $B$ as well as the opening angles of the two conical accretion columns. In this paper, the continuum model of {\tt BW22} is combined with the rotational and magnetic geometry of the neutron star in order to synthesize a new comprehensive model that can be used to compute pulse profiles and phase-averaged spectra for accretion-powered X-ray pulsars. The rotational and magnetic geometry of the spinning neutron star is specified by the rotational inclination angle, $\Psi$, the rotational latitudes of the two accretion columns, $\varphi_1$ and $\varphi_2$, and the corresponding rotational phase shifts, $\Delta\beta_1$ and $\Delta\beta_2$, respectively. The radiation emitted from the two accretion columns is propagated through the Schwarzschild metric to simulate the X-ray spectrum measured by a distant observer. 

The implementation of the model is based on a novel unitary emission framework, in which the radiation escaping through the walls and tops of the two columns is expanded over a set of ``laser-like'' emission directions. The resulting sets of expansion coefficients are determined using a nonlinear least-squares method to fit the theoretical pulse profile to the pulse profile data for a specific source. We investigated two possibilities regarding the relationship between the two accretion columns. In Scenario~1, the two columns are assumed to have identical angular distributions for the escaping radiation, and in Scenario~2, these distributions are treated as independent. Our search for the best fit sets of geometry parameters involved the computation of thousands of candidate models. In order to quantitatively compare the overall quality of each candidate model, we have developed a new composite statistic, $\Gamma$, defined in Equation~(\ref{GammaStat}). This statistic provide a metric that combines the reduced $\chi^2_\text{red}$ statistic (Equation~(\ref{chiSquaredStat})) for the pulse-profile fit with the $\zeta_\text{wall}$ statistic (Equation~(\ref{zetaStat})), which describes the smoothness of the angular distribution (beaming pattern) for the radiation emitted through the walls of the two accretion columns. By minimizing the value of $\Gamma$, we are able to identify the best-fit models under either Scenario~1 (identical columns) or Scenario~2 (independent columns). While we are able to identify unique best-fit models for each of the two scenarios, it must be emphasized that our approach is phenomenological since it depends on the values selected for the statistical weight parameters $A$ and $B$ appearing in Equation~(\ref{GammaStat}), which we set using $A = 0.5$ and $B = 0.5$.

The computation of a candidate model for comparison with the observational data begins with the selection of values for the physical input parameters, including $T_e$, $\dot M$, and $B$, required for the computation of the continuum spectrum using the {\tt BW22} model. These parameters are varied in order to obtain a reasonable qualitative fit to the observed phase-averaged X-ray spectrum for the selected source. With the continuum parameters set, we proceed to investigate the parameter space of the angles $(\Psi,\varphi_1,\varphi_2,\Delta\beta_1,\Delta\beta_2)$ describing the rotational and magnetic geometry of the neutron star with two accretion columns. The search of the multidimensional parameter space requires the evaluation of the quality of thousands of computed candidate models. The quality of each model is determined quantitatively using a composite metric $\Gamma$ (Equation~(\ref{GammaStat})) that combines the valued of the reduced chi-squared metric for the pulse-profile fit, $\chi^2_\text{red}$ (Equation~(\ref{chiSquaredStat})), with the $\zeta_\text{wall}$ metric (Equation~(\ref{zetaStat})) that quantifies the smoothness or ``spikiness'' of the angular beaming pattern for the emission radiated from the column walls.

In order to reduce the complexity of the model, we have assumed here that the two accretion columns have identical physical structures, with the same values for the fundamental physical input parameters such as the electron temperature, $T_e$, the magnetic field strength, $B$, and the accretion rate, $\dot M$. However, we have explored two variations of the model, in which we make different assumptions regarding the angular distributions (beaming patterns) of the radiation escaping through the walls and tops of the two columns. In Scenario~1, we assume that the two columns radiate with identical beaming patterns, and in Scenario~2, we relax this constraint and allow the beaming patterns for the two columns to be completely independent. The pulse profile fits obtained under Scenario~1 and 2 are plotted in Figures~\ref{fig:11a} and \ref{fig:11b}, respectively. As expected, the pulse profile fit obtained under Scenario~2 is somewhat more accurate, due to the doubled number of weight coefficients $\{W_{i}^\text{wall,1}, W_{i}^\text{wall,2}, W_{i}^\text{top,1}, W_{i}^\text{top,2}\}$ available for the expansion.

In this paper, we have made an initial application to Her X-1, which is one of the most widely studied accretion-powered X-ray pulsars \citep{Staubert_etal2019,Staubert_etal2020}. In Figures~\ref{fig:subprofilesidentical} and \ref{fig:herx1Ws}, we present the results obtained for the sets of angular weight coefficients $\{W_{i}^\text{wall,1}, W_{i}^\text{wall,2}, W_{i}^\text{top,1}, W_{i}^\text{top,2}\}$ in our application to Her X-1 based on Scenarios~1 and 2, respectively. The weight coefficients are free parameters that are varied in the process of fitting the observational pulse-profile data. These plots represent the angular beaming patterns for the walls and tops of the two columns, as viewed in the {\tt RM88} frame, which is a local stationary frame. In this sense, they are similar to the beaming patterns that \citet{Blum_Kraus_2000} obtained in their application of the method developed by \citet{Kraus_etal_1995} to Her X-1. However, there are some interesting differences between their method and ours. Specifically, we expand the local intensity in the {\tt RM88} frame using the angular $\delta$-functions indicated in Equations~(\ref{intensitysumwallNEW}) and (\ref{totalintensitydeftopNEW2}), which describe the angular beaming patterns for the wall and top emission, respectively. This is analogous to the method developed by \cite{Kraus_etal_1995}, however they focused on rings and hot spots on the stellar surface, rather than allowing for the possibility of extended accretion columns. Furthermore, we note that \citet{Blum_Kraus_2000} made no attempt to fit the phase-averaged spectrum of the source, whereas our model generates self-consistent theoretical predictions for both the pulse profile and the phase-averaged spectrum.

\citet{Blum_Kraus_2000} determined the geometry and the beaming pattern for Her X-1 by fitting their model using observation pulse profile data. Likewise, we also accomplish these same two objectives using our model, with the important distinction that we also generate a meaningful physical result for the phase-averaged X-ray spectrum that can also be compared with observational data. In our model, the theoretical phase-averaged X-ray spectrum is computed using the same angular weight coefficient values that are obtained via fitting the pulse profile data. Hence the resulting phase-averaged spectrum is in some sense independent of the pulse-profile fit, and therefore it provides a meaningful test of the self-consistency of the model. The theoretical results for the phase-averaged spectrum of Her X-1 obtained using our model based on Scenario~1 and 2 are plotted and compared with the observational data in Figures~\ref{fig:herx1IdenticalColumnSpec} and \ref{fig:specplot}, respectively. Both of these results exhibit good agreement with the spectral data, corresponding to Observation~II from \citet{Furst_et_al_2013}, which was also analyzed by \cite{Wolff_etal_2016}. Hence we conclude that our model yields good qualitative agreement with both the pulse profile and the phase-averaged spectrum for Her X-1, while providing important insight into the values of the physical parameters for the source.

The resulting pulse-profile fits are plotted in Figures~\ref{fig:11a} and \ref{fig:11b} for Scenario~1 (identical column beaming patterns) and Scenario~2 (independent column beaming patterns), respectively.  The results agree reasonably well with the data taken from Observation~II of \cite{Furst_et_al_2013}. The fitting procedure is based on variation of the sets of angular weight coefficients $\{W_{i}^\text{wall,1}, W_{i}^\text{wall,2}, W_{i}^\text{top,1}, W_{i}^\text{top,2}\}$ in Equation~(\ref{pulseprofilefrombasis}), which describes the variation of the theoretical pulse profile count-rate function, $S(\beta)$. The associated sets of weight coefficients, describing the angular beaming patterns for the radiation emitted from the walls and tops of the two accretion columns, are depicted in Figures~\ref{fig:herx1IdenticalColumnWs} and \ref{fig:herx1Ws}. The theoretical formalism is based on the propagation of a set of ``laser-like'' emission vectors generated in the {\tt RM88} frame and transmitted through the Schwarzschild metric to the distant observer.

Our model for Her X-1 incorporates the emission generated from two separate accretion columns, with rotational latitudes $\varphi_1$ and $\varphi_2$, and with rotational phase shifts $\Delta\beta_1$ and $\Delta\beta_2$. The column locations are treated as independent, and therefore the values of the angles ($\varphi_1,\varphi_2,\Delta\beta_1,\Delta\beta_2)$ are determined as part of the fitting procedure. Consequently, the magnetospheric structure is allowed to deviate from a pure dipole configuration. Our computational results are based on a canonical neutron star, with mass $M_\text{star} = 1.4\,M_\odot$ and radius $R_\text{star} = 10^6\,$cm. In order to reduce the number of free parameters in the model, we have assumed here that the two columns have identical physical properties, such as the accretion rate per column in the stellar frame, $\dot M$, the radius at the column top, $R_\text{top}$, the electron scattering cross sections, $(\sigma_\perp,\sigma_{||},\bar\sigma)$, and the solid angle subtended by the conical column, $\Omega_*$. We review our primary results here, and we also compare our conclusions with previous work.

\subsection{Comparison with Previous Pulse Profile Models}

\begin{deluxetable}{clcccccccccccc}
\tabletypesize{\scriptsize}
\tablecaption{Her X-1 Column Geometry Parameters\label{tbl-2new}}
\tablewidth{0pt}
\tablehead{
\colhead{Model\!\!\!\!\!}
& \colhead{$\Psi$}
& \colhead{$\varphi_1$}
& \colhead{$\varphi_2$}
& \colhead{$\Delta\beta_1$}
& \colhead{$\Delta\beta_2$}
& \colhead{$\Xi$}
& \colhead{$\delta$}
}
\startdata
Identical Columns (this paper)
&32.0$\degree$
&48.0$\degree$
&110.0$\degree$
&196.0$\degree$
&4.0$\degree$
&155.8$\degree$
&24.2$\degree$
\\
Independent Columns (this paper)
&77.0$\degree$
&63.0$\degree$
&111.0$\degree$
&196.0$\degree$
&4.0$\degree$
&167.5$\degree$
&12.5$\degree$
\\
\citet{Leahy_1991} Model 1
&85.4$\degree$
&36.0$\degree$
&$\cdots$
&$\cdots$
&$\cdots$
&169.1$\degree$
&10.9$\degree$
\\
\citet{Leahy_1991} Model 2
&79.6$\degree$
&14.3$\degree$
&$\cdots$
&$\cdots$
&$\cdots$
&174.8$\degree$
&5.2$\degree$
\\
\citet{Leahy2004a} Model
&64.2$\degree$
&52.1$\degree$
&330.8$\degree$
&187.3$\degree$
&197.1$\degree$
&$\cdots$
&$\cdots$
\\
\citet{Leahy2004b} Model
&73.0$\degree$
&23.0$\degree$
&$\cdots$
&$\cdots$
&$\cdots$
&173.0$\degree$
&7.0$\degree$
\\
\citet{Blum_Kraus_2000} Model
&83.0$\degree$
&18.0$\degree$
&159.0$\degree$
&$\cdots$
&$\cdots$
&$>175\degree$
&$<5\degree$
\\
\enddata


\end{deluxetable}

The results presented here include a complete phenomenological solution for the rotational and magnetic geometry of the accreting neutron star, as well as the detailed angular distributions (beaming patterns) for the radiation escaping through the walls and tops of the two accretion columns, and the determination of the values of the physical input parameters, including $T_e$, $\dot M$, and $B$, that are used to calculate the continuum spectrum based on the {\tt BW22} model. In order to gain additional physical insight, we have explored two distinct scenarios here regarding the relationship between the two accretion columns. In Scenario~1, we assume that the emitted angular intensity distributions (beaming patterns) in the {\tt RM88} frame are the same for the two columns, and in Scenario~2 we assume that they are independent. However, we emphasize that in each scenario, the physical structures of the two columns (determined by the physical parameters, including $T_e$, $\dot M$, and $B$) are assumed to be identical. These two scenarios for the treatment of the beaming patterns yield different solutions for the rotational and magnetic geometry parameters, which can be compared with the results obtained by previous authors who also studied the geometry of the accreting neutron star in Her X-1. The detailed results obtained under Scenarios~1 and 2 were presented in Sections~\ref{sec:IdenticalColumns} and \ref{sec:IndependentColumns}, respectively. We compare our results with those obtained by previous authors here.

\citet{Leahy1990} developed a model for generating pulse profiles based on configurations of flat radiating rings lying on the stellar surface, centered on the magnetic poles. His model is non-relativistic and therefore it utilizes ray tracing in flat spacetime to simulate pulse profiles. \citet{Leahy_1991} used the model of \citet{Leahy1990} to generate fits to the Her X-1 pulse profile. The angular distribution of the emission from the rings was assumed to have a specified dependence on the local polar angle $\theta_0$. Based on the work of \citet{MeszarosandNagel1985b}, in Model~1 from \citet{Leahy_1991} it was assumed that the intensity $I$ had the angular dependence $I \propto \cos^2\theta_0$, whereas in his Model 2 it was assumed that $I \propto \cos^4\theta_0$. These distributions peak at small angles, indicating radially beamed emission. While the flat rings assumed by \citet{Leahy_1991} cannot be used to model the emission from the walls of the extended accretion column in our model, his results can be compared with our results for the emission from the column top. We find that Leahy's angular dependence is qualitatively similar to the intensity distributions we obtained for the column top emission, plotted in Figures~\ref{fig:IdenticalIntensity} and \ref{fig:IndepIntensity}. The differences probably reflect the fact that our model utilizes extended accretion columns and is fully relativistic.

It is also interesting to compare the values for the geometrical angles that Leahy obtained with those found here. This comparison is summarized in Table~\ref{tbl-2new}. In our Scenario~1, our best qualitative fit is obtained using the parameters $\Psi = 32\degree$, $\varphi_1=48\degree$, $\varphi_2 = 110\degree$, $\Delta\beta_1 = 196\degree$, and $\Delta\beta_2 = 4\degree$, which yield for the column separation angle $\Xi = 155.8\degree$ according to Equation~(\ref{eq:Xi}). Conversely, in Scenario~2, we find that $\Psi = 77\degree$, $\varphi_1 = 63\degree$, $\varphi_2 = 111\degree$, $\Delta\beta_1 = 196\degree$, and $\Delta\beta_2 = 4\degree$, which yields $\Xi = 167.5\degree$. In his Model~1, \citet{Leahy_1991} obtains the values $\Psi = 85.4\degree$, $\varphi_1 = 36.1\degree$, and $\Xi = 169.1\degree$, whereas in his Model~2, he finds that $\Psi = 79.6$, $\varphi_1 = 14.3\degree$, and $\Xi = 174.8\degree$. The geometry parameters obtained by \citet{Leahy_1991}, and particularly the rotational inclination angle $\Psi$, are reasonably close to our results for Scenario~2, which is depicted graphically in Figure~\ref{fig:12b}. Once again, the slight deviations in the geometrical values are not unexpected since our model is relativistic, and includes emission from both the column walls and tops, whereas the non-relativistic model of \citet{Leahy_1991} only considered rings of emission on the stellar surface.

\citet{Leahy2004a} also analyzed the Her X-1 pulse profile by finding the best-fit parameters for a relativistic, ray-tracing accretion column model using \textit{RXTE} data, with the goal of constraining the relationship between the mass and radius of the neutron star. This was carried out by treating the ratio $R_\text{star}/M_\text{star}$ as a free parameter that is varied during the fitting procedure. It is therefore interesting to compare our results with those obtained by \citet{Leahy2004a}. This comparison is carried out in Table~\ref{tbl-2new}. Assuming a neutron star with canonical mass $M_\text{star} = 1.4\,M_\odot$, \citet{Leahy2004a} obtains for the stellar radius $R_\text{star} = 11.25\,$km, with the best-fit values for the various angles given by $\Psi = 64.2\degree$, $\varphi_1 = 52.1\degree$, $\varphi_2 = 330.8\degree$ (transposed to our coordinate system), $\Delta\beta_1 = 187.3\degree$, and $\Delta\beta_2 = 197.1\degree$. A similar relativistic ray-tracing model for Her X-1 was investigated by \citet{Leahy2004b}, but without the mass-radius constraint adopted by \citet{Leahy2004a}. This model yielded the best-fit angular values $\Psi = 73.0\degree$, $\varphi_1 = 23.0\degree$, and $\delta = 7.0\degree$, where $\delta = 180\degree - \Xi$ and the angle between the magnetic poles, $\Xi$, is computed using Equation~(\ref{eq:Xi}). It is worth noting that \citet{Leahy2004b} uses the same canonical neutron star mass that we adopt, $M_\text{star} = 1.4\,M_\odot$, but he allows the stellar radius to vary between $R_\text{star} = 10\,$km (our canonical value) and $R_\text{star} = 13.65\,$km. Despite the difference in the radius constraints used by \citet{Leahy2004a} and \citet{Leahy2004b}, both of these models yield values for the Her X-1 geometry angles that are reasonably consistent with the values we obtain under the assumption of independent columns, corresponding to our Scenario 2.

\citet{Blum_Kraus_2000} analyzed the geometry of the radiating neutron star in Her X-1 using the formalism developed by \citet{Kraus_etal_1995}, which includes the gravitational distortions introduced as the radiation propagates through the curved spacetime surrounding the neutron star. In some ways, their approach resembles ours, because they also compute sub-profiles resulting from precise emission components, which in their model correspond to idealized emission rings on the stellar surface. Similarly, we also utilize an expansion, but in our case the expansion is over a set of ``laser-like'' emission directions in the {\tt RM88} frame. The resulting phase-dependent ``sub-profiles,'' denoted by $h_{i}^\text{wall,1}(\beta)$, $h_{i}^\text{wall,2}(\beta)$, $h_{i}^\text{top,1}(\beta)$, and $h_{i}^\text{top,2}(\beta)$, are plotted in Figures~\ref{fig:subprofilesidentical} and \ref{fig:subprofiles} for Scenarios~1 and 2, respectively. These sub-profiles form quasi-orthogonal sets that can be used as the basis functions for the theoretical expansion of the pulse profile function, $S(\beta)$, using Equation~(\ref{pulseprofilefrombasis}).

Another interesting similarity between our work and that of \citet{Blum_Kraus_2000} is that their method includes a reconstruction of the angular distribution (beaming pattern) for the radiation escaping from the rings on the stellar surface, which is analogous to the beaming pattern we obtain using our angular expansion. Based on the similarities between the two expansion methods, it is instructive to compare our results with those obtained by \citet{Blum_Kraus_2000} in their analysis of Her X-1. As indicated in Table~\ref{tbl-2new}, their computed geometry parameters are given by $\Psi=83\degree$, $\varphi_1=18\degree$, $\varphi_2=159\degree$. Assuming that the two poles have opposite phases with respect to the rotational angle $\beta$ in their model, this would imply that $\Delta\beta_1-\Delta\beta_2=180\degree$ in Equation~(\ref{eq:Xi}), which yields $\Xi = 177\degree$. Their result for the inclination angle $\Psi$ is relatively consistent with the analysis of \citet{Leahy_1991}, and it is also in reasonable agreement with the results we obtain using our model, under the Scenario~2 assumption of independently radiating columns. In particular, we note that the beaming patterns depicted in Figures~\ref{fig:herx1Ws} and \ref{fig:IndepIntensity} for our Scenario~2 are roughly consistent with those obtained by \citet{Blum_Kraus_2000} and \citet{Leahy2004b}, which provides further support for our formalism.

\begin{deluxetable}{clcccccccccccc}
\tabletypesize{\scriptsize}
\tablecaption{Her X-1 Continuum Model Input Parameters\label{tbl-3new}}
\tablewidth{0pt}
\tablehead{
\colhead{Model\!\!\!\!\!}
& \colhead{$\alpha$}
& \colhead{$\xi$}
& \colhead{$\psi$}
& \colhead{$\thetamax (\degree)$}
& \colhead{$\thetamin (\degree)$}
& \colhead{$k_0$}
& \colhead{$k_\infty$}
& \colhead{$y_\text{top}$}
& \colhead{$\!\!\!\dot M_\text{tot} ({\rm g\,s}^{-1})$}
& \colhead{\!\!\!$T_e {\rm (K)}$}
& \colhead{\!\!\!$B {\rm (G)}$}
}
\startdata
BW22
&0.35
&1.14
&1.42
&0.315
&0.0
&1.0
&0
&2.15
&$1.90 \times 10^{17}$
&$5.5 \times 10^{7}$
&$3.26 \times 10^{12}$
\\
Our model
&0.35
&1.14
&1.42
&0.240
&0.0
&1.0
&0
&2.15
&$3.4 \times 10^{17}$
&$6.7 \times 10^{7}$
&$4.25 \times 10^{12}$
\enddata


\end{deluxetable}

\begin{deluxetable}{ccccccccccc}
\tabletypesize{\scriptsize}
\tablecaption{Computed Model Parameters
\label{tbl-4new}}
\tablewidth{0pt}
\tablehead{
\colhead{Model}
& \colhead{$\sigperp/\sig$}
& \colhead{$\sigpar/\sig$}
& \colhead{$\sigbar/\sig$}
& \colhead{$r_\text{col}\,$(m)}
& \colhead{$z_\text{th}\,$(cm)}
& \colhead{$T_\text{th}\,$(K)}
& \colhead{$\vel_\text{th} / c$}
& \colhead{$\tau_\text{th}$}
& \colhead{$\tau_\text{top}$}
}
\startdata
BW22
&$0.537$
&$6.68 \times 10^{-5}$
&$5.90 \times 10^{-4}$
&$55.0$
&692
&$6.07 \times 10^7$
&-0.031
&0.08
&2.91
\\
Our Model
&$0.601$
&$4.33\times10^{-5}$
&$3.14\times10^{-4}$
&$41.9$
&776
&$6.96 \times 10^7$
&-0.031
&0.09
&2.91
\\
\enddata


\end{deluxetable}

\subsection{Comparison with {\tt BW22}}

In Table~\ref{tbl-3new}, we list the values of the input parameters that were used to compute the Her X-1 continuum spectrum based on the {\tt BW22} analytical model, which provides the emission components generated from the walls and tops of the two accretion columns. The same parameter values were used in both Scenarios~1 and 2. These emission components are subsequently combined with our relativistic geometrical formalism to compute the observed pulse profile and phase-averaged X-ray spectrum. For comparison, we also include the parameter values used by {\tt BW22} in their treatment of Her X-1, which was non-relativistic, and essentially computed the spectrum in the frame of the neutron star. We note that several of the continuum model parameters remain the same, reflecting the invariance of the power-law shape of the spectrum. However, there are also some interesting modifications to the electron temperature, $T_e$, the magnetic field strength, $B$, and the total accretion rate, $\dot M_\text{tot}$. Specifically, we observe that our parameters have shifted from those used by {\tt BW22} in order to properly account for the redshifting effect of the star's gravitational field, in a manner consistent with the Schwarzschild metric, which describes the spacetime curvature around the star. We discuss various aspects of this comparison in detail below.

\subsection{Temperature and Magnetic Field}
\label{sec:TempMagField}
In general, the redshift $z$ relating the frame of a local stationary observer at radius $R$ to the frame of a distant stationary observer is computed using \citep{Weinberg1972}
\begin{equation}
    z(R) \equiv \left(1-\frac{2 G M_\text{star}}{c^2 R}\right)^{-1/2} - 1 \ .
    \label{eq:RedShift1}
\end{equation}
The radiation spectrum observed from an X-ray pulsar is composed of emission that escapes from the accretion column at all values of the radius $R$ within the domain $R_\text{star} \le R \le R_\text{top}$. Since the redshift varies with the radius $R$, this implies that there is no global redshift factor that characterizes all of the emission. However, detailed analysis shows that most of the radiation escapes from the lower region of the accretion column, and therefore the surface redshift, $z_\text{star}$, computed using the relation
\begin{equation}
    z_\text{star} \equiv \left(1-\frac{2 G M_\text{star}}{c^2 R_\text{star}}\right)^{-1/2} - 1 = 0.306 \ .
    \label{eq:starZ}
\end{equation}
should provide a reasonable approximation of the global redshift \citep{BeckerandWolff2007}. Adopting this approximation, we conclude that the shifts in the magnetic field and temperature values between our model and that of {\tt BW22} should satisfy the relations
\begin{equation}
T_e \approx (1+z_\text{star}) \, T_{\tt BW22} \ ,
\qquad    
B \approx (1+z_\text{star}) \, B_{\tt BW22} \ ,
\label{eq:RedTB}
\end{equation}
where $T_{\tt BW22}$ and $B_{\tt BW22}$ denote the electron temperature and magnetic field values used by {\tt BW22}, respectively, and the gravitational redshift at the surface of the star, $z_\text{star}$, is given by Equation~(\ref{eq:starZ}).

While the relations introduced in Equation~(\ref{eq:RedTB}) should provide a reasonable description of the shifts in the values of the temperature and magnetic field between the two models, we must keep in mind that this is only an approximation when comparing our parameters with those used by {\tt BW22}. Referring to the values listed in Table~\ref{tbl-3new}, we note that the electron temperature has been increased by a factor of 1.22 going from {\tt BW22} to the present paper, and likewise the magnetic field strength has been increased by a factor of 1.30. Both of these factors of increase are close to the gravitational redshift factor $1+z_\text{star} = 1.31$ predicted using Equation~(\ref{eq:RedTB}), and therefore the structure of our model is consistent with the expected general relativistic relations.

\subsection{Luminosity and Accretion Rate}
\label{SecMdot}

The standard relationship between the X-ray luminosity, $L_\text{X}$, and the observer-frame accretion rate, $\dot M_\text{obs}$, is given by Equation~(\ref{eq:mdotFirst}). However, in relativistic models, one must distinguish between the frame of the distant observer, who measures luminosity $L_\text{X}$, and the frame of an observer at the stellar surface, who measures apparent luminosity $L_\text{X,0}$. The corresponding accretion rate per column measured in the local frame is denoted by $\dot M$, which is the value used in the {\tt BW22} model to compute the continuum spectrum. The two luminosities are related via the general relativistic transformation 
\citep{Thorne_1977,LewinEtal1993,GuichandutEtal2021,ZhenEtal2023}
\begin{equation}
L_\text{X} = (1 + z_\text{star})^{-2} \, L_\text{X,0} \ ,
\label{eq:LxGR}
\end{equation}
where $z_\text{star}$ is the redshift at the stellar surface given by Equation~(\ref{eq:starZ}).

It is also interesting to examine the relationship between the total accretion rate in the stellar frame in our model, denoted by $\dot M_\text{tot}$, and the observed accretion rate, $\dot M_\text{obs}$. These two quantities are connected by the relativistic expression \citep{Michel1972,ThorneEtal1981,ZhenEtal2023}
\begin{equation}
\dot M_\text{tot} = (1 + z_\text{star}) \,
\dot M_\text{obs} \ .
\label{eq:RelMdotTotFirst}
\end{equation}
Combining Equations~(\ref{eq:mdotFirst}) and (\ref{eq:RelMdotTotFirst}) yields
\begin{equation}
\dot M_\text{tot} = (1 + z_\text{star}) \,
\frac{L_\text{X} R_\text{star}}{G M_\text{star}} \ .
\label{eq:RelMdotTot}
\end{equation}
In accretion-powered X-ray pulsars, the total accretion rate in the frame of the star, $\dot M_\text{tot}$, will be distributed between the two accretion columns, each centered on one of the magnetic poles. In our work here, we make the simplifying assumption that the matter transferred from the binary companion to the neutron star is equally distributed between the two accretion columns, and therefore the accretion rate per column in the frame of the neutron star, $\dot M$, used to compute the {\tt BW22} continuum spectrum in the local {\tt RM88} frame, is related to the total accretion rate, $\dot M_\text{tot}$, via
\begin{equation}
\dot M = \frac{1}{2} \, \dot M_\text{tot} \ ,
\label{eq:MdotPerCol1}
\end{equation}
which can be combined with Equation~(\ref{eq:RelMdotTot}) to obtain
\begin{equation}
\dot M = \frac{1}{2} \, (1 + z_\text{star}) \,
\frac{L_\text{X} R_\text{star}}{G M_\text{star}} \ .
\label{eq:RelMdotTot2}
\end{equation}

In light of the discussion presented above, it interesting to compare the value for the Her X-1 accretion rate adopted by {\tt BW22} with the value used here. In the {\tt BW22} model, the accretion rate, $\dot M_\text{obs}$, is derived directly from the observed X-ray luminosity, $L_\text{X}$, using Equation~(\ref{eq:mdotFirst}). Furthermore, in the {\tt BW22} model, the emission is associated with a single accretion column, whereas in the new model developed here, there are two separate columns. The {\tt BW22} model also includes no general relativistic correction between the two frames, and therefore {\tt BW22} set the accretion rate per column using $\dot M = \dot M_\text{obs}$ for one single accretion column. By contrast, in the new model developed here, the gravitational redshift is included, and therefore the total accretion rate in the frame of the star, $\dot M_\text{tot}$, and the accretion rate per column in the frame of the star, $\dot M$, are computed using Equations~(\ref{eq:RelMdotTot}) and (\ref{eq:MdotPerCol1}), respectively.

Based on Equation~(\ref{eq:mdotFirst}), the accretion rate used by {\tt BW22}, which we refer to as $\dot M_{\tt BW22}$, is given by
\begin{equation}
\dot M_{\tt BW22} = \frac{L_\text{X} R_\text{star}}{G M_\text{star}} \ .
\label{eq:MdotPerColBW}
\end{equation}
The corresponding relation that we use to set the value of the total accretion rate in the stellar frame in our relativistic model, $\dot M_\text{tot}$, is provided by Equation~(\ref{eq:RelMdotTot}). Hence, Equations~(\ref{eq:RelMdotTot}) and (\ref{eq:MdotPerColBW}) provide a quantitative basis for comparing the value of the accretion rate used here with that used by {\tt BW22} in their treatment of Her X-1. We therefore conclude that
\begin{equation}
\dot M_\text{tot} = (1+z_\text{star}) \, \dot M_{\tt BW22} \ ,
\label{eq:MdotTest}
\end{equation}
which provides an interesting test of the model parameters. Based on the observations of Her X-1 considered here, and assuming isotropic emission with a source distance of 6.6\,kpc, \citet{Wolff_etal_2016} obtained for the 0.1-100\,keV X-ray luminosity $L_\text{X} = 4.9 \times 10^{37}\,{\rm erg\,s}^{-1}$. Substituting this luminosity into Equation~(\ref{eq:RelMdotTot}), we find that the total accretion rate in our model is given by $\dot M_\text{tot} = 3.4\times10^{17}\,{\rm g\,s}^{-1}$. The accretion rate assumed by {\tt BW22} is $\dot M_{\tt BW22} = 1.9\times 10^{17}{\rm g\,s}^{-1}$. The values of the two accretion rates are approximately related via the redshift expression given in Equation~(\ref{eq:MdotTest}), with the remaining discrepancy due to the smaller luminosity value assumed by {\tt BW22}, which was $L_\text{X} = 3.73\times 10^{37}\,{\rm erg\,s}^{-1}$.

We can also compare the values for the density $\rho$ obtained in our Her X-1 model with those obtained by {\tt BW22}. The basis for the comparison is provided by the mass conservation relation (Equation~(\ref{eq:MdotDyn})), which yields
\begin{equation}
    \rho(R_0) = n_e(R_0) \, m_p = \frac{\Dot{M}}{\Omega \, R_0^2 \, \abs{\vel(R_0)}}
    \label{densityeq}
\end{equation}
This allows us to compute the density at a given radius $R_0$, as a function of the accretion rate per column, $\dot M$, the solid angle $\Omega$, and bulk accretion speed $\abs{\vel}$. Equation~(\ref{densityeq}) applies in both models. Furthermore, the velocity profile $\vel(R_0)$ is the same in both models, since they both use the values $k_0 = 0$ and $k_\infty = 1$ (see Equation~(\ref{eq:velocity})). We can therefore relate the densities in the two models by writing
\begin{equation}
    \frac{\rho(R_0)}{\rho_{\tt BW22}(R_0)} = \frac{\dot M/\Omega}{\dot M_{\tt BW22}/\Omega_{\tt BW22}}
    = 1.54 \ ,
    \label{eq:MdotComp}
\end{equation}
where $\dot M = (1/2) \dot M_\text{tot}$ is the accretion rate per column in our model, and $\rho_{\tt BW22}$, $\Omega_{\tt BW22}$, and $\dot M_{\tt BW22}$ denote {\tt BW22} parameter values. The numerical result in Equation~(\ref{eq:MdotComp}) follows from the quantities listed in Table~\ref{tbl-3new}. Given the decrease in the column opening angle, $\thetamax$, from $0.315\degree$ in {\tt BW22} to $0.240\degree$ in our model, the radius of the accretion column at the stellar surface, $r_\text{col} = R_\text{star} \, \thetamax$, has decreased slightly in our model. This fact, combined with the decrease in the accretion rate per column, leads to the increase in the computed density by a factor of 1.5. The density increase in turn causes the height of the thermal mound, $z_\text{th}$, and its corresponding optical depth, $\tau_\text{th}$, to each increase slightly, relative to the {\tt BW22} values, as indicated in Table~\ref{tbl-4new}. Finally, we note that the scattering cross sections $\sigma_\perp$, $\sigpar$, and $\sigbar$ have all shifted slightly relative to the {\tt BW22} values, but the variations are not physically significant.

\subsection{Pair-Production Effects}

In highly magnetized, high-temperature plasmas, such as those present in the accretion columns of XRPs, the production of electron-positron pairs can potentially play an important role in both enhancing the productions of bremsstrahlung seed photons, and also in the subsequent Comptonization of those seed photons. Both of these processes tend to cool the plasma, and they also tend to increase the scattering opacity \citep{Suleimanov2022MagneticOpacity}. In the high-temperature, strongly magnetized environment inside the accretion column, pairs may be created as a result of particle-particle collisions, particle-photon collisions, and photon-photon collisions. These processes, and their importance in various astrophysical situations, have been studied in detail by \citet{Svensson1982Pairs} and \citet{Mushtukov2019PairsAccretionColumn}. In this section we will review the fundamental expressions and use them to determine the relevance of pair production in the context of the XRP accretion columns studied here.

\citet{Mushtukov2019PairsAccretionColumn} and \citet{CanutoAndChiu} obtained expressions describing the number density of positrons and electrons in a hot, magnetized plasma in thermal equilibrium, including all production mechanisms. Their results can be used to estimate the number density of pairs in the thermal mound at the base of the XRP accretion column, where full thermodynamic equilibrium is expected to prevail. The expressions obtained can also be used to provide an upper limit on the pair number density throughout the remainder of the column, where the gas is more tenuous and full thermodynamic equilibrium is not established. Using results from \citet{Mushtukov2019PairsAccretionColumn} and \citet{CanutoAndChiu}, we find that the number densities of the electrons and positrons, denoted by $n_-$ and $n_+$, respectively, are given by
\begin{equation}
    n_{\mp} = \mathscr{A}_0 \, \frac{B}{B_\text{crit}} \, \sum_{n=0}^\infty g_n\int_{-\infty}^\infty f_{\mp}(p_z) \, dp_z \ \ \text{cm}^3 \ ,
    \label{nplusminus}
\end{equation}
where $p_z$ is the particle momentum in the $\vec B$-field direction in units of $m_e c$, the particle momentum distribution function is denoted by $f_\mp(p_z)$, the quantity $g_n$ is the spin degeneracy of the $n$th Landau level, $B_\text{crit} \equiv 2\pi m_e^2 c^3/(e h) = 4.415\times10^{13}\,$G denotes the critical magnetic field strength, and the constant $\mathscr{A}_0$ is computed using
\begin{equation}
    \mathscr{A}_0 = \left(\frac{2 \pi m_e c}{h}\right)^3 \frac{1}{4 \pi^2} = 4.415 \times 10^{29} \ .
    \label{nplusminus2}
\end{equation}
The pair momentum distribution is computed using
\begin{equation}
    f_\mp(p_z) = \left\{\exp\left[\frac{E_n(p_z)\mp\mu}{kT}\right]+1\right\}^{-1} \ ,
    \label{fplusminus}
\end{equation}
where $\mu$ is the chemical potential, and
\begin{equation}
    E_n(p_z) = m_e c^2 \left(1 + p_z^2 + 2\,n\,\frac{B}{B_\text{crit}}\right)^{1/2}
\end{equation}
represents the total energy of an electron or positron with dimensionless momentum $p_z$ in the $n$th Landau level.

In order to ensure local charge neutrality, the number densities of the electrons, positrons, and protons must satisfy the constraint
\begin{equation}
    n_- = n_p + n_+ \ ,
    \label{eq:61}
\end{equation}
where $n_p$ denotes the proton number density, and we have assumed that aside from the pairs, the accreting gas is composed of pure, fully-ionized hydrogen. In the context of our model, we can compute the proton number density at radius $R_0$ using the relation (see Equation~(\ref{eq:MdotDyn}))
\begin{equation}
    n_p = \frac{\Dot{M}}{\Omega \, R_0^2 \, m_p \abs{\vel}} \ .
    \label{eq:npDen}
\end{equation}
The closure relation required to compute the value of the chemical potential $\mu$ is obtained by combining Equations~(\ref{eq:61}) and (\ref{eq:npDen}).

The efficiency of pair production in the accretion column is maximized within the thermal mound, and we will therefore evaluate the pair number density at the radius corresponding to the top of the thermal mound, denoted by $R_\text{th}$. In our application, we will adopt physical values appropriate for the Her X-1 model considered here, and reported in Tables~\ref{tbl-3new} and \ref{tbl-4new}. Hence we set the magnetic field strength $B = 4.25 \times 10^{12} \,$G, the accretion rate per column $\dot M_0 = (1/2) \dot M_\text{tot} = 1.70 \times 10^{17}\,{\rm g\,s}^{-1}$, and the electron temperature $T_e = 6.7 \times 10^7\,$K. We calculate the solid angle of the accretion column, $\Omega$, using Equation~(\ref{eq:Omega}), with $\thetamax = 0.240\degree$ and $\thetamin = 0.0\degree$. This yields $\Omega = 5.51 \times 10^{-5}\,$ster. The radius $R_0$ in Equation~(\ref{eq:npDen}) is set equal to the value at the top of the thermal mound, $R_\text{th} = R_\text{star} + z_\text{th}$, with $z_\text{th} = 776\,$cm, and the velocity $\vel$ in Equation~(\ref{eq:npDen}) is set equal to the value at the top of the thermal mound, $\vel_\text{th} = - 0.031\,c$. Based on these parameter values, we can now use Equation~(\ref{eq:npDen}) to show that the proton number density at the top of the thermal mound is given by $n_p = 1.98 \times 10^{24}\,\text{cm}^{-3}$. Combining this with the charge-neutrality constraint introduced in Equation~(\ref{eq:61}), we can determine the value of the chemical potential, obtaining $\mu = 0.902$. Finally, by evaluating Equation~(\ref{nplusminus}), we can show that the positron number density is given by $n_+ = 8.58 \times 10^{-46}\,{\rm cm}^{-3}$, which is obviously completely negligible. Similar results are obtained throughout the accretion column. We further argue that photon-photon and photon-particle pair production are also negligible due to the complete absence of gamma-rays with energy exceeding the pair production threshold $\sim 511\,$keV \citep{BeckerAndKafatos1995,GouldEtalAndSchreder1967}.

\subsection{Model Self-Consistency Checks}

In our model, the sets of angular weight coefficients $\{W_{i}^\text{wall,1}, W_{i}^\text{wall,2}, W_{i}^\text{top,1}, W_{i}^\text{top,2}\}$ are computed via fitting the theoretical pulse-profile function $S(\beta)$ computed using Equation~(\ref{pulseprofilefrombasis}) using observational pulse profile data. The resulting phase-averaged theoretical photon number spectrum, $\langle F_\#(\epsilon) \rangle$, is then computed using the same coefficients by evaluating Equations~(\ref{eq:PhaseDepSpec}) and (\ref{eq:PhaseAveSpec}). The availability of both the pulse-profile function $S(\beta)$ and the phase-averaged photon spectrum $\langle F_\#(\epsilon) \rangle$ presents us with the opportunity to confirm the self-consistency of our results. In particular, one crucial value that can be computed separately using the two theoretical functions is the counts per pulse, which can be obtained by either integrating the phase-averaged spectrum with respect to the photon energy, or by integrating the pulse profile with respect to time. If the theoretical approach is valid, then we would expect reasonable agreement between the values obtained using these two methods.

The pulse profile function $S(\beta)$ gives the count rate in counts per second as a function of the rotation angle $\beta$. Therefore, we can compute the counts per pulse detected in a single spin period using the integral
\begin{equation}
    N_\text{counts}^\text{pulse} = \frac{T_\text{spin}}{2\pi} \int_0^{2\pi}S(\beta) \, d\beta \ ,
    \label{ncountspp}
\end{equation}
where $T_\text{spin}$ is the spin period. We can also calculate the counts per pulse by integrating the phase-averaged photon spectrum, $\langle F_\#(\epsilon) \rangle$ with respect to the photon energy $\epsilon$, taking into account the effect of the detector response function, represented by $\mathscr{D}(\epsilon)$. Hence the counts per pulse can be computed from the phase-averaged spectrum using the integral
\begin{equation}
    N_\text{counts}^\text{spec}= T_{\text{spin}}\int_{\epsilon_\text{min}}^{\epsilon_\text{max}}F_\#(\epsilon)\,\mathscr{D}(\epsilon) \, d\epsilon \ .
    \label{ncountsspec}
\end{equation}
It is also instructive to apply Equation~(\ref{ncountspp}) to the data from \citet{Furst_et_al_2013} plotted  in Figure~\ref{fig:furstetal}, which yields the observational value for the counts per pulse. We obtain
\begin{equation}
    N_\text{counts}^\text{data}=113.59 \  \text{counts} \ .
    \label{ncountsdata}
\end{equation}
Hence this provides another useful cross-check between the theoretical model and the observations. For the Scenario~1 fit that was presented in Section~\ref{sec:IdenticalColumns}, we obtain
\begin{equation}
    N_\text{counts}^\text{pulse}=113.63 \ \text{counts} \ , \ \ \ \ \ \ N_\text{counts}^\text{spec}=115.55 \ \text{counts} \ ,
\end{equation}

and for the Scenario~2 fit presented in Section~\ref{sec:IndependentColumns}, we obtain
\begin{equation}
    N_\text{counts}^\text{pulse}=113.57 \ \text{counts} \ , \ \ \ \ \ \ N_\text{counts}^\text{spec}=115.88 \ \text{counts} \ ,
\end{equation}
where we have used the {\sl NuSTAR} detector response function given by \citet{Ballhausen_2021}.
The relative error between the two methods in each case is $\sim 2\%$, which we deem acceptable given the complexity of the calculation. We therefore conclude that the results obtained for the counts per pulse using the pulse profile and the phase-averaged spectrum agree, which helps to validate the self-consistency of the theoretical framework we have developed. Furthermore, both Scenario 1 and Scenario 2 show excellent agreement with the counts-per-pulse calculated using the data, given in Equation~(\ref{ncountsdata}). Hence, we have verified that the model is self-consistent, and that it also agrees with the pulse profile data.

\subsection{Conclusion}

The new relativistic model developed here provides for the first time a successful simultaneous calculation of the pulse profile and the phase-averaged spectrum of an accretion-powered X-ray pulsar that agrees closely with the observational data. The model utilizes the analytical formalism of {\tt BW22} to calculate the dynamical and radiative structures of two accretion columns powering the emission from an X-ray pulsar. The resulting {\tt BW22} continuum emission components are used to describe the radiation spectra escaping through the walls and tops of the two accretion columns. These radiation components are then linked with a realistic model for the rotational and magnetic geometry of the neutron star, and incorporated into the {\tt RM88} formalism to establish the angular, energy, and time dependence of the radiation field measured by a distant observer, including the effects of the stellar rotation, gravitational lensing, redshifting, and time dilation. The fits are carried out across a large region of the parameter space of the angles $(\Psi,\varphi_1,\varphi_2,\Delta\beta_1,\Delta\beta_2)$, and the quality of the fits is ranked using the statistical analysis procedure discussed in Section~\ref{sec:parameterspacesearch}. The best fit models obtained under Scenario~1 (identical column beaming patterns) and Scenario~2 (independent column beaming patterns) are presented in Sections~\ref{sec:IdenticalColumns} and \ref{sec:IndependentColumns}, respectively.

The new model allows the determination of the primary physical parameters such as the electron temperature, $T_e$, the magnetic field strength, $B$, and the accretion rate, $\dot M$, as well as the inclination of the neutron star spin axis, $\Psi$, and the rotational latitudes of the two accretion columns, $\varphi_1$, and $\varphi_2$, and the associated rotational phase shifts, $\Delta\beta_1$ and $\Delta\beta_2$, by performing fits to the pulse profile data for a given source. In the future, we plan to apply the model to additional sources, and we will also explore the possibility of porting the software into a high-level programming language for implementation in the {\sl XSPEC} environment, which would facilitate real-time data analysis. We also anticipate the development of a multidimensional version of our model, as well as the relaxation of the conical geometry assumed here, which deviates somewhat from the expected dipolar geometry. The conical geometry is likely to distort the beaming patterns relative to those obtained using dipole-shaped accretion columns, and therefore it would be very interesting to investigate this further using a dipole model. Here, we have focused on an initial application to Her X-1, but the model will be further tested in future work in which we plan to analyze the data for a wide variety of X-ray pulsars, including low luminosity sources such as X Per and high luminosity sources such as Cen X-3.

\begin{acknowledgments}

We would like to thank the Office of Research Computing at George Mason University (URL: https://orc.gmu.edu) for the partial support of computing resources, the Office of the Provost and the Department of Physics and Astronomy at George Mason University for the funding. We are also grateful to the anonymous referee who provided several useful comments that led to significant improvements in the manuscript. The source code used for the numerical simulations and data reduction in this paper is available upon reasonable request emailed to the authors.

\end{acknowledgments}

\begin{appendix}

\section{Calculation of Flux in Local Detector Frame for Column Wall Emission}
\label{sec:appendixA}

In order to calculate the energy flux spectrum for the wall emission in the frame of a local detector, $F_{\epsilon,\text{star}}^\text{wall}$, we must make a transformation from the {\tt RM88} frame to the detector frame, as discussed in Section \ref{sec:columnwallintensity}. The two frames are related by a 90\degree rotation, as indicated in Figure~\ref{fig:detector}. For the column wall the emitted flux in the local detector frame is given by
\begin{equation}
    F_{\epsilon,\text{star}}^\text{wall} = \iint_{4\pi} I_{\epsilon,\text{star}}^\text{wall} \,\cos\tilde\theta \, \sin\tilde\theta \, d\tilde\theta \, d\tilde\phi
    \label{appendixflux1}
\end{equation}
where the $(\tilde\theta,\tilde\phi)$ represents the photon propagation direction as measured in the spherical polar system of the local detector frame. In Equation~(\ref{appendixflux1}), the factor $\cos\tilde\theta$ represents the dot product between the propagation direction and the normal to the detector plane, which is the polar axis for the $\tilde\theta$ angle. The coordinate transformation from the local detector frame to the {\tt RM88} coordinate system is accomplished by rotating the coordinate system by 90\degree. This leads to the following relationships between coordinates
\begin{align}
    \cos\theta_0&= \sin\tilde\theta \cos\tilde\phi
    \ ,\label{defeq4} \\
    \sin\theta_0 \sin\phi_0&= \sin\tilde\theta \sin\tilde\phi \ ,\label{defeq5}\\
    \sin\theta_0\cos\phi_0&= \cos\tilde\theta \ .
    \label{defeq6}
\end{align}
To transform the integral in Equation~(\ref{appendixflux1}) into the {\tt RM88} coordinates, we must compute the Jacobian for this transformation, which is given by
\begin{equation}
    J(\theta_0,\phi_0) = \abs{\pdv{\tilde\theta}{\theta_0}\pdv{\tilde\phi}{\phi_0}-\pdv{\tilde\theta}{\phi_0}\pdv{\tilde\phi}{\theta_0}} \ .
    \label{jacobianappendix}
\end{equation}
In terms of the Jacobian, the integral in Equation~(\ref{appendixflux1}) can now be written as
\begin{equation}
    F_{\epsilon,\text{star}}^\text{wall} = \iint_{4\pi} I_{\epsilon,\text{star}}^\text{wall} \,\cos\tilde\theta \, \sin\tilde\theta \, J(\theta_0,\phi_0) \, d\theta_0 \, d\phi_0
    \label{appendixflux2}
\end{equation}

To calculate the required derivatives, we must first derive explicit expressions for the angles $(\tilde\theta,\tilde\phi)$ as functions of the angles $(\theta_0,\phi_0)$, obtaining
\begin{equation}
    \tilde\theta = \cos^{-1}\left(\sin\theta_0 \, \cos\phi_0\right) \ ,\qquad\qquad \tilde\phi = \tan^{-1}\left(\tan\theta_0 \, \sin\phi_0\right) \ .
\end{equation}
Carrying out the derivatives in Equation~(\ref{jacobianappendix}) and simplifying, we arrive at
\begin{equation}
    J(\theta_0,\phi_0) = \frac{\sin\theta_0}{\sin\tilde\theta} \ .
    \label{jacobiananswer}
\end{equation}
Combining Equations~(\ref{appendixflux2}) and (\ref{jacobiananswer}) yields the final result
\begin{equation}
    F_{\epsilon,\text{star}}^\text{wall} = \iint_{4\pi}I_{\epsilon,\text{star}}^\text{wall}\, (\sin\theta_0 \, \cos\phi_0)\sin\theta_0 \, d\theta_0 \, d\phi_0 \ ,
    \label{eq:DetFlux}
\end{equation}
which represents the energy flux spectrum measured by a local detector oriented tangent to the column wall, with the integration carried out in the local {\tt RM88} frame.

\section{Average Intensity Distribution in Co-moving Frame}
\label{sec:appendixB}

We have obtained results for the  angular intensity weight coefficients $\{W_{i}^\text{wall,1}$, $W_{i}^\text{wall,2}$, $W_{i}^\text{top,1}$, $W_{i}^\text{top,2}\}$ by fitting the observational pulse profile data using the theoretical pulse-profile function $S(\beta)$ computed using Equation~(\ref{pulseprofilefrombasis}). We remind the reader that the four sets of weight coefficients determine the angular variation of the column-wall and column-top intensity distributions measured in the local {\tt RM88} frame via Equations~(\ref{intensitysumwallNEW2}) and (\ref{totalintensitydeftopNEW}), respectively. Since the local {\tt RM88} frame is stationary with respect to the neutron star, it is also of interest to compute the energy and angular distribution of the radiation intensity measured by a co-moving observer who is advecting downward toward the stellar surface with the local flow velocity. Since the flow has a variable velocity, the aberration between the frame of the co-moving observer and the local {\tt RM88} frame, which is stationary with respect to the star, is variable. Hence we will seek to compute the time-average of the intensity distribution measured in the co-moving frame. We carry out the general derivation here, and present the quantitative results in Section~\ref{sec:results}.

\subsection{Co-moving Average Intensity for Column Wall Emission}
\label{sec:deaberrationwall}

In this section we focus on establishing the quantitative transformation of the radiation field specified in the local {\tt RM88} frame into the frame co-moving with the bulk plasma flow at the same physical location. These two frames are related via a Lorentz transformation. Once the transformation is established, we can perform the time average of the intensity field measured in the co-moving frame.

Consider an arbitrary intensity distribution $I_\epsilon(R_0,\mu_0,\phi_0,\epsilon_0)$, defined in the local {\tt RM88} frame at radius $R_0$. We seek to compute the corresponding co-moving frame intensity distribution $I_\epsilon'(R_0,\mu_0',\phi_0',\epsilon_0')$. Based on the Lorentz invariance of the photon occupation number, we find that the relation between the intensity distributions in the two frames is given by \citep{RybickiandLightman1979}
\begin{equation}
\frac{I_\epsilon'(R_0,\mu_0',\phi_0',\epsilon_0')}{\epsilon_0^{'3}} = \frac{I_\epsilon(R_0,\mu_0,\phi_0,\epsilon_0)}{\epsilon_0^3} \ ,
\label{eq:Lorentz}
\end{equation}
where the photon energy in the local {\tt RM88} frame, $\epsilon_0$, is related to the corresponding energy in the co-moving frame, $\epsilon_0'$, via \citep{RybickiandLightman1979}
\begin{equation}
    \epsilon_0' = \epsilon_0 \, \gamma(R_0) \, (1+\mu_{0} \, \abs{\vel(R_0)/c}) \ ,
    \label{eq:EnLor}
\end{equation}
with $\gamma$ denoting the bulk Lorentz factor for the advecting plasma, defined by
\begin{equation}
    \gamma(R_0) \equiv \frac{1}{\sqrt{1-\vel^2(R_0)/c^2}} \ ,
    \label{LorentzFactor}
\end{equation}
and $\vel(R_0) < 0$ representing the flow velocity at radius $R_0$.

In order to complete the transformation of the intensity between the two frames, we must also consider the aberration of the emission direction, represented by $(\mu_0,\phi_0)$ in the {\tt RM88} frame and by $(\mu_0',\phi_0')$ in the co-moving frame. The polar axes of the two frames are both aligned with the local bulk flow velocity, and therefore the azimuthal angles are equal, so that $\phi_0' = \phi_0$. The transformation of the polar angle is given by the aberration formula, which in our application can be written as \citep{RybickiandLightman1979}
\begin{equation}
    \mu_0' = \frac{\mu_0+\abs{\vel(R_0)/c}}{1+\mu_0 \, \abs{\vel(R_0)/c}} \ .
    \label{aberration1}
\end{equation}
We note that this formula deviates from the standard form because in our application the polar angle $\theta_0 = \cos^{-1}\mu_0$ is measured with respect to the outward radial vector, which is in the opposite direction compared with the flow velocity.

The radiation intensity distribution emitted from the column wall in the {\tt RM88} frame in our model is represented using an expansion over ``laser-like'' emission directions, given by Equation~(\ref{intensitysumwallNEW2}). The propagation of radiation is a linear process, and therefore the radiation generated in each {\tt RM88} direction $(\mu_{0i},\phi_{0j})$ is independently transformed into the co-moving frame. It is therefore sufficient to focus on a single unitary emission vector in the {\tt RM88} frame. We begin by restating the fundamental sum for the column wall intensity distribution measured in the {\tt RM88} frame, given by (see Equation~(\ref{intensitysumwallNEW}))
\begin{equation}
    I_{\epsilon,\text{star}}^\text{wall}(R_0,\mu_0,\phi_0,\epsilon_0) = \sum_{i=1}^{N_\text{wall}} \sum_{j=1}^{M_\text{wall}} \frac{W_{i}^\text{wall} \, \epsilon_0\Dot{N}^\text{tot}_\epsilon(R_0,\epsilon_0)}{2\pi R_{0}\sin\thetamax} \, \delta(\mu_{0}-\mu_{0i})\delta(\phi_{0}-\phi_{0j}) \ .
    \label{eq:B1new}
\end{equation}
Next, we recognize that each term in this expansion can be written in the form
\begin{equation}
    I_\epsilon(R_0,\mu_0,\phi_0,\epsilon_0) = U_{i,\text{wall}}(R_0,\epsilon_0) \, \delta(\mu_0-\mu_{0i}) \, \delta(\phi_{0}-\phi_{0j}) \ ,
    \label{RM88intensitydef}
\end{equation}
where $U_{i,\text{wall}}(R_0,\epsilon_0)$ is the {\tt RM88} intensity amplitude function for the column wall emission, defined by
\begin{equation}
    U_{i,\text{wall}}(R_0,\epsilon_0) \equiv \frac{W_i^\text{wall}\epsilon_0\Dot{N}_\epsilon^\text{tot}(R_0,\epsilon_0)}{2\pi R_0\sin\thetamax} \ .
    \label{unitaryemissionamplitude}
\end{equation}
Equation~(\ref{RM88intensitydef}) expresses the intensity in the {\tt RM88} frame, and the corresponding intensity in the frame of the co-moving observer is given by
\begin{equation}
    I_\epsilon'(R_0,\mu_0',\phi_0',\epsilon_0') = U_{i,\text{wall}}'(R_0,\epsilon_0') \, \delta(\mu_0'-\mu_{0i}') \, \delta(\phi_{0}'-\phi_{0j}') \ ,
    \label{primedintensitydef}
\end{equation}
where $\phi_{0j}' = \phi_{0j}$, and $\mu_{0i}'$ is related to $\mu_{0i}$ via (see Equation~(\ref{aberration1}))
\begin{equation}
    \mu_{0i}' = \frac{\mu_{0i}+\abs{\vel(R_0)/c}}{1+\mu_{0i} \, \abs{\vel(R_0)/c}} \ .
    \label{aberration1b}
\end{equation}

According to Equation~(\ref{eq:Lorentz}), the co-moving amplitude, $U_{i,\text{wall}}'$, is related to $U_{i,\text{wall}}$ via
\begin{equation}
    U_{i,\text{wall}}'(R_0,\epsilon_0') = U_{i,\text{wall}}(R_0,\epsilon_0) \, \left(\frac{\epsilon_0'}{\epsilon_0}\right)^3 \, 
    \frac{d\mu_0'}{d\mu_0} \, \frac{d\phi_0'}{d\phi_0} \ ,
    \label{eq:Uprime}
\end{equation}
where the derivatives $d\mu_0'/d\mu_0$ and $d\phi_0'/d\phi_0$ appear due to the transformations between the $\delta$-functions in Equations~(\ref{RM88intensitydef}) and (\ref{primedintensitydef}).
The azimuthal angles $\phi_0'$ and $\phi_0$ are equal and therefore $d\phi_0'/d\phi_0 = 1$. The required derivative of the polar coordinate is obtained by differentiating Equation~(\ref{aberration1}) with respect to $\mu_0$, which yields
\begin{equation}
    \frac{d\mu_0'}{d\mu_0} = \frac{1}{\gamma^2(R_0) \, (1+\mu_0 \, \abs{\vel(R_0)/c})^2} \ ,
    \label{muDeriv2}
\end{equation}
where $\gamma(R_0)$ denotes the bulk Lorentz factor for the advecting plasma at radius $R_0$, defined by Equation~(\ref{LorentzFactor}).

Combining Equations~(\ref{eq:EnLor}), (\ref{eq:Uprime}), and (\ref{muDeriv2}), we obtain upon simplification
\begin{equation}
    U_{i,\text{wall}}'(R_0,\epsilon_0') = \gamma(R_0) \, (1+\mu_{0} \, \abs{\vel(R_0)/c}) \, U_{i,\text{wall}}(R_0,\epsilon_0) \ ,
    \label{eq:Uprime2}
\end{equation}
where $\epsilon_0$ is computed from the value of the co-moving frame energy $\epsilon_0'$ using Equation~(\ref{eq:EnLor}), and $\mu_0$ is a function of $\mu_0'$ and $R_0$ via the relation
\begin{equation}
    \mu_0 = \frac{\mu_0'-\abs{\vel(R_0)/c}}{1-\mu_0' \, \abs{\vel(R_0)/c}} \ ,
    \label{aberration2}
\end{equation}
obtained by inverting Equation~(\ref{aberration1}). We can now use Equation~(\ref{eq:Uprime2}) to substitute for $U_{i,\text{wall}}'$ in Equation~(\ref{primedintensitydef}) to obtain
\begin{equation}
    I_\epsilon'(R_0,\mu_0',\phi_0',\epsilon_0') = \gamma(R_0) \, (1+\mu_{0} \, \abs{\vel(R_0)/c}) \, U_{i,\text{wall}}(R_0,\epsilon_0) \, \delta(\mu_0'-\mu_{0i}') \, \delta(\phi_{0}'-\phi_{0j}') \ ,
    \label{primedintensitycolumnwall}
\end{equation}
which describes the intensity distribution measured by a co-moving observer at radius $R_0$ due to unitary emission with amplitude $U_{i,\text{wall}}(R_0,\epsilon_0)$ in the $(\mu_{0i},\phi_{0j})$ direction in the local {\tt RM88} frame, where $\mu_{0i}'$ is computed from $\mu_{0i}$ using Equation~(\ref{aberration1b}), and $\phi_{0j}' = \phi_{0j}$.

The velocity of the accretion flow onto the magnetic pole of the neutron star varies as the material first accelerates towards the neutron star, and then eventually decelerates to come to rest at the stellar surface (see Figure~\ref{fig:BW22HerX1_spect}b). Hence there is a different bulk flow velocity, $\vel(R_0)$, at each value of $R_0$. The co-moving frame intensity computed using Equation~(\ref{primedintensitycolumnwall}) therefore gives distinct results at each radius. In order to obtain a global representation of the co-moving frame intensity distribution, we will therefore perform a time average of Equation~(\ref{primedintensitycolumnwall}) for an observer starting at the column top, located at radius $R_\text{top}$, and advecting downward at the local flow velocity, with the trip terminating at radius $R_\text{star}$, the stellar surface.

The time-weighted average of the co-moving frame intensity distribution, denoted by $\mathscr{I}_\epsilon(\mu_0',\phi_0',\epsilon_0')$, is defined by the integral
\begin{equation}
    \mathscr{I}_\epsilon(\mu_0',\phi_0',\epsilon_0') \equiv \frac{1}{t_\text{tot}'} \int_{0}^{t'_\text{tot}} I_\epsilon'(R_0,\mu_0',\phi_0',\epsilon_0') \, dt' \ ,
    \label{timeweightedintegral}
\end{equation}
where $t'_\text{tot}$ is the total co-moving transit time for the journey, and the radius $R_0$ is a function of $t'$ due to the accretion onto the neutron star. Based on the time dilation relation
\begin{equation}
dt' = \frac{dt}{\gamma(R_0)} = \frac{dR_0}{\gamma(R_0) \, \abs{\vel(R_0)}} \ ,
\label{eq:TimeDilat}
\end{equation}
we can compute the co-moving transit time using the integral
\begin{equation}
    t_\text{tot}' = \int_0^{t_\text{tot}'}dt' = \int_{R_\text{star}}^{R_\text{top}} \frac{dR_0}{\gamma(R_0) \, \abs{\vel(R_0)}} \ .
    \label{eq:TotTime}
\end{equation}
We can now combine Equations~(\ref{primedintensitycolumnwall}), (\ref{timeweightedintegral}), and (\ref{eq:TimeDilat}) to obtain
\begin{align}
    \mathscr{I}_\epsilon(\mu_0',\phi_0',\epsilon_0') &= \frac{1}{t_\text{tot}'} \int_{R_\text{star}}^{R_\text{top}} \, \gamma(R_0) \, (1+\mu_{0} \, \abs{\vel(R_0)/c}) \, U_{i,\text{wall}}(R_0,\epsilon_0)\nonumber\\
    &\times \delta(\mu_0'-\mu_{0i}') \, \delta(\phi_{0}'-\phi_{0j}') \, \frac{dR_0}{\gamma(R_0) \, \abs{\vel(R_0)}} \ ,
    \label{timeweightedintegral2}
\end{align}
where the co-moving frame emission direction $(\mu_{0i}',\phi_{0j}')$ is a function of $R_0$ due to the variation of the coordinate transformation with the changing flow velocity.

At this point, we recall that the variable $\mu_0$ appearing in Equation~(\ref{timeweightedintegral2}) is a function of $\mu_0'$ and $R_0$ via Equation~(\ref{aberration2}), from which we can derive the useful identity
\begin{equation}
1+\mu_{0} \, \abs{\vel(R_0)/c} = \frac{1}{\gamma^2(R_0) \, (1-\mu_0' \, \abs{\vel(R_0)/c})} \ .
    \label{eqB19}
\end{equation}
This relation can be combined with Equation~(\ref{timeweightedintegral2}) to obtain
\begin{align}
    \mathscr{I}_\epsilon(\mu_0',\phi_0',\epsilon_0') &= \frac{1}{t_\text{tot}'} \int_{R_\text{star}}^{R_\text{top}} \, \frac{1}{\gamma(R_0) \, (1-\mu_0' \, \abs{\vel(R_0)/c})} \, U_{i,\text{wall}}(R_0,\epsilon_0)\nonumber\\
    &\times \delta(\mu_0'-\mu_{0i}') \, \delta(\phi_{0}'-\phi_{0j}') \, \frac{dR_0}{\gamma(R_0) \, \abs{\vel(R_0)}} \ ,
    \label{eqB20}
\end{align}
Our next task is to evaluate the integral in Equation~(\ref{eqB20}) analytically. We can use  Equation~(\ref{aberration1b}) to substitute for $\mu_{0i}'$ in the $\delta$-function in Equation~(\ref{eqB20}), which yields
\begin{align}
    \mathscr{I}_\epsilon(\mu_0',\phi_0',\epsilon_0') &= \frac{1}{t_\text{tot}'} \, \delta(\phi_{0}'-\phi_{0j}') \int_{R_\text{star}}^{R_\text{top}} \frac{1}{\gamma(R_0) \, (1-\mu_0' \, \abs{\vel(R_0)/c}} \, U_{i,\text{wall}}(R_0,\epsilon_0)\nonumber\\
    &\times \delta\left(\mu_0'-\frac{\mu_{0i}+\abs{\vel(R_0)/c}}{1+\mu_{0i} \, \abs{\vel(R_0)/c}}\right) \frac{dR_0}{\gamma(R_0) \, \abs{\vel(R_0)}} \ ,
    \label{timeweightedintegral3}
\end{align}
where the $\phi_0'$-dependent $\delta$-function has been pulled out of the integral since it does not depend on $R_0$.

In order to carry out the integration in Equation~(\ref{timeweightedintegral3}), it is necessary to transform the argument of the delta function in the integrand so that it depends directly on $R_0$. Making this transformation gives the result
\begin{align}
    \mathscr{I}_\epsilon(\mu_0',\phi_0',\epsilon_0') &= \frac{1}{t_\text{tot}'} \, \delta(\phi_{0}'-\phi_{0j}') \int_{R_\text{star}}^{R_\text{top}} \frac{1}{\gamma(R_0) \, (1-\mu_0' \, \abs{\vel(R_0)/c}} \, U_{i,\text{wall}}(R_0,\epsilon_0)\nonumber\\
    &\times \delta(R_0 - R_0^*) \, \frac{(1+\mu_{0i}\abs{\vel(R_0)/c})^2}{(1-\mu_{0i}^2)\abs{\vel'(R_0)/c}} \, \frac{dR_0}{\gamma(R_0) \, \abs{\vel(R_0)}} \ ,
    \label{timeweightedintegral5}
\end{align}
where $\vel'(R_0)$ denotes the radial derivative of the velocity, and the parameter $R_0^*$ is the root of the equation
\begin{equation}
    \mu_0' = \frac{\mu_{0i}+\abs{\vel(R_0^*)/c}}{1+\mu_{0i} \, \abs{\vel(R_0^*)/c}} \ ,
    \label{eq:Rroot}
\end{equation}
or, equivalently,
\begin{equation}
    \frac{\abs{\vel(R_0^*)}}{c} = \frac{\mu_{0}'-\mu_{0i}}{1-\mu_{0i} \, \mu_0'} \ .
    \label{eq:Rroot2}
\end{equation}
We can use Equation~(\ref{eq:velocity}) to substitute for the function $\abs{\vel(R_0^*)/c}$ on the left-hand side of Equation~(\ref{eq:Rroot2}). This yields an explicit form for the root equation, given by
\begin{equation}
    \left(\frac{2R_g}{R_\text{star}}\right)^{1/2} \left[k_\infty^2
    \left(\frac{R_0^*}{R_\text{star}}\right)^{-1} + (k_0^2-k_\infty^2)\left(\frac{R_0^*}{R_\text{star}}\right)^{-4}\right]^{1/2} = \frac{\mu_{0}'-\mu_{0i}}{1-\mu_{0i} \, \mu_0'} \ .
    \label{eq:Rroot3}
\end{equation}
This equation can be solved numerically to determine the value of $R_0^*$ for given values of $\mu_{0i}$ and $\mu_0'$. Physically, $R_0^*$ represents the radius at which radiation with propagation direction $\mu_{0i}$ in the {\tt RM88} frame has propagation direction $\mu_0'$ as viewed in the co-moving frame. In general, Equation~(\ref{eq:Rroot3}) may have zero, one, or two acceptable physical roots for $R_0^*$ within the radial domain of the accretion column, which is $R_\text{star} \le R_0^* \le R_\text{top}$.

We can now evaluate the integral in Equation~(\ref{timeweightedintegral5}) exactly, obtaining
\begin{align}
    \mathscr{I}_\epsilon(\mu_0',\phi_0',\epsilon_0') &= \frac{1}{t_\text{tot}'} \, \delta(\phi_{0}'-\phi_{0j}') \, (1+\mu_{0i} \, \abs{\vel(R_0^*)/c}) \, U_{i,\text{wall}}(R_0^*,\epsilon_0)\nonumber\\
    &\times \frac{(1+\mu_{0i}\abs{\vel(R_0^*)/c})^2}{(1-\mu_{0i}^2)\abs{\vel'(R_0^*)/c}} \, \frac{1}{\abs{\vel(R_0^*)}} \ ,
    \label{timeweightedintegral6}
\end{align}
where the value of $\epsilon_0$ is computed using
\begin{equation}
    \epsilon_0 = \epsilon_0' \, \gamma(R_0^*) \, (1-\mu_{0}' \, \abs{\vel(R_0^*)/c}) \ .
    \label{eq:EnLor2}
\end{equation}
We remind the reader that the value of $R_0^*$ in Equation~(\ref{timeweightedintegral6}) is a function of $\mu_0'$ and $\mu_{0i}$ via Equation~(\ref{eq:Rroot3}). Using Equation~(\ref{unitaryemissionamplitude}) to substitute for $U_{i,\text{wall}}(R_0^*,\epsilon_0)$ in Equation~(\ref{timeweightedintegral6}) and simplifying, we obtain
\begin{equation}
    \mathscr{I}_\epsilon(\mu_0',\phi_0',\epsilon_0') = \frac{1}{t_\text{tot}'} \, \delta(\phi_{0}'-\phi_{0j}') \, \frac{W_i^{\rm wall}\epsilon_0\Dot{N}_\epsilon^{\rm tot}(R_0^*,\epsilon_0)}{2\pi R_0^*\sin\thetamax} \, \frac{(1+\mu_{0i}\abs{\vel(R_0^*)/c})^3}{(1-\mu_{0i}^2)\abs{\vel'(R_0^*)/c}} \, \frac{1}{\abs{\vel(R_0^*)}} \ ,
    \label{timeweightedintegral7}
\end{equation}
which describes the time-average of the intensity distribution measured in the frame of a co-moving observer resulting from radiation emitted in the $(\mu_{0i},\phi_{0j})$ direction in the {\tt RM88} frame.

Our next step is to generalize Equation~(\ref{timeweightedintegral7}) so that we can transform the complete expansion for the {\tt RM88}-frame intensity given by Equation~(\ref{eq:B1new}) into the frame of a co-moving observer. Because the expansion in Equation~(\ref{eq:B1new}) is a linear sum comprising individual terms that each resemble Equation~(\ref{RM88intensitydef}), it follows that we can use Equation~(\ref{timeweightedintegral7}) to treat each term in Equation~(\ref{eq:B1new}) separately. This immediately yields an expansion for the total time-averaged, co-moving wall intensity, denoted by $\mathscr{I}^\text{tot}_{\epsilon,\text{wall}}(\mu_0',\phi_0',\epsilon_0')$. The result obtained is
\begin{align}
    \mathscr{I}^{\rm tot}_{\epsilon,\text{wall}}(\mu_0',\phi_0',\epsilon_0') &= \sum_{i=1}^{N_{\rm wall}} \sum_{j=1}^{M_{\rm wall}} \frac{1}{t_\text{tot}'} \, \delta(\phi_{0}'-\phi_{0j}') \, \frac{W_i^{\rm wall}\epsilon_0\Dot{N}_\epsilon^{\rm tot}(R_0^*,\epsilon_0)}{2\pi R_0^*\sin\thetamax} \nonumber\\
    &\times \frac{(1+\mu_{0i}\abs{\vel(R_0^*)/c})^3}{(1-\mu_{0i}^2)\abs{\vel'(R_0^*)/c}} \, \frac{1}{\abs{\vel(R_0^*)}} \ ,
    \label{eq:B25newA}
\end{align}
where $R_0^*$ depends on $\mu_0'$ and $\mu_{0i}$ via Equation~(\ref{eq:Rroot3}), and the value of $\epsilon_0$ is computed using Equation~(\ref{eq:EnLor2}).
Equation~(\ref{eq:B25newA}) can also be reorganized to obtain
\begin{align}
    \mathscr{I}^{\rm tot}_{\epsilon,\text{wall}}(\mu_0',\phi_0',\epsilon_0') &= \sum_{i=1}^{N_{\rm wall}} \frac{1}{t_\text{tot}'} \, \frac{W_i^{\rm wall}\epsilon_0\Dot{N}_\epsilon^{\rm tot}(R_0^*,\epsilon_0)}{2\pi R_0^*\sin\thetamax}\nonumber\\
    &\times \frac{(1+\mu_{0i}\abs{\vel(R_0^*)/c})^3}{(1-\mu_{0i}^2)\abs{\vel'(R_0^*)/c}} \, \frac{1}{\abs{\vel(R_0^*)}} \sum_{j=1}^{M_{\rm wall}} \, \delta(\phi_{0}'-\phi_{0j}') \ .
    \label{eq:B25new}
\end{align}
We will use Equation~(\ref{eq:B25new}) when applying our model to the analysis of the Her X-1 data in Section~\ref{sec:results}.

\subsection{Co-moving Intensity for Column Top Emission}
\label{sec:deaberrationtop}

The case of the column-top emission is more straightforward to treat than the column-wall emission because only one value of the radius is involved, since the radiation escapes from the upper surface of the accretion column, at radius $R_0 = R_\text{top}$. Hence in this situation, we can compute the intensity in the co-moving frame by performing a Lorentz transformation from the stationary {\tt RM88} frame into the co-moving frame at a single value of the radius, and no time average is required. The treatment of the transformation of the column-top emission is consequently much simpler than the transformation of the column-wall emission considered in Section~\ref{sec:deaberrationwall}.

We begin with the expansion for column-top intensity in the {\tt RM88} frame, given by (see Equation~(\ref{totalintensitydeftopNEW2}))
\begin{equation}
    I_{\epsilon,\text{star}}^\text{top}(\mu_0,\phi_0,\epsilon_0) = \sum_{i=1}^{N_\text{top}}\sum_{j=1}^{M_\text{top}}\frac{W^\text{top}_{i}\epsilon_0\,\Dot{\mathcal{N}}^\text{tot}_\epsilon(\epsilon_0)}{\Omega R_\text{top}^2} \, \delta(\mu_0-\mu_{0i})\delta(\phi_0-\phi_{0j}) \ .
    \label{totalintensitydeftop2}
\end{equation}
In similar fashion to the case of the column-wall emission, described by Equation~(\ref{eq:Lorentz}), the Lorentz transformation of the column-top emission implies that
\begin{equation}
    \frac{I_\epsilon'(\mu_0',\phi_0',\epsilon_0')}{\epsilon_0^{'3}} = \frac{I_\epsilon(\mu_0,\phi_0,\epsilon_0)}{\epsilon_0^3} \ ,
    \label{eq:B28new}
\end{equation}
where the angular variables $\mu_0$ and $\mu_0'$ are related by the aberration formula
\begin{equation}
    \mu_0' = \frac{\mu_0+\abs{\vel(R_\text{top})/c}}{1+\mu_0 \, \abs{\vel(R_\text{top})/c}} \ ,
    \label{aberration1again}
\end{equation}
and the photon energy in the co-moving frame is computed using
\begin{equation}
    \epsilon_0' = \epsilon_0 \, \gamma(R_\text{top}) \, (1+\mu_{0} \, \abs{\vel(R_\text{top})/c}) \ ,
    \label{eq:B29new}
\end{equation}
with the column-top Lorentz factor, $\gamma(R_\text{top})$, defined by
\begin{equation}
    \gamma(R_\text{top}) \equiv \frac{1}{\sqrt{1-\vel^2(R_\text{top})/c^2}} \ .
    \label{LorentzFactorRop}
\end{equation}

By analogy with Equation~(\ref{eq:B1new}), we can write each term in Equation~(\ref{totalintensitydeftop2}) in the form
\begin{equation}
    I_\epsilon(\mu_0,\phi_0,\epsilon_0) = U_{i,\text{top}}(\epsilon_0) \, \delta(\mu_0-\mu_{0i}) \, \delta(\phi_{0}-\phi_{0j}) \ ,
    \label{eq:B31new}
\end{equation}
where $U_{i,\text{top}}(\epsilon_0)$ represents the {\tt RM88} intensity amplitude function for the column-top emission, defined by
\begin{equation}
    U_{i,\text{top}}(\epsilon_0) \equiv \frac{W^\text{top}_{i}\epsilon_0\,\Dot{\mathcal{N}}^\text{tot}_\epsilon(\epsilon_0)}{\Omega R_\text{top}^2} \ .
    \label{eq:B32new}
\end{equation}
We can transform the {\tt RM88} frame intensity given by Equation~(\ref{eq:B31new}) by making use of Equation~(\ref{eq:B28new}), which yields
\begin{equation}
    I_\epsilon'(\mu_0',\phi_0',\epsilon_0') = U_{i,\text{top}}'(\epsilon_0') \, \delta(\mu_0'-\mu_{0i}') \, \delta(\phi_{0}'-\phi_{0j}') \ ,
    \label{eq:B33new}
\end{equation}
where $\mu_{0i}'$ is related to $\mu_{0i}$ via (see Equation~(\ref{aberration1again}))
\begin{equation}
    \mu_{0i}' = \frac{\mu_{0i}+\abs{\vel(R_\text{top})/c}}{1+\mu_{0i} \, \abs{\vel(R_\text{top})/c}} \ ,
    \label{eq:B34new}
\end{equation}
and the co-moving amplitude, $U_{i,\text{top}}'(\epsilon_0')$, is given by
\begin{equation}
    U_{i,\text{top}}'(\epsilon_0') = U_{i,\text{top}}(\epsilon_0) \, \left(\frac{\epsilon_0'}{\epsilon_0}\right)^3 \, 
    \frac{d\mu_0'}{d\mu_0} \, \frac{d\phi_0'}{d\phi_0} \ .
    \label{eq:B35new}
\end{equation}
We note that $\phi_{0j}' = \phi_{0j}$, and therefore $d\phi_0'/d\phi_0 = 1$. Furthermore, the result for $d\mu_0'/d\mu_0$ obtained by differentiating Equation~(\ref{aberration1again}) is
\begin{equation}
    \frac{d\mu_0'}{d\mu_0} = \frac{1}{\gamma^2(R_\text{top}) \, (1+\mu_0 \, \abs{\vel(R_\text{top})/c})^2} \ ,
    \label{eq:B36new}
\end{equation}
where $\gamma(R_\text{top})$ denotes the bulk Lorentz factor at the column top, computed using Equation~(\ref{LorentzFactorRop}).

We can now combine Equations~(\ref{eq:B29new}), (\ref{eq:B33new}), (\ref{eq:B35new}), and (\ref{eq:B36new}) to obtain
\begin{equation}
    U_{i,\text{top}}'(\epsilon_0') = \gamma(R_\text{top}) \, (1+\mu_{0} \, \abs{\vel(R_\text{top})/c}) \, U_{i,\text{top}}(\epsilon_0) \ .
    \label{eq:Uprime2TEMP}
\end{equation}
Next we substitute for $\mu_0$ in Equation~(\ref{eq:Uprime2TEMP}) using the relation
\begin{equation}
    \mu_0 = \frac{\mu_0'-\abs{\vel(R_\text{top})/c}}{1-\mu_0' \, \abs{\vel(R_\text{top})/c}} \ ,
    \label{eq:B39newNEW}
\end{equation}
which yields
\begin{equation}
    U_{i,\text{top}}'(\epsilon_0') = \frac{1}{\gamma(R_\text{top})} \, \frac{1}{1-\mu_{0}' \, \abs{\vel(R_\text{top})/c}} \, U_{i,\text{top}}(\epsilon_0) \ .
    \label{eq:Uprime4TEMP}
\end{equation}
Making use of Equation~(\ref{eq:Uprime4TEMP}), we can rewrite Equation~(\ref{eq:B33new}) for $I_\epsilon'$ in the form
\begin{equation}
    I_\epsilon'(\mu_0',\phi_0',\epsilon_0') = \frac{1}{\gamma(R_\text{top})} \, \frac{1}{1-\mu_{0}' \, \abs{\vel(R_\text{top})/c}} \, U_{i,\text{top}}(\epsilon_0) \, \delta(\mu_0'-\mu_{0i}') \, \delta(\phi_{0}'-\phi_{0j}') \ ,
    \label{eq:B38new}
\end{equation}
which can be combined with Equation~(\ref{eq:B32new}) to obtain
\begin{equation}
    I_\epsilon'(\mu_0',\phi_0',\epsilon_0') = \frac{1}{\gamma(R_\text{top})} \, \frac{1}{1-\mu_{0}' \, \abs{\vel(R_\text{top})/c}} \, \frac{W^\text{top}_{i}\epsilon_0\,\Dot{\mathcal{N}}^\text{tot}_\epsilon(\epsilon_0)}{\Omega R_\text{top}^2} \, \delta(\mu_0'-\mu_{0i}') \, \delta(\phi_{0}'-\phi_{0j}') \ .
    \label{eq:B40new}
\end{equation}
This result expresses the contribution to the co-moving frame intensity made by the {\tt RM88}-frame intensity distribution given by Equation~(\ref{eq:B31new}).

Since the transformation of the radiation field between the {\tt RM88} frame and the co-moving frame is a linear process, we can use Equation~(\ref{eq:B40new}) to separately treat each term in the expansion given by Equation~(\ref{totalintensitydeftop2}). This allows us to compute the total co-moving frame intensity distribution measured at the column top, denoted by $\mathscr{I}_{\epsilon,\text{top}}^\text{tot}(\mu_0',\phi_0',\epsilon_0')$, taking into account the entire grid of $(\mu_{0i},\phi_{0j})$ emission directions in the {\tt RM88} frame. The result obtained is
\begin{align}
    \mathscr{I}_{\epsilon,\text{top}}^\text{tot}(\mu_0',\phi_0',\epsilon_0') = \sum_{i=1}^{N_\text{top}}\sum_{j=1}^{M_\text{top}} &\,\frac{1}{\gamma(R_\text{top})} \, \frac{1}{1-\mu_{0}' \, \abs{\vel(R_\text{top})/c}}\nonumber\\
    &\times \frac{W^\text{top}_{i}\epsilon_0\,\Dot{\mathcal{N}}^\text{tot}_\epsilon(\epsilon_0)}{\Omega R_\text{top}^2} \, \delta(\mu_0'-\mu_{0i}') \, \delta(\phi_{0}'-\phi_{0j}') \ ,
    \label{eq:B41newA}
\end{align}
where the value of $\epsilon_0$ is given by
\begin{equation}
    \epsilon_0 = \epsilon_0' \, \gamma(R_\text{top}) \, (1-\mu_{0}' \, \abs{\vel(R_\text{top})/c}) \ .
    \label{eq:B42new}
\end{equation}
We note that Equation~(\ref{eq:B41newA}) can also be rewritten in the form
\begin{align}
    \mathscr{I}_{\epsilon,\text{top}}^\text{tot}(\mu_0',\phi_0',\epsilon_0') = \sum_{i=1}^{N_\text{top}} &\,\frac{1}{\gamma(R_\text{top})} \, \frac{1}{1-\mu_{0}' \, \abs{\vel(R_\text{top})/c}}\nonumber\\
    &\times \frac{W^\text{top}_{i}\epsilon_0\,\Dot{\mathcal{N}}^\text{tot}_\epsilon(\epsilon_0)}{\Omega R_\text{top}^2} \, \delta(\mu_0'-\mu_{0i}') \, \sum_{j=1}^{M_\text{top}} \delta(\phi_{0}'-\phi_{0j}') \ .
    \label{eq:B41new}
\end{align}
Equation~(\ref{eq:B41new}) will be utilized in our application of the new relativistic model to Her X-1 in Section~\ref{sec:results}.

\section{Flux Components in Frame of Distant Observer}
\label{AppendixC}

The calculation of the contributions to the energy flux spectrum, $F_{\epsilon,\text{obs}}$, measured by a distant observer due to emission from the walls and tops of the two accretion columns is a central result obtained using our formalism. We develop the required closed-form expressions here, and apply them to the analysis of the observational data for Her X-1 in Section~\ref{sec:results}.

\subsection{Column Wall Emission}

First we calculate the phase-dependent observational energy flux spectrum resulting from emission escaping through the column wall, denoted by $F_{\epsilon,\text{obs}}^\text{wall}(\epsilon,\beta)$. By substituting the intensity given by Equation~(\ref{intensitysumwallNEW}) into Equation~(\ref{eq:WallFlux}), we obtain
\begin{align}
    F_{\epsilon,\text{obs}}^\text{wall}(\epsilon,\beta) = \sum_{i=1}^{N_\text{wall}} \sum_{j=1}^{M_\text{wall}} &\int_{0}^{2\pi}\int_{-1}^{1}\frac{\epsilon^3}{\epsilon_0^3} \, \mu D_{\text{wall}}(\mu,\phi;\mu_0,\phi_0)\nonumber\\
    &\times \frac{W_{i}^\text{wall}\epsilon_0\Dot{N}^\text{tot}_\epsilon(R_0,\epsilon_0)}{2\pi R_{0}\sin\thetamax} \, \delta(\mu_{0}-\mu_{0i})\delta(\phi_{0}-\phi_{0j}) \, d\mu_{0} \, d\phi_{0} \ .
    \label{eq:WallFlux1}
\end{align}
The dependence on the rotational phase angle, $\beta$, enters implicitly on the right-hand side because the emission radius, $R_0$, and also the observer-frame photon propagation direction $(\mu,\phi)$ corresponding to the {\tt RM88}-frame emission direction $(\mu_{0i},\phi_{0j})$, all vary as the star rotates, according to the ray-tracing procedure for determining the null geodesics outlined in Section~\ref{sec:colwallgeo}. For geodesics with roots for $R_0$ that are not within the radial extent of the accretion column, the contribution to the integral is zero. The evaluation of the integrals in Equation~(\ref{eq:WallFlux1}) is trivial, and we obtain
\begin{equation}
    F_{\epsilon,\text{obs}}^\text{wall}(\epsilon,\beta) = \sum_{i=1}^{N_\text{wall}} \sum_{j=1}^{M_\text{wall}} W_{i}^\text{wall} \, g_{ij}^\text{wall}(\epsilon,\beta) \ ,
    \label{obsfluxwall1}
\end{equation}
where the expansion functions for the column wall emission, $g_{ij}^\text{wall}(\epsilon,\beta)$, are defined by
\begin{equation}
    g_{ij}^\text{wall}(\epsilon,\beta)\equiv\frac{\epsilon^3}{\epsilon_0^2} \, \mu D_{\text{wall}}(\mu,\phi;\mu_{0i},\phi_{0j}) \,\frac{\Dot{N}^\text{tot}_\epsilon(R_0,\epsilon_0)}{2\pi R_0\sin\thetamax} \ ,
    \label{gfunwall}
\end{equation}
and $\epsilon_0$ is related to $\epsilon$ via Equation~(\ref{eq:redshift1}).

As discussed in Section~\ref{sec:AziIso}, based on the angular dependence of the cyclotron scattering cross sections, we assume that the intensity weight coefficients for the wall emission, $W_{i}^\text{wall}$, are the same for all values of $\phi_{0j}$ in our angular grid. This implies that Equation~(\ref{obsfluxwall1}) can be written in the equivalent form
\begin{equation}
    F_{\epsilon,\text{obs}}^\text{wall}(\epsilon,\beta) = M_\text{wall} \sum_{i=1}^{N_\text{wall}} W_{i}^\text{wall} \, g_{i}^\text{wall}(\epsilon,\beta) \ ,
    \label{obsfluxwall}
\end{equation}
where the set of single-index basis functions, $g_i^\text{wall}(\epsilon,\beta)$, are obtained by performing an average in the azimuthal direction, so that
\begin{equation}
    g_i^\text{wall}(\epsilon,\beta) \equiv \frac{1}{M_\text{wall}} \sum_{j=1}^{M_\text{wall}}g_{ij}^\text{wall}(\epsilon,\beta) \ ,
    \label{gFunAvgWallPreCyc}
\end{equation}
with $M_\text{wall}$ representing the number of $\phi_0$ values in the azimuthal grid for the wall emission, defined in Equation~(\ref{anglegridwall}).

\subsection{Column Top Emission}

The phase-dependent observational energy flux spectrum resulting from emission escaping through the column top, $F_{\epsilon,\text{obs}}^\text{top}(\epsilon,\beta)$, can be computed by substituting the intensity given by Equation~(\ref{totalintensitydeftopNEW2}) into Equation~(\ref{eq:TopFlux}), which yields
\begin{align}
    F_{\epsilon,\text{obs}}^\text{top}(\epsilon,\beta) = \sum_{i=1}^{N_\text{top}} \sum_{j=1}^{M_\text{top}} &\int_{0}^{2\pi}\int_{-1}^{1}\frac{\epsilon^3}{\epsilon_0^3} \, \mu D_{\text{top}}(\mu,\phi;\mu_0,\phi_0)\nonumber\\
    &\times \frac{W_{i}^{\text{top}}\epsilon_0\,\Dot{\mathcal{N}}^\text{tot}_\epsilon(\epsilon_0)}{\Omega R_{\text{top}}^2} \, \delta(\mu_{0}-\mu_{0i})\delta(\phi_{0}-\phi_{0j}) \, d\mu_{0}\, d\phi_{0} \ .
    \label{eq:TopFlux1}
\end{align}
The right-hand side has an implicit dependence on the rotational phase angle $\beta$ because the spin of the star causes a variation in the observer-frame photon propagation direction $(\mu,\phi)$ for a fixed emission direction $(\mu_{0i},\phi_{0j})$ in the {\tt RM88} frame. Carrying out the integrations in Equation~(\ref{eq:TopFlux1}) yields
\begin{equation}
    F_{\epsilon,\text{obs}}^\text{top}(\epsilon,\beta) = \sum_{i=1}^{N_\text{top}} \sum_{j=1}^{M_\text{top}} W^{\text{top}}_{i} \, g_{ij}^{\text{top}}(\epsilon,\beta)
    \label{obsfluxtop1} \ ,
\end{equation}
where the expansion functions for the column top emission, denoted by $g_{ij}^{\text{top}}(\epsilon,\beta)$, are defined by
\begin{equation}
    g^\text{top}_{ij}(\epsilon,\beta)\equiv\frac{\epsilon^3}{\epsilon_0^2} \, \mu D_\text{top}(\mu,\phi;\mu_{0i},\phi_{0j}) \, \frac{\Dot{\mathcal{N}}^\text{tot}_{\epsilon}(\epsilon_0)}{\Omega R_{\text{top}}^2} \ ,
    \label{gfuntop}
\end{equation}
and the value of the {\tt RM88}-frame energy, $\epsilon_0$, is related to the observer-frame energy, $\epsilon$, via Equation~(\ref{eq:redshift1}).

The assumption of azimuthal isotropy introduced in Section~\ref{sec:AziIso} implies that the intensity weight coefficients for the column top emission, $W_{i}^\text{top}$, are the same for all values of $\phi_{0j}$ in the angular grid. Hence, by analogy with Equation~(\ref{obsfluxwall}), we can rewrite Equation~(\ref{obsfluxtop1}) in the equivalent form
\begin{equation}
    F_{\epsilon,\text{obs}}^\text{top}(\epsilon,\beta) = M_\text{top} \sum_{i=1}^{N_\text{top}}
    W^{\text{top}}_{i} \, g_{i}^{\text{top}}(\epsilon,\beta)
    \label{obsfluxtop} \ ,
\end{equation}
where the single-index basis functions for the top emission are defined by performing the azimuthal average given by
\begin{equation}
    g_i^\text{top}(\epsilon,\beta) \equiv \frac{1}{M_\text{top}} \sum_{j=1}^{M_\text{top}}g_{ij}^\text{top}(\epsilon,\beta) \ ,
    \label{gFunAvgTopPreCyc}
\end{equation}
with $M_\text{top}$ denoting the number of $\phi_0$ values in the azimuthal grid for the top emission, defined in Equation~(\ref{anglegridtop}).

Each of the azimuthally-averaged functions $g_i^\text{wall,1}(\epsilon,\beta)$, $g_i^\text{wall,2}(\epsilon,\beta)$, $g_i^\text{top,1}(\epsilon,\beta)$, and $g_i^\text{top,2}(\epsilon,\beta)$ computed using either Equation~(\ref{gFunAvgWallPreCyc}) or (\ref{gFunAvgTopPreCyc}) represents the contribution to the observed X-ray spectrum due to radiation generated in a specific set of emission directions in the {\tt RM88} frame. We will refer to these energy- and phase-dependent functions as ``sub-spectra,'' and they will be used subsequently in our astrophysical applications in two different ways. First, we will we integrate the sub-spectra with respect to the photon energy $\epsilon$ to compute the associated contributions to the pulse profile. Once the expansion coefficients are obtained by fitting the pulse profile data, we will then use the energy-dependent sub-spectra to compute the theoretical prediction for the phase-averaged X-ray spectrum of the source.

\end{appendix}

\clearpage

\bibliographystyle{aasjournal}
\bibliography{bwbiblio}


\clearpage

\end{document}